\documentclass[preprint]{elsarticle}

\usepackage[]{hyperref}
\usepackage[margin=2.0cm]{geometry}
\usepackage{amsmath}
\usepackage{graphicx}
\usepackage{subfigure}
\usepackage{enumitem}
\usepackage{tikz}
\usepackage{fixltx2e} 
\usepackage{textcomp}
\usepackage{float}
\usepackage{setspace}
\usepackage{nth}
\usepackage{accents}
\usepackage{titlesec}
\usepackage{placeins}
\usepackage{algorithmic}
\usepackage{algorithm}

\usepackage{fancyhdr}
\usepackage{nth}
\usepackage{breqn}

\newcommand{\mbb}[1]{\mathbb{#1}}
\newcommand{\mcal}[1]{\mathcal{#1}}
\newtheorem{remark}{Remark}

\newcommand{\scalef}{f_{0}^{\alpha}}

\usepackage{amsfonts}
\usepackage{amssymb}
\usepackage{longtable}
\usepackage{tensor}
\usepackage{color}   

\begin{document}

\begin{frontmatter}
\title{Moment Method for the Boltzmann Equation of Reactive Quaternary Gaseous Mixture}
\author[rwth]{Neeraj Sarna}
\ead{sarna@mathcces.rwth-aachen.de}
\author[ices]{Georgii Oblapenko}
\ead{georgii.oblapenko@austin.utexas.edu}
\author[rwth]{Manuel Torrilhon}
\ead{mt@mathcces.rwth-aachen.de}

\address[rwth]{Center for Computational Engineering, Department of Mathematics, RWTH Aachen University, Schinkelstr 2, 52062, Germany}
\address[ices]{Oden Institute for Computational Engineering and Sciences, The University of Texas at Austin, 2201 E 24th St, Stop C0200, Austin, TX 78712}


\begin{abstract}
We are interested in solving the Boltzmann equation of chemically reacting rarefied gas flows using the Grad's-14 moment method. We first propose a novel mathematical model that describes the collision dynamics of chemically reacting hard spheres. Using the collision model, we present an algorithm to compute the moments of the Boltzmann collision operator. Our algorithm is general in the sense that it can be used to compute arbitrary order moments of the collision operator and not just the moments included in the Grad's-14 moment system. For a first-order chemical kinetics, we derive reaction rates for a chemical reaction outside of equilibrium thereby, extending the Arrhenius law that is valid only in equilibrium. We show that the derived reaction rates (i) are consistent in the sense that at equilibrium, we recover the Arrhenius law and (ii) have an explicit dependence on the scalar fourteenth moment, highlighting the importance of considering a fourteen moment system rather than a thirteen one. Through numerical experiments we study the relaxation of the Grad's-14 moment system to the equilibrium state. 
\end{abstract}

\end{frontmatter}

\section{Introduction}
The Boltzmann equation (BE) accurately models gas flows in all thermodynamic regimes. It governs the evolution of a probability density function that is defined on a seven dimensional space-time-velocity domain. We consider the BE of a gaseous mixture of four mono-atomic gases that react via a chemical reaction given as
\begin{gather}
A_1 + A_2 \leftrightarrows A_3 + A_4.  \label{chemical reaction}
\end{gather}
In a concise form, the BE for such a gaseous mixture can be written as 
\begin{gather}
d_t f_{\alpha}(c_\alpha,t) = \mcal Q_\alpha(f_1(c_1,t),f_2(c_2,t),f_3(c_3,t),f_4(c_4,t)).\label{BE}
\end{gather}
Above, $\alpha\in\{1,2,3,4\}$ is an index for the different gases in the mixture, $f_\alpha$ represents the probability density function of the $\alpha$-th gas, and $c_{\alpha}\in\mbb R^3$ is the molecular velocity. For simplicity, we restrict our study to a spatially homogeneous BE. The collision operator $\mcal Q_\alpha$ models the inter-particle interaction and contains contributions from both the chemical and the mechanical interactions. It is noteworthy that a collision between molecules that can undergo a chemical reaction ($A_1$ and $A_2$, for instance) does not necessarily result in a chemical reaction---the kinetic energy should be large enough to trigger a chemical reaction. Therefore, the part of $\mcal Q_\alpha$ that models mechanical collisions contains interactions between all the gas molecules. The explicit form of $\mcal Q_\alpha$ discussed later provides further clarification.

We want to solve the above equation using a deterministic method i.e., with a method that does not require Monte-Carlo sampling, which introduces an undesirable stochastic noise in the numerical approximation. We consider a Galerkin type approach where we approximate $f_\alpha(\cdot,t)$ in a span of the basis functions $\{\phi_{\alpha,i}\}_{i}$. This provides the approximation
\begin{gather}
f_\alpha(\cdot,t)\approx f_{\alpha,N}(\cdot,t) =\sum_{i=0}^N \alpha_i(t)\phi_{\alpha,i}f_0^{\alpha}. \label{Grad exp}
\end{gather}
We choose $\phi_{\alpha,i}$ as polynomials in the velocity variable $c_\alpha$ and scale them with a Gaussian distribution function $\scalef$---the exact form of $\phi_{\alpha,i}$ and $\scalef$ is discussed later in \autoref{sec: mom apprx}. The above approximation was first proposed by Grad \cite{Grad1949} and has several desirable properties, it (i) preserves the Galelian invariance of the BE, (ii) results in mass, momentum and energy conservation, (iii) at least in the linearized regime, converges to the BE \cite{SarnaJanTor2018,SchmeiserProof,Torrilhon2015}, (iv) results in a hyperbolic moment system under appropriate regularization \cite{Cai2014}, etc.  We refer to the review article \cite{ReviewTor} and the references therein for an exhaustive discussion. 

To compute the approximation $f_{\alpha,N}$, we replace $f_\alpha$ by $f_{\alpha,N}$ in the BE, multiply by the test functions $\{\phi_{\alpha,i}\}_i$ and integrate over the velocity-domain $\mbb R^3$. This results in a time-dependent ordinary-differential-equation (ODE) for the expansion coefficients $\{\alpha_i\}_i$, or the so-called moment equations. To well-define the moment equations, we need integrals of the form 
\begin{gather}
\mcal I_{\alpha}(t) := \int_{\mbb R^3}\phi_{\alpha,i}(c_{\alpha},t) \mcal Q_\alpha(f_1(c_1,t),f_2(c_2,t),f_3(c_3,t),f_4(c_4,t))dc_{\alpha}.
\end{gather}
The main objective of the paper is to compute the above integral. To this end, we study the following three questions related to chemically reactive collisions.
\begin{enumerate}
\item Through which molecular potential do the molecules interact?
\item For a given molecular potential and for binary collisions, how to relate the pre and the post collision relative velocities using mass, momentum and energy conservation?
\item For a given relation between the post and the pre collisional relative velocities, how to compute $\mcal I_\alpha(t)$?
\end{enumerate}
Note that we only consider the contribution from binary collision in the collision operator $Q_\alpha$---a standard assumption in the kinetic gas theory \cite{Struchtrupbook}. Furthermore, all the above questions have been extensively studied in the context of mechanical collisions, see \cite{Chapman1941,Gupta2012,GuptaBinary,Struchtrupbook,Grad1949}.

We use the Direction Simulation Monte Carlo (DSMC) method as a motivation to answer the first question and consider a hard sphere interaction potential for both the chemical and the mechanical interactions \cite{boyd2017nonequilibrium}. Intuitively, the hard sphere potential treats molecules like billiard balls, which interact with each other only when they "touch" each other. To answer the second question, we assume that the collisions are symmetric that is, the pre collisional relative velocity changes in equal proportions along the normal and the tangential direction of collision. We refer to \autoref{sec: bin col} for a precise relation between the pre and the post collisional relative velocities. An answer to the third question follows by extending the framework for a mono-atomic binary gas mixture (proposed in \cite{GuptaBinary,Gupta2012}) to a mixture of chemically reacting gases. 

To the best of our knowledge, only the work in \cite{Bisi2002} proposes an algorithm to compute the integral $\mcal I_\alpha(t)$. It differs from our work in the following sense. Authors in \cite{Bisi2002} accommodate the chemically reacting molecules at the Maxwell-Boltzmann distribution function corresponding to the thermal equilibrium. This largely simplifies the expression for $\mcal I_\alpha(t)$, making the computations simpler.  In contrast, we do not make any such assumption. Our framework allows for chemical reactions that occur outside of a thermal equilibrium, which is of particular interest for rarefied gas flows. Here, we emphasis that a thermal equilibrium is different from a chemical equilibrium. In a thermal equilibrium, the probability density function is a Maxwell-Boltzmann distribution function but the number densities do not need to be related by the law of mass action, see \cite{Kremerbook,Bisi2002} or \autoref{sec: num exp} for details. Furthermore, authors in \cite{Bisi2002} restrict to a Grad's-13 approximation i.e., to a particular choice of $N$ in $f_{\alpha,N}$. The framework we propose is not restricted to a particular $N$. However, for demonstration purposes, we focus particularly on the Grad's-14 approximation. 

Let us mention that in DSMC computations, simple scattering models are used to simulate chemically reactive collisions~\cite{boyd2017nonequilibrium}. For instance, see the isotropic variable hard sphere model (VHS) or the anisotropic variable soft sphere model (VSS)~\cite{Bird}. The DSMC literature focuses largely on incorporating effects of preferential dissociation from high-energy internal states into the chemical cross-section models~\cite{gimelshein2019bird}. Moreover, it has been argued that the microscopic details of chemically reactive collisions do not have a significant influence on the flow dynamics~\cite{lilley2004macroscopic}. However, to realize/well-define a deterministic method, it is imperative to compute the integral $\mcal I_\alpha(t)$, which requires the microscopic details of chemically reactive collisions. Our work proposes one possible way to define these microscopic details.

The BE of chemically reacting gases has not received much attention from a numerical approximation standpoint. Nevertheless, several works in the previous decades have contributed to its theoretical understanding. For the sake of completeness, we briefly recall some of these works. One can show that similar to a single mono-atomic gas, the BE exhibits the \textit{H}-theorem \cite{rykov1972kinetic,Rossani1999}. Furthermore, the equation has a unique equilibrium that is stable and can be given in terms of the Maxwell-Boltzmann distribution function \cite{Rossani1999}. Using the Chapman-Enksog expansion, one can also understand the behaviour of the transport coefficients of the gas mixture \cite{Silva2007}. Furthermore, with a generalized Chapman--Enksog method, one can provide viscous corrections to the non-equilibrium process rates and study the effect of flow compressibility on the reaction rate coefficients \cite{kustova2014chemical,kustova2016mutual}.

\section{Dynamics of reactive binary collisions}\label{sec: bin col}
We start with discussing the dynamics of reactive binary collisions. The dynamics of mechanical binary collisions is well studied in the literature \cite{Chapman1941} and is a special case of reactive collision---details included in \autoref{rem: mechanical col}. With 
$m_{\alpha}$ we denote the mass of the molecule $A_\alpha$ and with $\epsilon_{\alpha}$
we denote its energy of formation. We consider chemical reactions of the type $ A_1 + A_2 \leftrightarrows A_3 + A_4.$
With $\epsilon_f$ and $\epsilon_r$ we represent the activation energy of the forward and the reverse reaction, respectively. The heat of the reactions is represented by $Q$ and reads $$Q = \epsilon_f - \epsilon_r = \left(\epsilon_3 + \epsilon_4\right) - (\epsilon_1 + \epsilon_2).$$
\subsection{Mass, momentum and energy conservation}	\label{laws}
For brevity, we only consider the forward reaction---the relations for the reverse reaction are similar. Throughout the article, we ignore the angular momentum of the molecules. Consider two molecules with masses $m_1$ and $m_2$ travelling with
 velocities $c_1$ and $c_2$, respectively. The molecules collide to form two molecules with masses $m_3$ and $m_4$, respectively, and with the post collisional velocities $c_3$ and $c_4$, respectively. We define the pre and the post collisional relative velocities as
\begin{gather}
g_{12} = c_1 - c_2,\quad g_{34} = c_3 - c_4. \label{def g}
\end{gather}
We assume that $A_1$ and $A_2$ have enough kinetic energy to trigger a chemical reaction i.e.,
\begin{align}
\frac{m_{12}g_{12}^2}{2}\geq \epsilon_f,\quad \frac{m_{34}g_{34}^2}{2}\geq \epsilon_r, \label{collide cond}
\end{align}
where $m_{\alpha\beta} = \frac{m_{\alpha}m_{\beta}}{m_{\alpha}+m_{\beta}}$. 

Since mass, momentum and energy are conserved in a binary collision, we find \cite{Kremerbook}
\begin{equation}
\begin{gathered}
m_1+m_2=m_3+m_4,\quad  m_1c_1 + m_2c_2=m_3c_3+m_4c_4,\quad 
 \frac{m_{12}|g_{12}|^2}{2} = \frac{m_{34}|g_{34}|^2}{2} + Q.	 \label{balance laws}
\end{gathered}
\end{equation}
Above, $|\cdot|$ represents the Eucledian norm of a vector. For later convenience, we use the energy balance to derive the following two relations
\begin{gather}
|g_{34}|^2=\hat Q_{12}g_{12}^2, \quad |g_{12}|^2=\hat Q_{34}|g_{34}|^2, \label{energy balance molecular 2}
\end{gather}
where
\begin{align}
\hat Q_{12} = \frac{m_{12}}{m_{34}}\left(1-\frac{2 Q}{m_{12}g_{12}^2}\right),\quad 
\hat Q_{34} = \frac{m_{34}}{m_{12}}\left(1+\frac{2 Q}{m_{34}g_{34}^2}\right). \label{def Q}
\end{align}
\begin{remark}
The constraint on the kinetic energies to trigger a chemical reaction given in \eqref{collide cond} does not take into account the geometry of collision---see \cite{Kremerbook,Light1969,Present}. Taking these geometrical effects into account is beyond the scope of the present work and we hope to discuss them in one of our future works. 
\end{remark}
\subsection{Hard sphere interaction potential}	\label{interaction_potential}
With $r_{\alpha\beta}$ we represent the distance between the centres of the two molecules $A_\alpha$ and $A_\beta$. We assume that these two molecules interact via a molecular potential  $\varPsi_{\alpha\beta}\left(r_{\alpha\beta}\right)$ that depends solely on $r_{\alpha\beta}$. We consider a hard sphere interaction potential given as
\begin{align}
 \varPsi_{\alpha\beta}\left(r_{\alpha\beta}\right) = \left\{
      \begin{array}{lr}
       0 & \forall r_{\alpha\beta} > r^{\min}_{\alpha\beta}\\
       \infty & \forall r_{\alpha\beta} \leq r^{\min}_{\alpha\beta},
       \end{array}
      \right.
\end{align}
where $r_{\alpha\beta}^{\min}$ is the minimum distance between the 
two molecules. When molecules undergo a mechanical collision, $r_{\alpha\beta}^{\min}$ is the average of the molecular diameters $d_\alpha$ and $d_\beta$ i.e., $r^{\min}_{\alpha\beta} = \frac{1}{2}\left(d_\alpha + d_\beta\right).$
For the molecules that undergo a chemical collision, 
we define $r_{\alpha\beta}^{\min}$ separately for the 
forward and the reverse reaction as 
\begin{gather}
r_{12,f}^{\min} = d_f,\quad r_{34,r}^{\min} = d_r. \label{def d}
\end{gather}
Here, $d_f$ and $d_r$ are the reactive diameters for the forward 
and the reverse reactions, respectively. Similar to \cite{Atkins}, we assume that $d_f$ and $d_r$ are 
linearly related to $d_{12}$ and $d_{34}$ in the following way $ d_f = s_f d_{12}$ and $d_r = s_r d_{34}
$, 
where $s_f$ and $s_r$ are the so called steric factors---a relation between $s_f$ and $s_r$ follows from the law of mass action and is given later in \eqref{relation steric}. 

\begin{figure}[ht!]
\begin{center}
\includegraphics[width=3in]{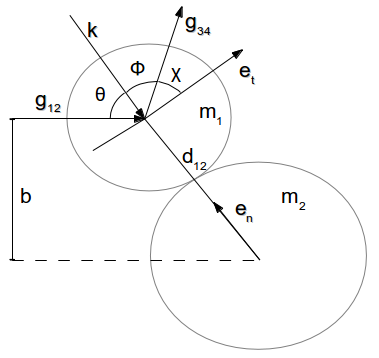} 
\end{center}
\hfill
\caption{Binary collision for a hard sphere interaction potential. $g_{12}$: relative pre-collisional velocity, $g_{34}$: relative post-collisional velocity, $b$: impact parameter, $k$: a unit vector that passes through the centres of the two colliding spheres, $e_n$: normal direction of collision, and $e_t$: tangential direction of collision.} \label{binary_collision}
\end{figure}

For clarity of the following discussion, we visualize a collision between two hard spheres in \autoref{binary_collision}. The vector $k$ is a unit vector (i.e., $|k| = 1$) that passes through the line joining the centres of the two interacting molecules and $b$ is the so-called impact parameter.
The angles $\theta$ and $\chi$ are
the angle-of-attack and the scattering angle, respectively. 
\subsection{Velocity transformations}\label{transform vel}
The aim of the following discussion is to express the post-collisional velocities $c_3$ and $c_4$ in terms of the pre-collisional $c_1$ and $c_2$ ones and vice-versa. We will use these relations later to find the moments of the collision operators given in \eqref{BE}. We start with relating the relative velocity $g_{12}$ to $g_{34}$. 

\subsubsection{Relating relative velocities: restitution coefficients} 
Let $e^n$ and $e^t$ be two orthonormal unit-vectors that define the normal and the tangential direction of collision, respectively. Both of these vectors are shown in \autoref{binary_collision} and can be related to the vector $k$ via
$e^n  = -k$ and $e^t = (-k)_{\bot}$. Let $g_{12}^n = \left(g_{12}\cdot e^n\right) e^n$ and $g_{12}^t = \left(g_{12}\cdot e^t\right) e^t$. Here, $u\cdot v$ represents an Eucledian inner-product between two vectors $u$ and $v$. Similarly, define $g_{34}^n$ and $g_{34}^t$. Since the span of $e^n$ and $e^t$ approximates every vector in the collision plan, using the orthogonality of $e^n$ and $e_t$, we find 
\begin{align}
g_{12} = g_{12}^{n} + g_{12}^t, \quad g_{34} = g_{34}^n + g_{34}^t. \label{decomposition g}
\end{align}
Let $\mcal E_n$ and $\mcal E_t$ be the coefficients of restitution along the normal and the tangential direction, respectively, defined as
\begin{align}
g^{n}_{34} = \mcal E_n g^{n}_{12}, \quad  
g^{t}_{34} = \mcal E_t g^{t}_{12}.	\label{def restitution}
\end{align}

We want to find a relation for $\mcal E_n$ and $\mcal E_t$ in terms of the pre/post collisional velocities $g_{12}/g_{34}$ and the heat of the reaction $Q$. One possibility is to use the Hertzian contact theory of rough inelastic hard spheres \cite{Brilliantov1996}. However, involvement of chemical 
reactions makes the application of Hertzian contact theory complicated and, as yet, it is unclear how to extend this contact theory to chemically reacting molecules. In the present work, we propose to compute $\mcal E_n$ and $\mcal E_t$ by assuming that the pre-collisional relative velocity $g_{12}$ changes in equal proportions along the normal and the tangential direction of collision. Equivalently, 
\begin{gather}
\mcal E_n = \mcal E_t = \mcal E. \label{assume symmetry}
\end{gather}
To compute $\mcal E$, we first note that $|g_{34}|^2 = \mcal E^2|g_{12}|^2$. Then, the energy balance given in \eqref{energy balance molecular 2} provides 
\begin{gather}
\mcal E = \sqrt{\hat Q_{12}}.  \label{def epsilon}
\end{gather}
Note that we have only considered the positive root in the above expression because $\mcal E_n$ cannot be negative. Also note that $\hat Q_{12} \geq 0$ because of the constraint on the kinetic energy \eqref{collide cond}. 

We elaborate on the above assumption \eqref{assume symmetry}. A collision between chemically reactive molecules results in a destruction or a formation of chemical bonds. This results in (i) exchange of reaction heat that is quantified by $Q$, and (ii) a change in mass of the interacting molecules. As per the above assumption, both these effects influence the pre collisional relative velocity $g_{12}$ equally along the normal and the tangential direction of collision. Equivalently, there is no preferred direction of motion along which a chemical reaction changes the pre-collisional relative velocity differently.

A collision between two rough inelastic hard spheres also results in a change in both the tangential and the normal pre-collisional relative velocity. 
We discuss the differences of our collision model to the collision model of a rough inelastic hard sphere. We refer to \cite{Kremer2011a} for a detailed discussion on a rough inelastic hard sphere.
\begin{enumerate}
\item For an inelastic rough hard sphere, two different mechanisms change the pre-collisional relative velocity---friction changes the tangential velocity and the inelasticity changes the normal velocity. Since both these mechanisms are different, usually, $\mcal E_n\neq \mcal E_t$. In contrast, for a chemically reactive collision, a single mechanism---the chemical reaction---is responsible for a change in both the normal and the tangential pre-collisional velocity, which motivates our assumption $\mcal E_n = \mcal E_t$.
\item For a rough inelastic hard sphere, if the friction is sufficiently large then $\mcal E_t$ can be negative, resulting in a change in both the tangential and the normal pre-collisional velocity. In our model $\mcal E_t$ is positive thus, the pre-collisional relative velocity can only change in magnitude. 
 \end{enumerate} 

\subsubsection{Relating pre and post collisional velocities} \label{velocity transformations}
We now express $c_3/c_4$ in terms of $c_1/c_2$. We first summarise our results, and devote the rest of the discussion in proving it. The post-collisional velocities $c_{3/4}$ are related to the pre-collisional ones $c_{1/2}$ via
\begin{align}
 &c_3 = c_1 + g_{12}\left(\mu_{43}\sqrt{\hat Q_{12}}-\mu_{21}\right)-2\mu_{43}\sqrt{\hat Q_{12}}\left(k\cdot g_{12}\right)k,	\label{expression for C3 interms of C1}\\
 &c_4 = c_2 - g_{12}\left(\mu_{34}\sqrt{\hat Q_{12}}-\mu_{12}\right)+2\mu_{34}\sqrt{\hat Q_{12}}\left(k\cdot g_{12}\right)k. \label{expression for C4 interms of C1}
\end{align}
Above, $k$ is the unit vector shown in \autoref{binary_collision} and $\hat Q_{12}$ is as given in \eqref{energy balance molecular 2}. Furthermore, $\mu_{\alpha\beta}$ is the mass fraction defined as 
\begin{gather}\mu_{\alpha\beta} = m_{\alpha}/(m_\alpha+m_\beta).\label{mass fraction}
\end{gather}
 Expressions for $c_1$ and $c_2$ can be derived in a similar fashion and are given as
 \begin{align}
c_1 = c_3 + g_{34}\left(\mu_{21}\sqrt{\hat Q_{34}}-\mu_{43}\right)+2\mu_{21}\sqrt{\hat Q_{34}}\left(k\cdot g_{34}\right)k,\label{expression for C1 interms of C3}\\ 
c_2 = c_4 - g_{34}\left(\mu_{12}\sqrt{\hat Q_{34}}-\mu_{34}\right)-2\mu_{12}\sqrt{\hat Q_{34}}\left(k\cdot g_{34}\right)k.
\label{expression for C2 interms of C3}
\end{align}

We now derive the above relations. First recall that the angles $\theta$, $\phi$ and $\chi$ are as shown in \autoref{binary_collision}. Since we assume that the collision is symmetric (i.e., \eqref{assume symmetry} holds) we find that (i) $\theta=\phi$, (ii) $\chi=\pi-2\theta$, and (iii) $k$ is a vector that bisects $g_{12}$ and $g_{34}$ and is given as
\begin{align}
k =  \frac{\left(\sqrt{\hat Q_{12}}g_{12}-g_{34}\right)}{\left|\sqrt{\hat Q_{12}}g_{12}-g_{34}\right|}. \label{def k}
\end{align}
Taking a dot product of the above equation with $\sqrt{\hat Q_{12}}g_{12}-g_{34}$,
we obtain the following relation
\begin{align}
k\cdot \left(\sqrt{\hat Q_{12}}g_{12}-g_{34}\right) =  \left |\sqrt{\hat Q_{12}}g_{12}-g_{34}\right|	\label{dotk g12 g34}
\end{align}
Since $k$ bisects $g_{12}$ and $g_{34}$, we also have
\begin{align}
k\cdot g_{34} = -|g_{34}|\cos(\phi) = -\sqrt{\hat{Q}_{12}}|g_{12}|\cos(\theta) = -\sqrt{\hat{Q}_{12}}k\cdot g_{12} \label{dotk g34}.
\end{align}
Substituting \eqref{dotk g34} into \eqref{dotk g12 g34} we have 
\begin{align}
 \left|\sqrt{\hat Q_{12}}g_{12}-g_{34}\right| = 2\sqrt{\hat{Q}_{12}} k\cdot g_{12}
\end{align}
Substituting the above relation into the definition of $k$ given in \eqref{def k} we have
\begin{align}
g_{34} = \sqrt{\hat{Q}_{12}} g_{12}-2\sqrt{\hat{Q}_{12}}\left(k\cdot g_{12}\right) k.	\label{relation g34}
\end{align}
The velocity of the center of mass of the system, namely $h$, is given
as 
\begin{align}
h = \mu_{12}c_1 + \mu_{21}c_2  = \mu_{34}c_3 + \mu_{43}c_4 \label{def h}
\end{align}
Note that the second equality is a result of the momentum conservation given in \eqref{balance laws}. Using the center of mass velocity $h$ and the post collisional relative velocity $g_{34}$, the post-collisional velocities $c_3$ and $c_4$ can be written as
\begin{align}
c_3 = h+ \mu_{43} g_{34} \quad c_4 = h-\mu_{34} g_{34}, \label{relation c3 c4 h}
\end{align}
which is equivalent to
\begin{align}
  c_3 =\mu_{12}c_1 + \mu_{21}c_2 + \mu_{43} g_{34}, \quad c_4 = \mu_{12} c_1 + \mu_{21} c_2-\mu_{34}g_{34}. 
\end{align}
Finally, using \eqref{relation g34} in the above expression, we express $g_{34}$ in terms of $g_{12}$ to find the desired result.

\begin{remark}\label{rem: mechanical col}
For consistency, we show that the velocity transformations for a single mono-atomic gas and a binary mixture of (chemically neutral) mono-atomic gases follow from the velocity transformations of a chemically reactive gas mixture.
For a single gas and for all $\alpha\in\{1,2,3,4\}$, we have $ m_\alpha = m$ and $Q = 0$. 
As a result, the velocity transformations \eqref{expression for C3 interms of C1}--\eqref{expression for C2 interms of C3}, change to $c_3 = c_1-\left(k\cdot g_{12}\right)k$ and $c_4 = c_2 +\left(k\cdot g_{12}\right) k$. For a binary mixture, we have $m_1 = m_3,$ $m_2 = m_4,$ and $Q = 0$.
Using these physical parameters, the velocity transformations change to $
c_3 = c_1-2\mu_{21}\left(k\cdot g_{12}\right)k$ and $c_4 = c_2 +2\mu_{12}\left(k\cdot g_{12} 
\right)k$. One can easily check that these relations are the same as those given in \cite{Chapman1941,Gupta2012}. 
\end{remark}
\section{Moment Approximation}\label{sec: mom apprx}
We present our moment approximation for the BE given in \eqref{BE}. We start with giving the explicit form of the collision operator appearing in the BE \eqref{BE}.
\subsection{Collision operator}
We split the collision operator $\mcal Q_\alpha$ as 
\begin{gather}
\mcal Q_\alpha(f_1,f_2,f_3,f_4) = \sum_{\beta=1}^4 \mcal Q_{M}(f_\alpha,f_\beta) + \mcal Q_{\alpha,R}(f_1,f_2,f_3,f_4). \label{splitQ}
\end{gather}
Above, $\mcal Q_{M}(f_\alpha,f_\beta)$ models the mechanical collisions between $A_\alpha$ and $A_\beta$, and $\mcal Q_{\alpha,R}$ models the chemically reactive collisions. The expression for $\mcal Q_M$ is the same as that for a single gas and reads \cite{Chapman1941}
\begin{align}
\mcal Q_M(f_\alpha,f_\beta) = \int_{\mbb R^3}\int^{2\pi}_{0}\int_{0}^{\pi} \left(f_{\alpha}^{'}f^{'}_{\beta}-f_{\alpha}f_{\beta}\right)g_{\alpha\beta}\sigma_{\alpha\beta}\sin(\chi)d\chi d\epsilon d c_{\beta},	\label{expressionI}
\end{align}
where $f_{\alpha}^'$ and $f_{\beta}^{'}$ represent the phase densities corresponding to the post collisional 
velocities $c_{\alpha}^{'}$ and $c_{\beta}^'$, respectively, $\epsilon$ represents the angle made by the collisional plane, and $\sigma_{\alpha\beta}$ represents the differential cross-section. For a hard sphere interaction potential, $\sigma_{\alpha\beta}$ is given as \cite{Chapman1941,GuptaBinary}
$$
\sigma_{\alpha\beta} = \frac{1}{4}d^2_{\alpha\beta},\quad d_{\alpha\beta} = \frac{1}{2}(d_\alpha + d_\beta).
$$

The operator $\mcal Q_{\alpha,R}$ that accounts for chemical reactions can be given as \cite{boyd2017nonequilibrium,Kremerbook}
\begin{equation}
\begin{aligned}
\mcal Q_{\alpha,R}(f_1,f_2,f_3,f_4)d c_{\alpha} = &\nu_{\alpha} \int_{\mbb R^3}\int_{\mbb R^3}\int^{2\pi}_{0}\int_{0}^{\pi} f_1 f_2 g_{12} \sigma_{12}^f \sin (\chi)d\chi d\epsilon d c_1 d c_2\\
& - \nu_{\alpha} \int_{\mbb R^3}\int_{\mbb R^3}\int^{2\pi}_{0}\int_{0}^{\pi} f_3 f_4 g_{34} \sigma_{34}^r \sin(\chi)d\chi d\epsilon d c_3 d c_4, \label{expressionR}		
\end{aligned}
\end{equation}
where $\nu_{\alpha}$ is a stoichiometric coefficient given as 
$-\nu_1 = -\nu_2 = \nu_3 = \nu_4 = 1.$
Furthermore, $\sigma_{12}^f$
and $\sigma_{34}^r$ are the differential cross-sections corresponding to 
the forward and the reverse reaction, respectively. For a hard sphere interaction potential, both these cross-sections read
\begin{align}
 \sigma_{12}^f = &U\left(1-\frac{2\epsilon_f}{m_{12}g_{12}^2}\right)\frac{d_f^2}{4}\left(1-\frac{2\epsilon_f}{m_{12}g_{12}^2}\right), 	\label{sigma12}\\
 \sigma_{34}^r = &U\left(1-\frac{2\epsilon_r}{m_{34}g_{34}^2}\right)\frac{d_r^2}{4}\left(1-\frac{2\epsilon_r}{m_{34}g_{34}^2}\right).		\label{sigma34}
\end{align}
Above, $U(x)$ is a unit-step function that is one for $x \geq 0$ and zero otherwise. Furthermore, $d_f$ and $d_r$ are as given in \eqref{def d}.

\subsection{Definition of moments}
Let $i\in\mbb R^N$ denote a multi-index with each entry being a natural number smaller than the velocity space dimension $d$. Using $i$ and a scalar $a\in\mbb N$, we define a polynomial over $\mbb R^d$ as
\begin{gather}
\phi_{a,\langle i_1\dots i_N\rangle}(y) =  |y|^{2 a}y_{\langle i_{1}}\dots y_{i_{N}\rangle},\quad y\in\mbb R^d.\label{def phi complex}
\end{gather}
The tensor $y_{\langle i_1} \dots y_{i_N\rangle}$ represents the trace free part of $y_{i_1}\cdots y_{i_N}$ and its explicit form can be found in \cite{Struchtrupbook}. For completeness, we give this explicit form for $N=2$
\begin{gather}
B_{\langle i_1i_2\rangle} = \frac{1}{2}\left( B_{i_1i_2} + B_{i_2i_1}\right)-\frac{1}{3}\delta_{i_1 i_2}B_{i_ki_k},
\end{gather}
where $\delta_{i_1 i_2}$ is a Kronecker delta. Throughout the article, we use the Einstein's summation convention for the tensor notation.

Using $\phi_{a,\langle i_1\dots i_N\rangle}$, we define a general $(2a + N)$-th order moment of $f_{\alpha}$ as 
\begin{gather}
w^{\alpha}_{a,i_1\dots i_N}(t) = m_\alpha\int_{\mbb R^d}\phi_{a,\langle i_1\dots i_N\rangle}(C_\alpha)f_{\alpha}(c_\alpha,t)dC_\alpha.
\end{gather}
Above, $C_\alpha$ is the so-called peculiar velocity given as $C_{\alpha}=c_\alpha-v$. Here, $v$ is the macroscopic velocity of the gas mixture defined in the following way. Let the density $\rho_\alpha$ and the macroscopic velocity $v_{\alpha}$ of gas-$\alpha$ be defined as 
\begin{gather}
\rho_\alpha = w^{\alpha}_{0,0} ,\quad 
\rho_\alpha v^\alpha_{i_1}  = w^{\alpha}_{0,i_1} . 
\end{gather}
We define the mixture velocity $v$ and the mixture density $\rho$ as 
\begin{gather}
\rho = \sum_{\alpha=1}^4\rho_\alpha,\quad \rho v = \sum_{\alpha=1}^4 \rho_\alpha v_\alpha.
\end{gather}

The other moments of $f_\alpha$ that have a physical relevance are given as
\begin{equation}
\begin{gathered}
\frac{3}{2}k T_\alpha = \frac{3}{2}\rho_{\alpha} \theta_{\alpha} =\frac{1}{2}w^{\alpha}_{1,0}\quad 
 \sigma_{i_1i_2}^{\alpha}  = w^{\alpha}_{0,i_1i_2},\quad q^{\alpha}_{i_1} =\frac{1}{2}w^{\alpha}_{1,i_1}. \label{mom macro}
\end{gathered}
\end{equation}
Above, $\theta_\alpha = kT_\alpha/m_\alpha$ is the temperature in energy units with $k$ being the Boltzmann's constant, $\sigma_{i_1i_2}^{(\alpha)}$ is the stress-tensor and $q^{(\alpha)}_{i_1}$ is the heat flux. For convenience, we further define 
$$\Delta_\alpha = w^{\alpha}_{2,0}-15\rho_\alpha\theta_\alpha^2.$$
We also define the temperature $\theta$ and the number density $n$ of the gas mixture as
$$
kT = \rho\theta = \sum_{i=1}^4\rho_\alpha\theta_\alpha,\quad n=\sum_{i=1}^4n_\alpha,
$$
where $n_{\alpha}=\rho_\alpha/m_\alpha$.

\begin{remark}
Note that due to mass, momentum and energy conservation, the mixture velocity $v$ is constant over time resulting in a time-independent peculiar velocity $C_\alpha$. 
\end{remark}
\subsection{The fourteen moment system}
We are interested in solving for only the $14$ moments contained in the set 
$$w_\alpha^{[14]}=\{w^\alpha_{0,0},w^\alpha_{0,i_1},\frac{1}{2} w^\alpha_{1,0},w^\alpha_{0,i_1i_2},\frac{1}{2} w^\alpha_{1,i_1},w^\alpha_{2,0}\}.$$ 
Note that as per the above relations, this set contains one component for $\rho_\alpha$, three for $v_\alpha$, one for $\theta_\alpha$, six for $\sigma_\alpha$, three for $q_\alpha$ and one for $w^\alpha_{2,0}$---thus fourteen in total. 
Given the moment vector $w_\alpha^{[14]}$, we first want to approximate the distribution function $f_\alpha$. We consider the Grad's-14 (G14) moment approximation and approximate $f_\alpha$ by \cite{Grad1949,Kremer2011a}
\begin{align}
f_\alpha\approx f_{\alpha|G14} = f_0^{\alpha}\left[1+\frac{\Delta_{\alpha}}{8\rho_{\alpha}\theta_{\alpha}^2}\left(1-\frac{2}{3}\frac{|C_{\alpha}|^2}{\theta_{\alpha}}+\frac{1}{15}\frac{|C_{\alpha}|^4}{\theta_{\alpha}^2}\right)+\frac{q^{(\alpha)}_{i_1}C^{(\alpha)}_{i_1}}{5\rho_{\alpha}\theta_{\alpha}^{2}}\left(\frac{|C_{\alpha}|^2}{\theta_{\alpha}}-5\right)\right.\nonumber\\
 \left.+\frac{\sigma_{i_1i_2}^{(\alpha)}}{2\rho_{\alpha}\theta_{\alpha}^2}C_{i_1}^{(\alpha)}C_{i_2}^{(\alpha)}+\frac{u^{(\alpha)}_{i_1}C^{(\alpha)}_{i_1}}{\theta_{\alpha}}\right]. \label{Grad14}
 \end{align}
The function $f_0^{\alpha}$ is a Gaussian-distribution function defined as
\begin{align}
f_0^{\alpha} = n_\alpha \left(\frac{1}{2\pi\theta_{\alpha}}\right)^{\frac{d}{2}}\exp\left(-\frac{|C_{\alpha}|^2}{2\theta_{\alpha}}\right). 
\end{align}

To derive a governing equation for the moments $w_\alpha^{[14]}$, we replace $f_\alpha$ by $f_{\alpha|G14}$ in the BE \eqref{BE}, multiply the resulting equation by appropriate functions $\phi_{a,\langle i_1\dots i_N\rangle}(C_\alpha)$ and integrate over the velocity domain $\mbb R^d$. For all $\alpha\in\{1,2,3,4\}$, this results in the moment equations
\begin{equation}
\begin{gathered}
d_t n_{\alpha} = \mcal P^{\alpha}_{R,0},\quad  d_t \left(n_\alpha v^{\alpha}_{i_1}\right) = \mcal P^{\alpha}_{M,i_1} + \mcal P^{\alpha}_{R,i_1},\\ 
 \frac{3}{2}d_t  (n_{\alpha} T_{\alpha}) = \mcal P^{\alpha}_{M,1} + \mcal P^{\alpha}_{R,1},\quad   d_t \sigma^{(\alpha)}_{i_1i_2} = \mcal P^{\alpha}_{M,i_1i_2} + \mcal P^{\alpha}_{R,i_1i_2},\\
  d_t q^{(\alpha)}_{i_1} = \mcal P^{\alpha}_{M,1,i_1} + \mcal P^{\alpha}_{R,1,i_1},\quad 
 d_t u^{2(\alpha)} = \mcal P^{\alpha}_{M,2} + \mcal P^{\alpha}_{R,2}.  \label{mom eq explicit}
\end{gathered}
\end{equation}
The different $\mcal P^{\alpha}$'s are the moments of the collision operator defined as 
\begin{equation}
\begin{aligned}
\mcal P^{\alpha}_{M,a,i_1\dots i_N} = &m_\alpha\sum_{\beta=1}^4\int_{\mbb R^d} \phi_{a,\langle i_1\dots i_N\rangle}Q_M(f_{\alpha|G14},f_{\beta|G14})dC_\alpha,\\
 \mcal P^{\alpha}_{R,a,i_1\dots i_N} = &m_\alpha \int_{\mbb R^d} \phi_{a,\langle i_1\dots i_N\rangle} Q_{\alpha,R}(f_{1|G14},f_{2|G14},f_{3|G14},f_{4|G14})dC_\alpha.
\end{aligned}
\end{equation}
Above,  $Q_M$ and $Q_{\alpha,R}$ are the mechanical and the chemical parts of the collision operator given in \eqref{splitQ}. Note that the mechanical collisions do not contribute into the mass balance equation.

\begin{remark}
Strictly speaking, the Grad's-14 moment approximation is a special case of the approximation $f_{\alpha |G14}$ given in \eqref{Grad14}. In $f_0^\alpha$, if we replace the "local" gas temperature $\theta_\alpha$ by the temperature of the mixture $\theta$, we find the Grad's-14 moment approximation. We expect $f_{\alpha |G14}$ to be more accurate than a Grad's-14 moment approximation because it adapts better to the changes in the "local" temperature of the gas-$\alpha$. 
\end{remark}

\begin{remark}
The following two observations motivate our choice of a fourteen moment system rather than a thirteen one. Firstly, out of all the fourteen moments, only the non-equilibrium moment $\Delta_\alpha$ (and not the stress tensor and the heat flux) influences the reaction rates, which can result in a better approximation of the reaction rates outside of the equilibrium---see \eqref{explicit kf} and \eqref{explicit kr} derived later for explicit reaction rates. We emphasise that for a thirteen moment system, none of the non-equilibrium moments will appear in the reaction rates. Secondly, as discussed earlier, we treat a chemically reacting mixture similar to a granular gas in the sense that we allow both the normal and the tangential pre-collisional velocities to change. Previous works have shown that including $\Delta_\alpha$ in the set of moments better captures the cooling rate and the transport coefficients of a granular gas \cite{Kremer2011a,Risso2002}. We expect a similar result to hold true for a chemically reactive mixture. Our numerical experiments corroborate our expectations by demonstrating that non-equilibrium states where $\Delta_\alpha$ is initially zero can trigger non-zero perturbations in $\Delta_\alpha$, while all the other non-equilibrium moments remain zero. 
\end{remark}
\subsection{Moments of the collision operator}
To compute the moments of $\mcal Q_M$ we first derive moments of $\mcal Q_{\alpha,R}$ and interpret the mechanical collisions as a special case of the chemical collisions by using the physical parameters given in \autoref{rem: mechanical col}. This results in the same moments as those given in \cite{Gupta2012,GuptaBinary}. For brevity, we do not discuss these moments here again. 

To compute the moments of $\mcal Q_{\alpha,R}$, we restrict ourselves to $\alpha=1$, since computations for all the other gas components are the same. First, we decompose $\mcal Q_{1,R}$ into a gain $\mcal Q^g_{1,R}$ and a loss part $\mcal Q^f_{1,R}$ as
\begin{equation}
\begin{aligned}
\mcal Q_{1,R}(f_{1|G14},f_{2|G14},f_{3|G14},f_{4|G14})d C_{1} = &\underbrace{\int  f_{3|G14} f_{4|G14} g_{34} \sigma_{34}^r \sin(\chi)d\chi d\epsilon d C_3 d C_4}_{\mcal Q_g dC_1}\\
&\underbrace{-\int f_{1|G14} f_{2|G14} g_{12} \sigma_{12}^r \sin (\chi)d\chi d\epsilon d C_1 d C_2}_{=\mcal Q_l dC_1}.
\end{aligned}
\end{equation}
It is easy to conclude that $\mcal Q_g$ and $\mcal Q_l$ results in a gain and a loss in $f_1$, respectively. Note that because $f_{1|G14}$ is expressed in terms of $C_1$, we have changed the integration variable from $c_1$ to $C_1$. To perform the change we have used the relation $dc_1 = dC_1$, which immediately follows from the definition of $C_1$. 

To simplify the following notations, rather than computing the moments of $\mcal Q_{\alpha,R}$ with respect to the polynomial $\phi_{a,\langle i_1\dots i_N\rangle}$ given in \eqref{def phi complex}, we compute the moments with respect to the monomial $\phi_{i_1\dots i_N}$ defined as
\begin{gather}
\phi_{i_1\dots i_N}(C_\alpha) = C_{i_1}^{\alpha}\dots C_{i_N}^{\alpha}. \label{def phi}
\end{gather}
Using the definition of $C_{\langle i_1}\dots\cdot C_{i_N\rangle}$ and expressing $|C_\alpha|^{2a}$ in terms of the different components of $C_\alpha$, one can conclude that for any $a\in\mbb N$, the function $\phi_{a,\langle i_1\dots i_N\rangle}(C_\alpha)$ can be recovered via linear combinations of $\phi_{i_1\dots i_N}(C_\alpha)$, making it straightforward to express $\int_{\mbb R^d}\phi_{a,\langle i_1\dots i_N\rangle}\mcal Q_{\alpha,R}dC_\alpha$ in terms of $\int_{\mbb R^d}\phi_{i_1\dots i_N}\mcal Q_{\alpha,R}dC_\alpha$.

We define the moments of $\mcal Q_g$ and $\mcal Q_l$ with respect to $\phi_{i_1\dots i_N}(C_1)$ as 
\begin{equation}
\begin{aligned}
\mcal P^g_{i_1\dots i_N} = &m_1\int \phi_{i_1\dots i_N}(C_1) \mcal Q_g dC_1 = m_1\int  \phi_{i_1\dots i_N}(C_1) f_{3|G14} f_{4|G14} g_{34} \sigma_{34}^r \sin(\chi)d\chi d\epsilon d C_3 d C_4,\\
\mcal P^l_{i_1\dots i_N} = &m_1\int \phi_{i_1\dots i_N}(C_1) \mcal Q_l dC_1 =  -m_1\int \phi_{i_1\dots i_N}(C_1) f_{1|G14} f_{2|G14} g_{12} \sigma_{12}^f \sin (\chi)d\chi d\epsilon d C_1 d C_2. \label{lambda gl}
\end{aligned}
\end{equation}
Above, for brevity we have represented all the integrals with a single integration symbol.
To compute $\mcal P^g_{i_1\dots i_N}$, we follow the steps outlined in \autoref{algo}---the details of the algorithm are given thereafter. In the first step, we express $\phi_{i_1\dots i_N}(C_1)$ in terms of the integration variables $C_3$ and $C_4$. To compute $\mcal P^l_{i_1\dots i_N}$, this transformation is not needed. All the other steps remain the same as that for $\mcal P^g_{i_1\dots i_N}$ therefore, we only discuss the computation of $\mcal P^g_{i_1\dots i_N}$. Note that the production terms resulting from \autoref{algo} are included in the supplementary material.
 
\begin{algorithm}[ht!]
\caption{Algorithm to compute $\mcal P^g_{i_1\dots i_N}$}
\begin{algorithmic}[1] \label{algo}
\STATE Using the velocity transformations outlined in \autoref{transform vel}, express $\phi_{i_1\dots i_N}(C_1)$ in terms of $C_3$ and $C_4$. 
\STATE Perform integration over the angles that determine the collision orientation i.e., the angles $\chi$ and $\epsilon$ appearing in \eqref{lambda gl}. This results in an integral over $C_3$ and $C_4$.
\STATE Define new integration variables in terms of $C_3$ and $C_4$ such that the integral in the previous step can be written in terms of two separate integrals, with each integral expressed solely in terms of a single integration variable.
\STATE With a method-of-choice, perform integration over the two separate integrals resulting from the previous step. We use the symbolic integration from \texttt{mathematica} to compute these integrals.
\end{algorithmic}
\end{algorithm}

\begin{remark}
Note that \autoref{algo} can be used to compute an arbitrary order moment of the collision operator and not just the fourteen moments. However, for the sake of demonstration, during numerical experiments and in the production terms outlined in the supplementary material, we stick to the Grad's-14 moment approximation.  
\end{remark}

\subsubsection{Step-1:  \textit{Transformation of the test function}}
Let $\varpi_1 = \mu_{21}\sqrt{\hat{Q}_{34}}$, where $\mu_{21}$ is the mass fraction defined in \eqref{mass fraction} and $\hat Q_{34}$ is as given in \eqref{energy balance molecular 2}. We express the velocity transformations given in \eqref{expression for C1 interms of C3} as
\begin{align}
 C^{(1)}_{i_r} = C^{(3)}_{i_r} + (\varpi_1-\mu_{43})g^{(34)}_{i_r}-2\varpi_1 g_{34}\cos \theta k_{i_r} \quad (\because k\cdot g_{12} = |g_{12}| \cos (\theta)). \label{defC1}
\end{align}
Let $d=(\varpi_1-\mu_{43})g_{34}-2k \varpi_1 |g_{34}|\cos \theta $. Then, the monomial $\phi_{i_1\dots i_N}$ given in \eqref{def phi} can be expressed as
\begin{align}
 \phi_{i_1\dots i_N} &= m_1(C^{(3)}_{i_1} + d_{i_1})(C^{(3)}_{i_2} + d_{i_2})(C^{(3)}_{i_3} + d_{i_3}) \dots (C^{(3)}_{i_N} + d_{i_N}), \nonumber\\
  &= m_1\sum_{\beta_1=0}^{N}{{N}\choose{\beta_1}}d_{({i_1}}d_{i_2}d_{i_3} \dots d_{i_{\beta_1}}C^{(3)}_{i_{\beta_1+1}} \dots C^{(3)}_{{i_N})}. \label{expression phi}
\end{align}
We simplify $d_{{i_1}}d_{i_2}d_{i_3} \dots d_{i_{\beta_1}}$ as
\begin{align}
d_{i_1}d_{i_2}\dots d_{i_{\beta_1}} &= ((\varpi_1-\mu_{43})g^{(34)}_{i_1}-2\varpi_1 g_{34}\cos \theta k_{i_1})((\varpi_1-\mu_{43})g^{(34)}_{i_2}-2\varpi_1 g_{34}\cos \theta k_{i_2}) \nonumber\\
&\quad\dots((\varpi_1-\mu_{43})g^{(34)}_{w_{\beta_1}}-2\varpi_1 g_{34}\cos \theta k_{i_{\beta_1}})\\
&= \sum_{\beta_2=0}^{\beta_1}{{\beta_1}\choose{\beta_2}}(-1)^{\beta_2}(2g_{34}\cos \theta)^{\beta_2}\varpi_1^{\beta_2}(\varpi_1-\mu_{43})^{\beta_1-\beta_2}k_{(i_1}k_{i_2} \nonumber\\
&\quad\dots k_{i_{\beta_2}}g^{(34)}_{i_{\beta_2+1}}\dots g^{(34)}_{i_{\beta_1})}. \label{expression d}
\end{align}
Above and elsewhere, $B_{(i_1\dots i_r)}$ represents the symmetric part of a tensor $B$. For $r=2$, $B_{(i_1\dots i_2)} = (B_{i_1 i_2} + B_{i_2i_1})/2$, and for $r>2$, see \cite{Struchtrupbook}. 
Substituting the above simplification \eqref{expression d} into the expression for $\phi_{i_1\dots i_N}$ \eqref{expression phi}, we find
\begin{align}
 \phi_{i_1\dots i_N} &= m_1\sum_{\beta_1=0}^{N}\sum_{\beta_2=0}^{\beta_1}{{N}\choose{\beta_2, \beta_1-\beta_2}}(-1)^{\beta_2}(2g_{34}\cos \theta)^{\beta_2}\varpi_1^{\beta_2} \nonumber\\
 &\quad\times(\varpi_1-\mu_{43})^{\beta_1-\beta_2}k_{(i_1}k_{i_2} \dots k_{i_{\beta_2}}g^{34}_{i_{\beta_2+1}} \dots g^{(34)}_{i_{\beta_1}}C^{(3)}_{i_{\beta_1+1}} \dots C^{(3)}_{{i_N})}. \label{final phi}
\end{align}
In deriving the above expression, the following trivial property 
of the tensors has been used $A_{((i_1\dots i_r)}B_{i_{r+1}\dots i_{N})} = A_{(i_1\dots i_r}B_{i_{r+1}\dots i_{N})}$. With the above expression for $\phi_{i_1\dots i_N}$, the expression for $\mcal P^g_{i_1\dots i_N}$ transforms to
\begin{equation}
\begin{aligned}
\mcal P^g_{i_1\dots i_N}=&4m_1\sum_{\beta_1=0}^{N}\sum_{\beta_2=0}^{\beta_1}{{N}\choose{\beta_2, \beta_1-\beta_2}}(-1)^{\beta_2}2^{\beta_2}\\
&\times\int \underline{\left( \int_{0}^{2\pi}\int_{0}^{\frac{\pi}{2}}\sin \theta(\cos \theta)^{{\beta_2}+1}k_{(i_1}k_{i_2}\dots k_{i_{\beta_2}}d\theta d\epsilon\right)}\\
&\times g^{(34)}_{i_{{\beta_2}+1}}\dots g^{(34)}_{i_{\beta_1}}C^{(3)}_{i_{\beta_1+1}}\dots C^{(3)}_{{i_N})}f_{3|G14}f_{4|G14}\sigma^r_{34}g_{34}^{{\beta_2}+1}\varpi_1^{\beta_2}(\varpi_1-\mu_{43})^{\beta_1-{\beta_2}} dC_3dC_4.\label{int gain1}
\end{aligned}
\end{equation}
\subsubsection{Step-2: \textit{Integration over the collision orientations}}
We first compute the underlined integral appearing above in \eqref{int gain1}. Since the computation is the same as that discussed in \cite{Struchtrupbook}, we only discuss the main results. Consider a $\beta_2^{th}$ order tensor $F_{i_1i_2 \dots i_{\beta_2}}$ such that
\begin{align}
  F_{i_1i_2\dots i_{\beta_2}}=\int_{0}^{2\pi}\int_{0}^{\frac{\pi}{2}}k_{i_1}k_{i_2}\dots k_{i_{\beta_2}}\sin \theta (\cos \theta)^{{\beta_2}+1} d\theta d\epsilon
\end{align}
We can express $F$ as \cite{Gupta2012}
\begin{align}
   F_{i_1i_2\dots i_{\beta_2}} = \sum_{\gamma= 0}^{{\beta_2}/2}a_{\gamma}^{({\beta_2})}\delta_{(i_1i_2}\dots \delta_{i_{2\gamma-1}i_{2\gamma}}\frac{g^{(34)}_{i_{2\gamma+1}}}{g_{34}}\dots \frac{g^{(34)}_{i_{{\beta_2}})}}{g_{34}}, \label{expressionF}
\end{align}
where, for a hard sphere interaction potential, the coefficients $a_{\gamma}^{\beta_2}$ depend only on $g_{34}$. Replacing the underlined term in $\mcal P^g_{i_1\dots i_N}$ \eqref{int gain1} by the above expression, we find
\begin{align}
\mcal P^g_{i_1\dots i_N} &=4m_1\sum_{\beta_1=0}^{N}\sum_{\beta_2=0}^{\beta_1}{{N}\choose{\beta_2, \beta_1-\beta_2}}(-1)^{\beta_2}2^{\beta_2}\sum_{\gamma= 0}^{{\beta_2}/2}a_{\gamma}^{({\beta_2})}\int\delta_{(i_1i_2}\dots\delta_{i_{2\gamma-1}i_{2\gamma}}g^{(34)}_{i_{2\gamma+1}}\dots g^{(34)}_{i_{{\beta_2}}} \nonumber\\
&\quad \times g^{(34)}_{i_{{\beta_2}+1}}\dots g^{(34)}_{i_{\beta_1}}C^{(3)}_{i_{\beta_1+1}}\dots C^{(3)}_{{i_N})}f_{3|G14}f_{4|G14}\sigma_{34}^rg_{34}^{2\gamma+1}\varpi_1^{\beta_2}(\varpi_1-\mu_{43})^{\beta_1-{\beta_2}}d C_3d C_4.  \label{expression for P1g}
\end{align}
With the above expression, to complete our computation of $\mcal P^g_{i_1\dots i_N}$, we only need to perform integrals over $C_3$ and $C_4$. The following two steps take care of this.
\subsubsection{Step-3: \textit{Scales for non-dimensionalization}}
To make the following discussion simpler, we perform non-dimensionalization
\begin{equation}
\begin{gathered}
\hat{\sigma}_{12}^r = U\left(1-\frac{2\hat{\epsilon}_f}{\hat{g}_{12}^2}\right)\frac{d_f^2}{4}\left(1-\frac{2\hat{\epsilon}_f}{\hat{g}_{12}^2}\right),\hspace{0.5cm}  \hat{\sigma}_{34}^r = U\left(1-\frac{2\hat{\epsilon}_r}{\hat{g}_{34}^2}\right)\frac{d_r^2}{4}\left(1-\frac{2\hat{\epsilon}_r}{\hat{g}_{34}^2}\right),\label{scaling}\\
\hat{f}_{\alpha} = \frac{f_{\alpha}\theta_{\alpha}^{\frac{d}{2}}}{n_{\alpha}},\hspace{0.5cm} \hat{C}_{\alpha} = \frac{C_{\alpha}}{\sqrt{\theta_{\alpha}}},\hspace{0.5cm}
\hat{g}_{\alpha\beta} = \frac{g_{\alpha\beta}}{\sqrt{\theta_{\alpha\beta}}},\hspace{0.5cm}\hat\epsilon_r = \frac{\epsilon_r}{m_{34}\theta_{34}},\hspace{0.5cm}\hat\epsilon_f=\frac{\epsilon_f}{m_{12}\theta_{12}}\\
\hat{\varpi}_1 = \mu_{21}\sqrt{\frac{m_{34}}{m_{12}}\left(1+\frac{2\hat{Q}_1}{\hat{g}_{34}^2}\right)}, \hspace{0.5cm} \hat{Q}_1 = \frac{Q}{m_{34}\theta_{34}},
\end{gathered}
\end{equation}
where $\theta_{\alpha\beta} = \left(\theta_{\alpha} + \theta_{\beta}\right)/2$. Using the above scaling, $\mcal P^g_{i_1\dots i_N}$ given in \eqref{expression for P1g} reads
\begin{equation}
\begin{aligned}
 \mcal P^g_{i_1\dots i_N}&=4m_1n_3n_4\sum_{\beta_1=0}^{N}\sum_{\beta_2=0}^{\beta_1}{{N}\choose{\beta_2, \beta_1-\beta_2}}(-1)^{\beta_2}2^{\beta_2}\sum_{\gamma= 0}^{{\beta_2}/2}a_{\gamma}^{({\beta_2})}\int\delta_{(i_1i_2}\dots\delta_{i_{2\gamma-1}i_{2\gamma}}\hat g^{(34)}_{i_{2\gamma+1}}\dots\hat g^{(34)}_{i_{{\beta_2}}}\nonumber\\
&\quad\times\hat g^{(34)}_{i_{{\beta_2}+1}}\dots\hat g^{(34)}_{i_{\beta_1}}\hat C^{(3)}_{i_{\beta_1+1}}\dots\hat C^{(3)}_{{i_N})}\hat f_{3|G14}\hat f_{4|G14}\hat{\sigma}_{34}^r\hat g_{34}^{2\gamma+1}\\
&\quad \times\theta_{34}^{\frac{\beta_1+1}{2}}\theta_3^{\frac{n-\beta_1}{2}}\hat{\varpi}_1^{\beta_2}(\hat{\varpi}_1-\mu_{43})^{\beta_1-{\beta_2}}d\hat{C}_3d\hat{C}_4. \label{expressionP1}
\end{aligned}
\end{equation}
\subsubsection{Step-4: \textit{Separation of integrals}}
We want to define new integration variables in terms of $\hat C_3$ and $\hat C_4$ such that above integral \eqref{expressionP1} can be separated out into two separate integrals. The structure of the Gaussian $\scalef$ in \eqref{Grad14} motivates the transformation
\begin{align}
  \hat{h}_{34} = \frac{1}{2}\left(\sqrt{\frac{\theta_{4}}{\theta_{34}}}\hat{C}_{3} + \sqrt{\frac{\theta_{3}}{\theta_{34}}}\hat{C}_{4}\right), 
\quad \hat{g}_{34} = \sqrt{\frac{\theta_{3}}{\theta_{34}}}\hat{C}_{3} - \sqrt{\frac{\theta_{4}}{\theta_{34}}}\hat{C}_{4}. \label{def h34 g34}
\end{align}
Note that the above relation provides $\frac{1}{2}(|\hat{C}_3|^2+|\hat{C}_4|^2) = \left(\frac{1}{4}|\hat{g}_{34}|^2+|\hat{h}_{34}|^2\right)$. Furthermore, one can show that the determinant of the Jacobian of transformation given as
$\det(J) = \left|\det\left(\frac{\partial(\hat{C}_3,\hat{C}_4)}{\partial(\hat{h}_{34},\hat{g}_{34})}\right)\right|
 $, equals one. The above transformation of variables provides
 \begin{equation}
 \begin{aligned}
\mcal P^g_{i_1\dots i_N} = &\sum \limits_{p,q\leq N} \delta^{(p,q)} \underbrace{\int_{\mbb R^d} |\hat{h}_{34}|^{n} \hat{h}^{(34)}_{i_1}\dots \hat{h}^{(34)}_{i_p} \exp \left[-|\hat{h}_{34}|^2\right] d\hat{h}_{34}}_{I_h}\\
 &\quad\times  \underbrace{\int_{\mbb R^d} |\hat{g}_{34}|^{m}\varpi_1^l \hat{\sigma}^r_{34} \hat{g}^{(34)}_{i_1} \dots \hat{g}^{(34)}_{i_q} \exp\left[-\frac{|\hat{g}_{34}|^2}{4}\right] d\hat{g}_{34}}_{I_g}. \label{separate integral}
 \end{aligned}
 \end{equation}
 In the above expression, $\delta^{(p,q)}$ is some factor that results from the non-dimensionalization given in \eqref{scaling}, its precise form is not important here.
 From the above expression \eqref{separate integral} it is clear that we have separated out the 
 integration over $\hat{h}_{34}$ and $\hat{g}_{34}$. We now discuss the computation of $I_h$ and $I_g$.

\subsubsection{Step-5: \textit{Computation of $I_g$ and $I_h$}}
Expressing $\hat{h}_{34}$ in terms of spherical harmonics, $I_h$
can be given as \cite{Gupta2012}
\begin{align}
 I_h = \frac{1}{2} \Gamma\left(\frac{2N + p + 3}{2}\right) \frac{4\pi}{p+1}\delta_{i_1\dots i_p}
\end{align}
where $ \delta_{i_1\dots i_p}$ is the fully symmetric product of Kronecker deltas and $\Gamma(\cdot)$ denotes a gamma-function. Similarly, transforming $\hat{g}_{34}$ using spherical coordinates, $I_g$ can be expressed as
\begin{align}
  I_g = \underline{\int_0^{2\pi}\int_0^{\pi} \tilde{n}_{i_1} \dots \tilde{n}_{i_p} \sin \nu d\nu d\omega} \int_0^{\infty}\varpi_1^l\hat{\sigma}^{r}_{34}|\hat{g}_{34}|^{2m+q+2}\exp\left[-\frac{|\hat{g}_{34}|^2}{4}\right] d|\hat{g}_{34}|,	\label{exp Ig}
\end{align}
where $\tilde{n}_i = \hat{g}^{(34)}_i/\hat{g}_{34}$. Computing the underlined term in the above expression, we find \cite{Gupta2012}
\begin{align}
 I_g = \frac{4\pi}{p+1}\delta_{i_1 \dots i_p} \underline{\int_0^{\infty}\varpi_1^l\hat{\sigma}^{r}_{34}|\hat{g}_{34}|^{2m+p+2}\exp\left[-\frac{|\hat{g}_{34}|^2}{4}\right] d|\hat{g}_{34}|}. \label{expIg2}
\end{align}
Inserting the expression for $\hat{\sigma}_{34}^r$ in the above expression,
we find that the underlined integral in the above expression is of the form
$$ \Omega_{\epsilon_r}^{(l,r)} = \int_0^{\infty}U\left(1-\frac{2\hat{\epsilon}_r}{|\hat{g}_{34}|^2}\right)|\hat{g}_{34}|^r\hat{\varpi}_1^l \exp\left[{-\left(\frac{|\hat{g}_{34}|^2}{4}\right)}\right] d|\hat{{g}}_{34}|.$$

We first consider the case when $l=0$.
The definition of an incomplete Gamma function $\Gamma\left(a,b\right)$ provides $ \Omega_{\epsilon_r}^{(0,r)} = 2^{r}\Gamma\left(\frac{1+r}{2},\frac{\hat{\epsilon}_r}{2}\right)$. For $l \neq 0$, we simplify $\Omega_{\epsilon_r}^{(l,r)}$ in the following way and compute it explicitly using the symbolic integration in \texttt{mathematica}
 \begin{align}
 \Omega_{\epsilon_r}^{(l,r)} &= \int_0^{\infty} U\left(1-\frac{2\hat{\epsilon}_r}{|\hat{g}_{12}|^2}\right) |\hat{g}_{34}|^r \hat{\varpi}_1^{l}\exp\left[-\frac{|\hat{g}_{34}|^2}{4}\right]d|\hat{{g}}_{34}| \nonumber\\
  &= \int_0^{\infty} U\left(1-\frac{2\hat{\epsilon}_r}{|\hat{g}_{12}|^2}\right) |\hat{g}_{34}|^r \mu_{21}^l\left(\frac{m_{3}m_{4}}{ m_{1}m_{2}}\left(1+\frac{2(\hat{\epsilon}_f-\hat{\epsilon}_r)}{|\hat{g}_{34}|^2}\right)\right)^{\frac{l}{2}}\exp\left[-\frac{|\hat{g}_{34}|^2}{4}\right]d|\hat{{g}}_{34}| \nonumber\\
  &=\int_0^{\infty} U\left(1-\frac{2\hat{\epsilon}_r}{|\hat{g}_{12}|^2}\right) |\hat{g}_{34}|^{(r-l)} \mu_{21}^l\left(\frac{m_{3}m_{4}}{ m_{1}m_{2}}\left(|\hat{g}_{34}|^2+2\left(\hat{\epsilon}_f-\hat{\epsilon}_r\right)\right)\right)^{\frac{l}{2}}\exp\left[-\frac{|\hat{g}_{34}|^2}{4}\right]d|\hat{{g}}_{34}| .
 \end{align}
The second equality is a result of energy conservation given in \eqref{energy balance molecular 2}.
The different $ \Omega_{\epsilon_r}^{(l,r)}$ integrals used in the present work are given in the supplementary material.

\subsection{Non-equilibrium reaction rates}
Assuming that the chemical reaction \eqref{chemical reaction} follows first-order chemical kinetics, the forward $k_f$ and the backward $k_r$ reaction rates can be given as 
\begin{gather}
k_f = \frac{1}{n_1n_2}\int f_1f_2\sigma_{12}^fg_{12}\sin\chi d\chi d c_1d c_2, \quad k_r = \frac{1}{n_3n_4}\int f_3f_4\sigma_{34}^rg_{34}\sin\chi d\chi d c_3d c_4.
\end{gather}
Replacing the probability density functions by their Grad's-14 approximation and using technique outlined in \ref{algo}, we can compute the approximations $k_{f|G14}$ and $k_{r|G14}$. A concise form of both these rates can be given as
\begin{align}
 k_{f|G14} = 2\sqrt{2\pi\left(\theta_1+\theta_2\right)}\left(d_{12}s_f\right)^2\exp{\left[-\left(\frac{\epsilon_f}{m_{12}\left(\theta_1+\theta_2\right)}\right)\right]} + \tilde{\Delta}_f, \label{explicit kf}\\
 k_{r|G14} = 2\sqrt{2\pi\left(\theta_3+\theta_4\right)}\left(d_{34}s_r\right)^2\exp{\left[-\left(\frac{\epsilon_r}{m_{34}\left(\theta_3+\theta_4\right)}\right)\right]} + \tilde{\Delta}_r, \label{explicit kr}
\end{align}
where $\tilde{\Delta}_f$ and $\tilde{\Delta}_r$
represent the contribution from the non-equilibrium moment $\Delta_{\alpha}$ and are given as
\begin{equation}
\begin{aligned}
\tilde{\Delta}_r = &\frac{\beta_r \Delta_3 \theta_3^2 \mu_{12} \left(-\theta_{34}^2 m_{34}^2-2 \theta_{34} m_{34} \epsilon_r+\epsilon_r^2\right)}{120 m_1 n_3 n_4}+\frac{\beta_r \Delta_4 \theta_4^2 \mu_{12} \left(-\theta_{34}^2 m_{34}^2-2 \theta_{34} m_{34} \epsilon_r+\epsilon_r^2\right)}{120 m_1 n_3 n_4},\\
\tilde{\Delta}_f = &\frac{\beta_f \Delta_1 \theta_1^2 \mu_{12} \left(-\theta_{12}^2 m_{12}^2-2 \theta_{12} m_{12} \epsilon_f+\epsilon_f^2\right)}{120 m_1 n_1 n_2}+\frac{\beta_f \Delta_2 \theta_2^2 \mu_{12} \left(-\theta_{12}^2 m_{12}^2-2 \theta_{12} m_{12} \epsilon_f+\epsilon_f^2\right)}{120 m_1 n_1 n_2}.
\end{aligned}
\end{equation}
With $M= m_1 + m_2 = m_3 + m_4$, the coefficients $\beta_f$ and $\beta_r$ are given as
\begin{gather}
\beta_f = \left(\frac{\sqrt{\pi } d_{12}^2 M n_1 n_2 s_f^2 }{ \theta_{12}^{7/2} m_{12}^2} \right) \exp\left[{-\frac{\epsilon_f}{2 \theta_{12} m_{12}}}\right], \quad   \beta_r = \left(\frac{\sqrt{\pi } d_{34}^2  M n_3 n_4 s_r^2}{\theta_{34}^{7/2} m_{34}^2}\right)\exp\left[{-\frac{\epsilon_r}{2 \theta_{34} m_{34}}}\right].
\end{gather}
Furthermore, $s_f$ and $s_r$ represent the steric factors defined in \eqref{def d}. Note that, out of all the fourteen moments in the set $w_\alpha^{[14]}$, $\Delta_\alpha$ is the only non-equilibrium moment (i.e., a moment that vanishes in equilibrium) that influences the reaction rates. This highlights the importance of considering a 14-moment system instead of the 13 one. We expect that the presence of $\Delta_\alpha$ provides a better approximation to the true reaction rates outside of equilibrium. 

Our expressions for $k_{f|G14}$ and $k_{r|G14}$ are consistent in the sense that in equilibrium, we recover the Arrhenius law \cite{Atkins}. The details of the computations are as follows.
 In a chemical equilibrium, the following two conditions hold (i) all the gases have the same temperature i.e., $T_1 = T_2 = T_3 =T_4 = T^{eq}$,
and (ii) the non-equilibrium moment $\Delta_\alpha$ vanishes, providing $\tilde{\Delta}_f = \tilde{\Delta}_r = 0$---see \cite{Rossani1999,Bisi2002} for a detailed study of the equilibrium state. Substituting these two conditions into \eqref{explicit kf}-\eqref{explicit kr},
we obtain the Arrhenius law \cite{Atkins}
\begin{align}
k_{f,eq} = \sqrt{\frac{8\pi kT^{eq}}{m_{12}}} d_f^2 \exp\left[-\frac{\epsilon_f}{kT^{eq}}\right], \quad k_{r,eq} = \sqrt{\frac{8\pi kT^{eq}}{m_{34}}} d_r^2 \exp\left[-\frac{\epsilon_r}{kT^{eq}}\right] \label{rates equilibrium}
\end{align}
In equilibrium, the reaction rates satisfy the follow law of mass-action \cite{Atkins}
\begin{align}
 \frac{k_{f,eq}}{k_{r,eq}} = \left(\frac{m_3m_4}{m_1m_2}\right)^{\frac{3}{2}}\exp\left[-\frac{Q}{kT^{eq}}\right] = \frac{n_3 n_4}{n_1 n_2} \label{lma}
\end{align}
Substituting equilibrium reaction rates \eqref{rates equilibrium} into the law of mass action \eqref{lma}, we find a relation between the steric factors \cite{Kremerbook}
\begin{align}
s_f \sqrt{m_1m_2}d_{12} = s_r \sqrt{m_3m_4}d_{34}.	\label{relation steric}
\end{align}

\begin{remark}\label{remark: equi}
Using the law of mass action, we make a distinction between a thermal and a chemical equilibrium. At a thermal equilibrium, the contribution from mechanical collisions vanishes $\mcal Q_M = 0$, all the mixture components have the same temperature $T_1 = T_2 = T_3 =T_4 = T^{eq}$ and the probability density function is given by 
\begin{align}
f_{\alpha} = n_\alpha \left(\frac{1}{2\pi\theta_{\alpha}}\right)^{\frac{d}{2}}\exp\left(-\frac{m|C_{\alpha}|^2}{2kT^{eq}}\right). 
\end{align}
In a chemical equilibrium, the contribution from the chemical collisions vanishes $\mcal Q_{\alpha ,R} = 0$, all the mixture components have the same temperature and the probability density function is given by the above expression. In addition, the number densities satisfy the law of mass action given in \eqref{lma}. Note that a mixture in chemical equilibrium is also in thermal equilibrium but vice-versa does not necessarily holds.
\end{remark}

\section{Numerical Experiments}\label{sec: num exp}
Through numerical experiments we study how the Grad's-14 moment system (given in \eqref{mom eq explicit}) relaxes to the equilibrium state. In particular, we study the effect of the following parameters on the relaxation behaviour (i) initial conditions, (ii) steric factors, (iii) activation energies and (iv) the heat of the reaction. To solve the moment system, we use the \texttt{NDSolve} routine from \texttt{mathematica} with the default parameters. For the simplicity of exposition, we first non-dimensionalize the moment system. We use the same technique as that proposed in \cite{GuptaBinary}.
\subsection{Non-dimensionalization}
We non-dimensionalize the moment system \eqref{mom eq explicit} using the following scales
\begin{equation}
\begin{gathered}
 \hat{x}_i=\frac{x_i}{L},\hspace{0.5cm} \hat{t}=\frac{v_{o}}{L}t,\hspace{0.5cm}\hat{v}_i=\frac{v_i}{v_o},\hspace{0.5cm} \hat{n}_{\alpha}=\frac{n_{\alpha}}{n_{\alpha}^o},\hspace{0.5cm}\hat{T}_{\alpha}=\frac{T_{\alpha}}{T_o},\hspace{0.5cm}\hat{u}^{(\alpha)}_i=\frac{u^{(\alpha)}_i}{\sqrt{\theta_{\alpha}^o}}\\
 \hat{\sigma}^{(\alpha)}_{ij}=\frac{\sigma^{(\alpha)}_{ij}}{\rho_{\alpha}^o\theta_{\alpha}^o},\hspace{0.5cm}\hat{q}^{(\alpha)}_i=\frac{q^{(\alpha)}_i}{\rho^o_{\alpha}(\theta_{\alpha}^o)^{\frac{3}{2}}},\hspace{0.5cm}\hat{\Delta}_{\alpha} = \frac{\Delta_{\alpha}}{\rho_{\alpha}^o (\theta_{\alpha}^o)^2},\hspace{0.5cm}\hat{\epsilon}_f = \frac{\epsilon_f}{kT_o},\hspace{0.5cm}\hat{\epsilon}_r = \frac{\epsilon_r}{kT_o}
\end{gathered}
\end{equation}
where all the quantities with an 'o' represent the reference scale. 
In the following discussion, a hat over any 
quantity will denote its non-dimensionalized counterpart.
Defining the total relative number density as $n^{o} =  \sum_{\alpha=1}^4n^{o}_{\alpha}$, we define the mole-fraction of gas-$\alpha$ as $F_{\alpha} = n^{o}_{\alpha}/n^{o}$. Using the mole fraction, an averaged 
mass of the mixture and the velocity scale can be defined as $m = \sum_{\alpha=1}^4 F_{\alpha}m_{\alpha} $
and $ v_o = \sqrt{\frac{kT_o}{m}}$, respectively. We define the Knudsen number of the mixture as
\begin{align}
\mathrm{Kn} = \frac{5}{16\sqrt{\pi}n^{o}L\tilde{\Omega}}		\label{defvo}
\end{align}
where $L$ is the macroscopic length scale and the factor $\tilde{\Omega}$
can be looked upon as an average cross-section for the mixture. For a hard-sphere
interaction potential, an explicit expression for $\tilde{\Omega}$ can be given
 as $\tilde{\Omega} = \sum_{\alpha=1}^4(F_{\alpha}d_{\alpha}^2)$. 
Defining the Knudsen number as given in \eqref{defvo} introduces a factor $\Omega_{\alpha\beta} = \frac{\tilde{\Omega}}{d_{\alpha\beta}^2}$ in the dimensionless production terms. This factor can be looked upon as a scaled cross-section for every component in the mixture.

\subsection{Physical parameters}
For all the test cases, we consider the physical parameters given in \autoref{physical constants}. The parameters do not necessarily correspond to a realistic physical system and have been chosen only for demonstration purposes. Note that all the molecules are of the same size i.e., $\Omega_{\alpha\beta}=1$, and that all the components of the mixture are equally diluted i.e., $F_{\alpha}=0.25$.
\FloatBarrier
 \begin{table}[ht!]
\centering
  \begin{tabular}{ |c|c|c|c|c|c|c|c|c|c| }
    \hline
     parameters & $m_1$ & $m_2$ & $m_3$ & $m_4$ & Kn  & $\Omega_{\alpha\beta}$ & $F_{\alpha}$\\ \hline
     values & 11.7 & 3.6 & 8 & 7.3 & 0.1 & 1 & 0.25 \\ \hline
    \end{tabular}
  \caption{Physical parameters used in all the experiments.}	\label{physical constants}
  \end{table}

\subsection{Equilibrium state}
For all the coming experiments, we determine the equilibrium state as follows. We represent the initial number density by $\hat n_{\alpha,0}$, the initial temperature by $\hat T_{\alpha,0}$, and we start from a fluid at rest. The initial temperature of the mixture is denoted by $T_0$. In chemical equilibrium, all the gases have the same temperature $T^{eq}$. Furthermore, the density of gas-$\alpha$ at equilibrium is given as $\hat n_{\alpha,0} + x$. The value of $x$ and that of $T^{eq}$ follows by solving the equation \cite{Rossani1999}
\begin{align}
 \left(\frac{m_{3}m_4}{m_{1}m_{2}}\right)^{\frac{3}{2}}(n_{10}+x)(n_{20}+x)\exp\left(-1/\left[\frac{kT_0}{Q}+\frac{2x}{3n}\right]\right)=(n_{30}-x)(n_{40}-x), \label{solveequilibrium}
\end{align}
where $x=\frac{3}{2}nk(T^{(eq)}-T_0)/Q$. Recall that $n$ is the number density of the mixture and that it remains constant over time due to total mass conservation. 

\begin{remark}
For all the test cases, we initialize the flow velocity $\hat{u}_i^{(\alpha)} $, the stress tensor $\hat{\sigma}_{ij}^{(\alpha)}$, the heat-flux $\hat{q}_i^{(\alpha)}$ and $\hat{\Delta}_{\alpha}$ with zero. For such an initialization, using the moment system \eqref{mom eq explicit}, it can be shown that 
$\hat{u}^{(\alpha)}_i$, $\hat{q}^{(\alpha)}_i$ and $\hat{\sigma}_{ij}^{(\alpha)}$ remain zero for all time instances therefore, we do not discuss them further. Note that $\hat{\Delta}_{\alpha}$ is influenced by mass and energy production and is therefore, perturbed by the gas mixture being outside of equilibrium. The following experiments study the perturbations in $\hat \Delta_\alpha$ in detail.
\end{remark}
\subsection{Influence of initial conditions}
\subsubsection{Case-1: \textit{Exothermic Reaction}}
The initial conditions and the physical parameters 
corresponding to this test case are
given in \autoref{int cond exo} and \autoref{act energy exo}, respectively. Furthermore, the equilibrium state computed using \eqref{solveequilibrium} is given in \autoref{int cond exo}.

\begin{figure}[H]
\centering
\subfigure [number density $\hat n_\alpha$]{
\label{n exothermic}
\includegraphics[width=3in]{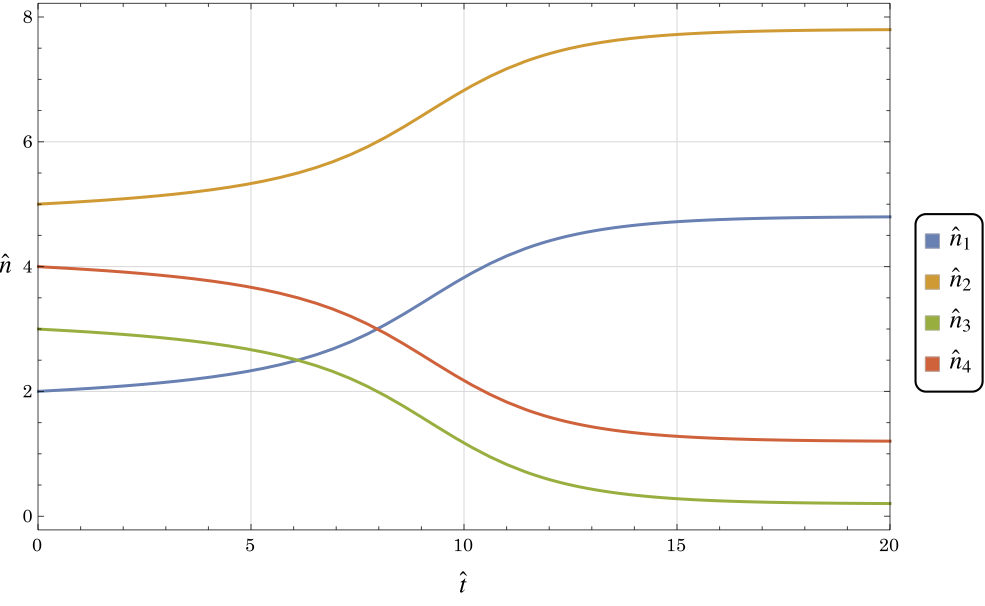} }
\hfill
\subfigure [Temperature $\hat{T}_\alpha$]{
\label{T exothermic}
\includegraphics[width=3in]{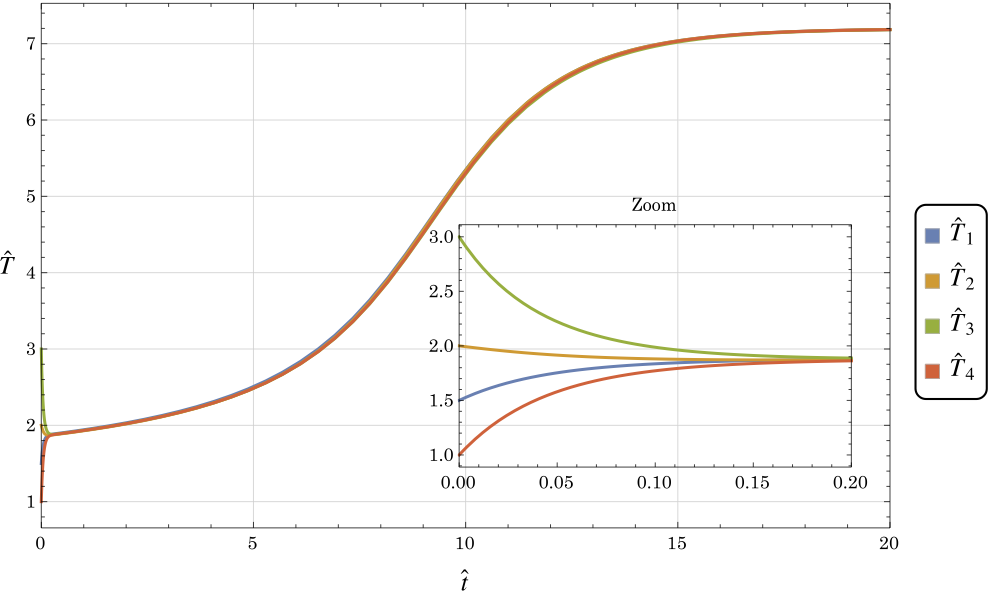} }
\hfill
\subfigure [Mixture temperature $\hat{T}_{mix}$]{
\label{Tmix exothermic}
\includegraphics[width=3in]{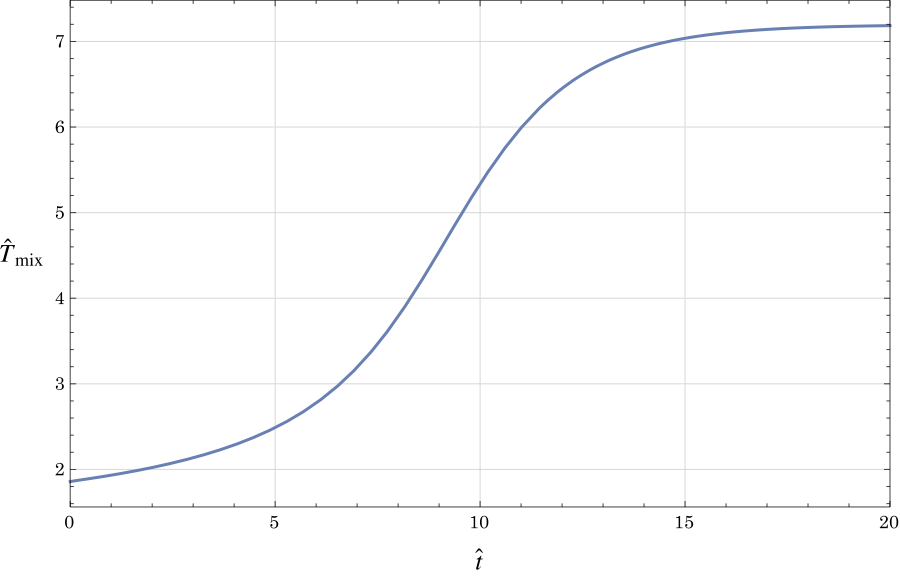} }
\hfill
\subfigure [The scalar moment $\hat{\Delta}_{\alpha}$]{
\label{Delta exothermic}
\includegraphics[width=3in]{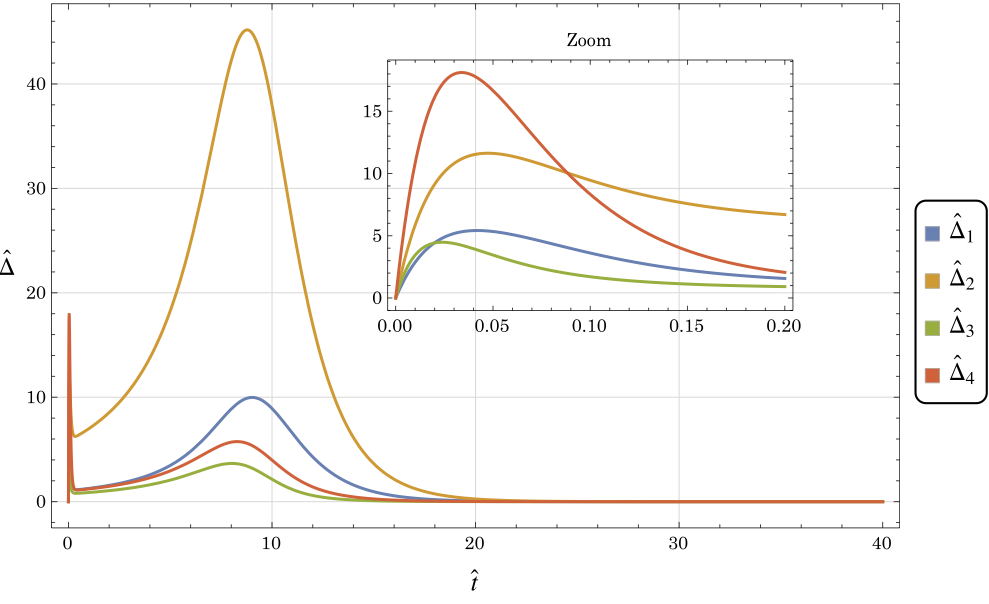} }
\hfill
\caption{Relaxation to equilibrium of the different moments.}	\label{variation exo}
\end{figure}

The time variation of $\hat n_\alpha$, $\hat T_\alpha$ and $\hat T$ is shown in \autoref{n exothermic}, \autoref{T exothermic} and \autoref{Tmix exothermic}, respectively. Clearly, the Grad-14 moment system relaxes to the correct equilibrium state. Furthermore, due to an exothermic nature of the reaction, the mixture temperature increases---see \autoref{Tmix exothermic}. In \autoref{T exothermic}
and \autoref{Delta exothermic}, the zoomed in plot shows the initial layer. The initial layer is a result of the initial temperature difference between the different gases, which results in a deviation from an equilibrium. 
\autoref{Delta exothermic} shows the time evolution of $\hat \Delta_\alpha$. Even though we start with a zero value of $\hat \Delta_\alpha$, it is perturbed to a non-zero value. It is noteworthy that all the different $\hat{\Delta}_{\alpha}$'s relax to zero at the same time scale as the number densities and the mixture temperatures.

One can identify the two time-scales from the temperature profile shown in \autoref{T exothermic}. At one time scale, which is much shorter than the other, the temperature of all the mixture components equalize to some value, say $\tilde T$. At the other time scale, the value $\tilde T$ increases and approaches the equilibrium temperature $T^{eq}$. 
These two distinct time-scales can also be identified in $\hat\Delta_{\alpha}$'s profile shown in \autoref{Delta exothermic} where in the first time-scale, $\hat\Delta_{\alpha}$ is pushed towards its equilibrium value of zero and in the second time-scale, $\hat\Delta_{\alpha}$ first increases and then approaches its equilibrium value of zero. Following is a possible explanation for the observance of these two time-scales. Initially, since the temperature of any mixture component is not sufficiently high, mechanical interactions dominate. These interactions try to push for a mechanical equilibrium where all the temperatures are equal and $\hat\Delta_\alpha$ is zero. This results in the first time-scale. The second time-scale is triggered by chemically reactive interactions that increase the mixture temperature and push $\hat\Delta_\alpha$ away from its equilibrium value of zero. The system eventually reaches equilibrium where all the components have the same temperature, the number densities satisfy the law of mass action \eqref{lma} and $\hat \Delta_\alpha = 0$.

\FloatBarrier
\begin{table}[!htbp]
\centering
  \begin{tabular}{ |c|c|c|c|c|c| }
    \hline
    \multicolumn{5}{|c|}{Initial conditions}\\
    \hline
     moments & Component-1 & Component-2 & Component-3 & Component-4  \\ \hline
     number density($\hat{n}_{\alpha}$) & 2 & 5 & 3 & 4   \\ \hline
     Temperature($\hat{T}_{\alpha}$) & 1.5 & 2 & 3 & 1   \\ \hline
    \multicolumn{5}{|c|}{Equilibrium values}\\
    \hline
     moments & Component-1 & Component-2 & Component-3 & Component-4  \\ \hline
     number density($\hat{n}_{\alpha}$) & 4.8 & 7.8 & 0.19 & 1.19   \\ \hline
     Temperature($\hat{T}_{\alpha}$) & 7.2 & 7.2 & 7.2 & 7.2   \\ \hline 
  \end{tabular}
  \caption{Initial data and the corresponding equilibrium state for
 the  exothermic reaction test case.}	\label{int cond exo}
  \end{table}
\FloatBarrier
\begin{table}[!ht]
\centering
  \begin{tabular}{ |c|c|c|c|c|c| }
    \hline
     Parameters & Forward Reaction & Reverse Reaction  \\ \hline
     $s_f, s_r$ & 0.18 & 0.5   \\ \hline
     $\epsilon_f, \epsilon_r$ & 50 & 10   \\ \hline
    \end{tabular}
  \caption{Activation energies and steric factors for the exothermic reaction test case.}	\label{act energy exo}
  \end{table}
 %
 %
 %
 %
\subsubsection{Case-2: \textit{Initial conditions in thermal equilibrium}}
We consider initial conditions that are in thermal equilibrium but not in chemical equilibrium. 
The initial conditions and the corresponding equilibrium states are given in \autoref{int cond thermal}. Since all the mixture components have the same temperature, the mixture is in thermal equilibrium. However, the mixture is not in chemical equilibrium since the law of mass action \eqref{lma} is not satisfied. We refer to \autoref{remark: equi} for a distinction between a thermal and a chemical equilibrium. 

\begin{figure}[H]
\centering
\subfigure [Temperature $\hat{T}^{(\alpha)}$]{
\label{T thermal equilibrium}
\includegraphics[width=3in]{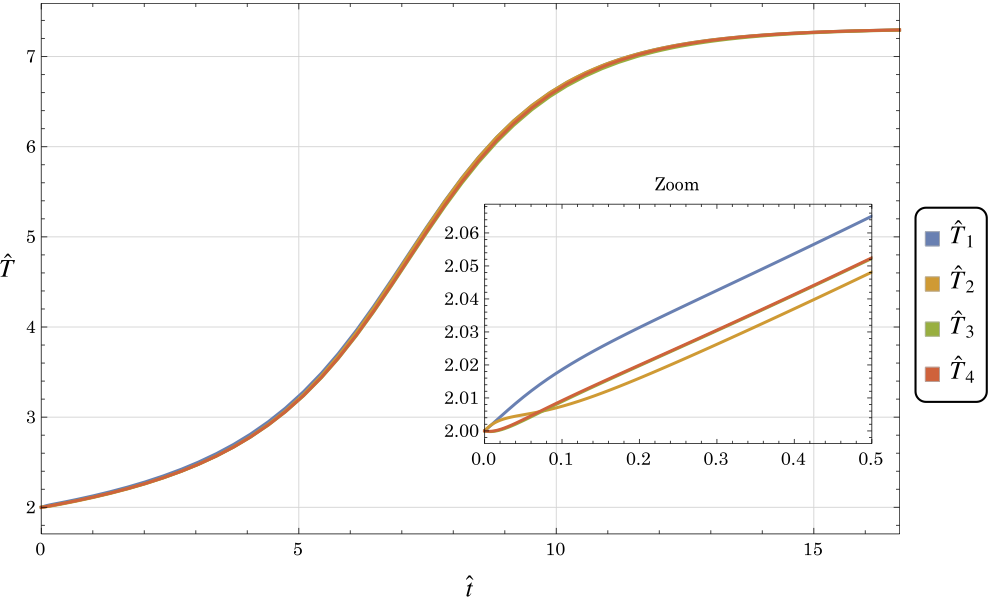} }
\hfill
\subfigure [The scalar moment $\hat{\Delta}^{(\alpha)}$]{
\label{Delta thermal equilibrium}
\includegraphics[width=3in]{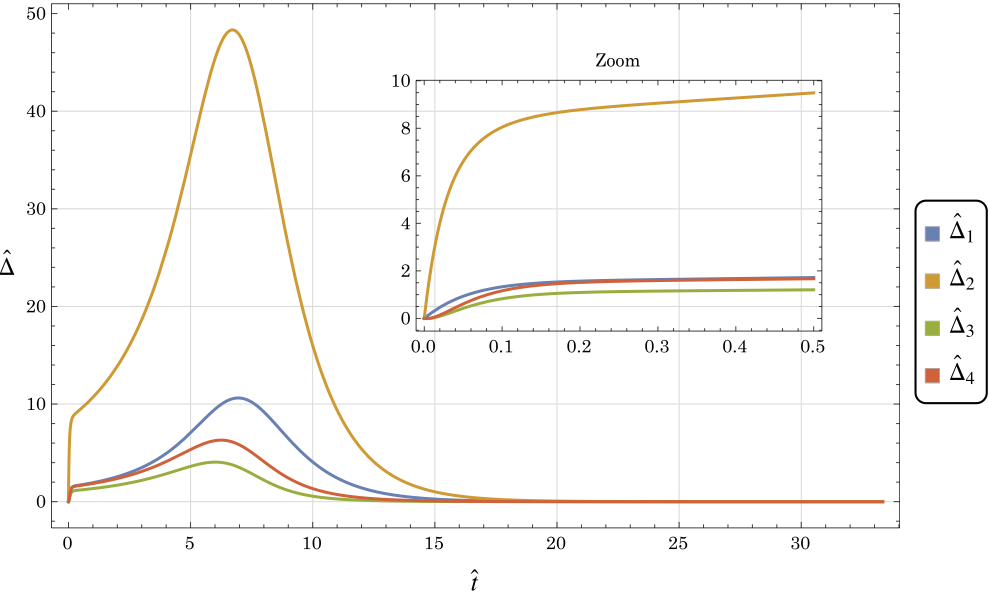} }
\hfill
\caption{Relaxation to equilibrium of the temperature and the scalar fourteenth moment $\hat\Delta_\alpha$.}
\end{figure}

Fig \autoref{T thermal equilibrium} shows the time variation of the different temperatures $\hat T_\alpha$. Even
though we start from the same temperatures, due to a chemical non-equilibrium, the values of the different temperatures initially
deviate from each other as the time progresses. Similar to the previous test case, $\hat{\Delta}_{\alpha}$ is perturbed from its initial value and we observe an initial layer which, due to a different initial data, is not as strong as the previous test case. 
\FloatBarrier
\begin{table}[!htbp]
\centering
  \begin{tabular}{ |c|c|c|c|c|c| }
    \hline
    \multicolumn{5}{|c|}{Initial conditions}\\
    \hline
     variable & Component-1 & Component-2 & Component-3 & Component-4  \\ \hline
     number density($\hat{n}_{\alpha}$) & 2 & 5 & 3 & 4   \\ \hline
     Temperature($\hat{T}_{\alpha}$) & 2 & 2 & 2 & 2   \\ \hline
    \multicolumn{5}{|c|}{Equilibrium values}\\
    \hline
     variable & Component-1 & Component-2 & Component-3 & Component-4  \\ \hline
     number density($\hat{n}_{\alpha}$) & 4.7 & 7.7 & 0.21 & 1.21   \\ \hline
     Temperature($\hat{T}_{\alpha}$) & 7.3 & 7.3 & 7.3 & 7.3   \\ \hline 
  \end{tabular}
  \caption{Initial conditions in thermal equilibrium and the 
  corresponding equilibrium state.}	\label{int cond thermal}
  \end{table}
\subsection{Influence of the steric factor}
\autoref{int cond exo} shows the initial data and the corresponding equilibrium state. The activation
energies and the steric factors are given in \autoref{act energy exo} and \autoref{steric factors therm equi}, respectively. Note that we only specify $s_r$, the value of $s_f$ follows from the relation in \eqref{relation steric}. 

\begin{figure}[H]
\begin{center}
\includegraphics[width=3in]{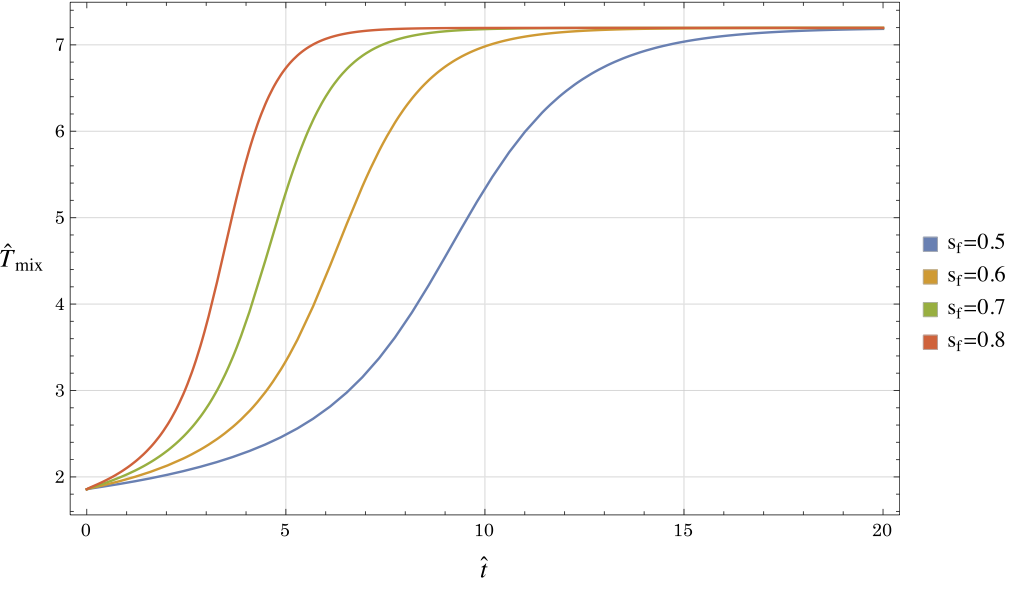} 
\hfill
\end{center}
\caption{Mixture temperature for different steric factors.} \label{Tmix steric}
\end{figure}

As is clear from \eqref{solveequilibrium}, the final equilibrium state is independent of the steric factors. However, the time-scale at which the system relaxes to the equilibrium decreases as the steric factor increases---see
\autoref{Tmix steric} that shows the variation of $\hat T$ for different steric factors. Following is a possible explanation for this behaviour.
From \eqref{sigma12} and \eqref{sigma34} we conclude 
that increasing $s_f$, which also increases $s_r$ due to \eqref{relation steric},
the chemical collision cross-section increases, resulting in an increased number of chemical collisions. Increase in the number of collisions results in a faster relaxation of the moment system. 

\FloatBarrier
\begin{table}[!ht]
\centering
  \begin{tabular}{ |c|c|c|c|c|c| }
    \hline
     parameters & Forward Reaction & Reverse Reaction  \\ \hline
     $s_f, s_r$ & 0.185 & 0.5   \\ \hline
     $s_f, s_r$ & 0.22 & 0.6   \\ \hline
     $s_f, s_r$ & 0.26 & 0.7   \\ \hline
     $s_f, s_r$ & 0.3 & 0.8   \\ \hline
    $\epsilon_f, \epsilon_r$ & 50 & 10   \\ \hline
    \end{tabular}
  \caption{Activation energies and the different steric factors.}	\label{steric factors therm equi}
  \end{table}
\subsection{Influence of the reaction heat ($Q$)}
The initial conditions and the equilibrium states are shown in \autoref{int cond exo} and \autoref{act energy heat}, respectively. Note that, unlike the steric factor, the equilibrium state changes with the heat of the reaction. We fix the steric factor to $s_f = 0.5$.

\begin{figure}[H]
\begin{center}
\includegraphics[width=3in]{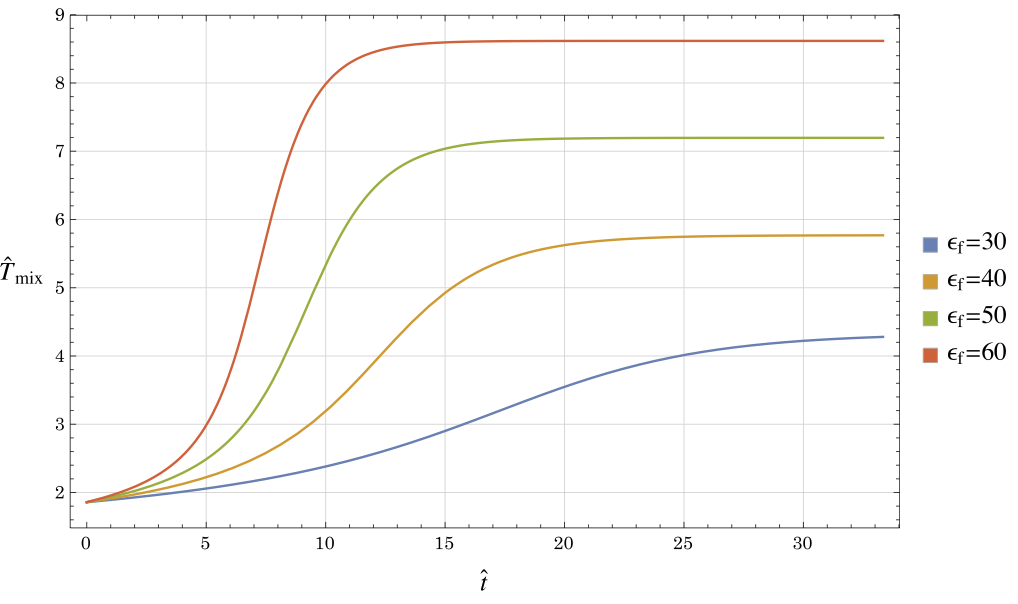} 
\hfill
\end{center}
\caption{Mixture temperature for different reaction heats.} \label{Tmix heat}
\end{figure}

\autoref{Tmix heat} shows the variation of the mixture temperature $\hat{T}$ for different 
values of the heat of the reaction. Since all the cases are exothermic, the 
equilibrium mixture temperature increases as the forward activation energy ($\epsilon_f$)
increases. Furthermore, increasing the heat of the reaction results in a faster increase in the mixture's temperature, leading to a faster relaxation of the moment system to the equilibrium.
\FloatBarrier
 \begin{table}[!ht]
\centering
  \begin{tabular}{ |c|c|c|c|c|c|c| }
    \hline
    Cases & $\epsilon_f$ & $\epsilon_r$ & $Q$ & $\mathrm{T}^{(eq)}$ \\ \hline
  1 & 30 & 10 & 20 & 4.2 \\ \hline
   2 & 40 & 10 & 30 & 5.8 \\ \hline
   3 & 50 & 10 & 40 & 7.2 \\ \hline
    4 & 60 & 10 & 50 & 8.7\\ \hline
    \end{tabular}
  \caption{Activation energies and the equilibrium values for temperatures.}	\label{act energy heat}
  \end{table}
\section{Conclusion}
We derived the Grad's--14 moment equations for a chemically reactive quaternary gaseous mixture. The main contribution was an algorithm to compute the moments of the Boltzmann's collision operator. The algorithm relied on a novel mathematical model that describes the microscopic details of a chemically reactive collision. To develop the model, we assumed that chemical reactions affect the tangential and the normal pre-collisional relative velocity equally, leading to a symmetric collision. Using the collision model, we derived relations between the pre and the post collisional velocities. These relations helped us extend the framework for the computation of the moments of a single-gas collision operator to a reactive gas mixture. For first-order chemical kinetics, we derived reaction rates for chemical reactions outside of equilibrium and extended the Arhenius law. The reaction rates were found to have an explicit dependence on the scalar fourteenth moment, highlighting the importance of considering a fourteen moment system rather than a thirteen one. Our numerical experiments showcased that the 14-moment system can correctly reproduce the equilibrium states under different initializations and physical parameters. 

\medskip
\bibliographystyle{unsrt}
\bibliography{../bib_files/papers,../bib_files/books}
\section{Supplementary material: Omega Integrals}
\subsection{Omega integrals for the reverse reaction} \label{omega reverse}
\begin{align}
 \Omega^{(3,2)}_{\epsilon_r} &= \frac{4 m_2^2 \left(2 \theta_{34} m_{34} \Gamma \left(2,\frac{\epsilon_r}{2 \theta_{34} m_{34}}\right)+Q \Gamma \left(1,\frac{\epsilon_r}{2 \text{$\theta_{34}$} m_{34}}\right)\right)}{\text{$\theta_{34}$} M^2 m_{12}}\\
 \Omega^{(5,2)}_{\epsilon_r} &= \frac{16 m_2^2 \left(2 \theta_{34} m_{34} \Gamma \left(3,\frac{\epsilon_r}{2 \theta_{34} m_{34}}\right)+Q \Gamma \left(2,\frac{\epsilon_r}{2 \theta_{34} m_{34}}\right)\right)}{\theta_{34} M^2 m_{12}}\\
 \Omega^{(5,4)}_{\epsilon_r} &= \frac{8 m_2^4 \left(Q^2 \Gamma \left(1,\frac{\epsilon_r}{2 \theta_{34} m_{34}}\right)+4 \theta_{34} m_{34} \left(\theta_{34} m_{34} \Gamma \left(3,\frac{\epsilon_r}{2 \theta_{34} m_{34}}\right)+Q \Gamma \left(2,\frac{\epsilon_r}{2 \theta_{34} m_{34}}\right)\right)\right)}{\theta_{34}^2 M^4 m_{12}^2}\\
 \Omega^{(7,2)}_{\epsilon_r} &= \frac{64 m_2^2 \left(2 \theta_{34} m_{34} \Gamma \left(4,\frac{\epsilon_r}{2 \theta_{34} m_{34}}\right)+Q \Gamma \left(3,\frac{\epsilon_r}{2 \theta_{34} m_{34}}\right)\right)}{\theta_{34} M^2 m_{12}}\\
 \Omega^{(9,2)}_{\epsilon_r} &= \frac{256 m_2^2 \left(2 \theta_{34} m_{34} \Gamma \left(5,\frac{\epsilon_r}{2 \theta_{34} m_{34}}\right)+Q \Gamma \left(4,\frac{\epsilon_r}{2 \theta_{34} m_{34}}\right)\right)}{\theta_{34} M^2 m_{12}}\\
 \Omega^{(7,4)}_{\epsilon_r} &= \frac{32 m_2^4 \left(Q^2 \Gamma \left(2,\frac{\epsilon_r}{2 \theta_{34} m_{34}}\right)+4 \theta_{34} m_{34} \left(\theta_{34} m_{34} \Gamma \left(4,\frac{\epsilon_r}{2 \theta_{34} m_{34}}\right)+Q \Gamma \left(3,\frac{\epsilon_r}{2 \theta_{34} m_{34}}\right)\right)\right)}{\theta_{34}^2 M^4 m_{12}^2}\\
 \Omega^{(9,4)}_{\epsilon_r} &= \frac{128 m_2^4 \left(Q^2 \Gamma \left(3,\frac{\epsilon_r}{2 \theta_{34} m_{34}}\right)+4 \theta_{34} m_{34} \left(\theta_{34} m_{34} \Gamma \left(5,\frac{\epsilon_r}{2 \theta_{34} m_{34}}\right)+Q \Gamma \left(4,\frac{\epsilon_r}{2 \theta_{34} m_{34}}\right)\right)\right)}{\theta_{34}^2 M^4 m_{12}^2}\\
 \Omega^{(11,2)}_{\epsilon_r} &= \frac{1024 m_2^2 \left(2 \theta_{34} m_{34} \Gamma \left(6,\frac{\epsilon_r}{2 \theta_{34} m_{34}}\right)+Q \Gamma \left(5,\frac{\epsilon_r}{2 \theta_{34} m_{34}}\right)\right)}{\theta_{34} M^2 m_{12}}\\
 \Omega^{(11,4)}_{\epsilon_r} &= \frac{512 m_2^4 \left(Q^2 \Gamma \left(4,\frac{\epsilon_r}{2 \theta_{34} m_{34}}\right)+4 \theta_{34} m_{34} \left(\theta_{34} m_{34} \Gamma \left(6,\frac{\epsilon_r}{2 \theta_{34} m_{34}}\right)+Q \Gamma \left(5,\frac{\epsilon_r}{2 \theta_{34} m_{34}}\right)\right)\right)}{\theta_{34}^2 M^4 m_{12}^2}
 \end{align}
%
%
 \subsection{Omega integrals for the forward reaction} \label{omega forward}
 \begin{align}
  \Omega^{(3,2)}_{\epsilon_f} &= \frac{4 m_4^2 \left(2 \theta_{12} m_{12} \Gamma \left(2,\frac{\epsilon_f}{2 \theta_{12} m_{12}}\right)-Q \Gamma \left(1,\frac{\epsilon_f}{2 \theta_{12} m_{12}}\right)\right)}{\theta_{12} M^2 m_{34}}\\
  \Omega^{(5,2)}_{\epsilon_f} &= \frac{16 m_4^2 \left(2 \theta_{12} m_{12} \Gamma \left(3,\frac{\epsilon_f}{2 \theta_{12} m_{12}}\right)-Q \Gamma \left(2,\frac{\epsilon_f}{2 \theta_{12} m_{12}}\right)\right)}{\theta_{12} M^2 m_{34}}\\
  \Omega^{(5,4)}_{\epsilon_f} &= \frac{8 m_4^4 \left(Q^2 \Gamma \left(1,\frac{\epsilon_f}{2 \theta_{12} m_{12}}\right)+4 \theta_{12} m_{12} \left(\theta_{12} m_{12} \Gamma \left(3,\frac{\epsilon_f}{2 \theta_{12} m_{12}}\right)-Q \Gamma \left(2,\frac{\epsilon_f}{2 \theta_{12} m_{12}}\right)\right)\right)}{\theta_{12}^2 M^4 m_{34}^2}\\
  \Omega^{(7,2)}_{\epsilon_f} &= \frac{64 m_4^2 \left(2 \theta_{12} m_{12} \Gamma \left(4,\frac{\epsilon_f}{2 \theta_{12} m_{12}}\right)-Q \Gamma \left(3,\frac{\epsilon_f}{2 \theta_{12} m_{12}}\right)\right)}{\theta_{12} M^2 m_{34}}\\
  \Omega^{(9,2)}_{\epsilon_f} &= \frac{256 m_4^2 \left(2 \theta_{12} m_{12} \Gamma \left(5,\frac{\epsilon_f}{2 \theta_{12} m_{12}}\right)-Q \Gamma \left(4,\frac{\epsilon_f}{2 \theta_{12} m_{12}}\right)\right)}{\theta_{12} M^2 m_{34}}\\
  \Omega^{(7,4)}_{\epsilon_f} &= \frac{32 m_4^4 \left(Q^2 \Gamma \left(2,\frac{\epsilon_f}{2 \theta_{12} m_{12}}\right)+4 \theta_{12} m_{12} \left(\theta_{12} m_{12} \Gamma \left(4,\frac{\epsilon_f}{2 \theta_{12} m_{12}}\right)-Q \Gamma \left(3,\frac{\epsilon_f}{2 \theta_{12} m_{12}}\right)\right)\right)}{\theta_{12}^2 M^4 m_{34}^2}\\
  \Omega^{(9,4)}_{\epsilon_f} &= \frac{128 m_4^4 \left(Q^2 \Gamma \left(3,\frac{\epsilon_f}{2 \theta_{12} m_{12}}\right)+4 \theta_{12} m_{12} \left(\theta_{12} m_{12} \Gamma \left(5,\frac{\epsilon_f}{2 \theta_{12} m_{12}}\right)-Q \Gamma \left(4,\frac{\epsilon_f}{2 \theta_{12} m_{12}}\right)\right)\right)}{\theta_{12}^2 M^4 m_{34}^2}\\
  \Omega^{(11,2)}_{\epsilon_f} &= \frac{1024 m_4^2 \left(2 \theta_{12} m_{12} \Gamma \left(6,\frac{\epsilon_f}{2 \theta_{12} m_{12}}\right)-Q \Gamma \left(5,\frac{\epsilon_f}{2 \theta_{12} m_{12}}\right)\right)}{\theta_{12} M^2 m_{34}}\\
  \Omega^{(11,4)}_{\epsilon_f} &= \frac{512 m_4^4 \left(Q^2 \Gamma \left(4,\frac{\epsilon_f}{2 \theta_{12} m_{12}}\right)+4 \theta_{12} m_{12} \left(\theta_{12} m_{12} \Gamma \left(6,\frac{\epsilon_f}{2 \theta_{12} m_{12}}\right)-Q \Gamma \left(5,\frac{\epsilon_f}{2 \theta_{12} m_{12}}\right)\right)\right)}{\theta_{12}^2 M^4 m_{34}^2}
 \end{align}
where $M = m_1 + m_2 = m_3 + m_4$.


\section{Supplementary material: Chemical production terms}\label{app: chem prod}
\subsection{Mass Balance}
\begin{itemize}
 \item Mass Balance
\begin{align}
\mcal P^{\alpha=1}_{R,0} =d_{r}^2 \tilde{n}_{34}\mu_{12} \exp\left[{-\frac{\epsilon_r}{2 \theta_{34} m_{34}}}\right]-d_{f}^2\tilde{n}_{12}\mu_{12} \exp\left[{-\frac{\epsilon_f}{2 \theta_{12} m_{12}}}\right] + \Delta^{r(1)0}	\label{pmass1}
\end{align}
\begin{align}
\mcal P^{\alpha=2}_{R,0} = d_{r}^2 \tilde{n}_{34}  \mu_{21} \exp\left[{-\frac{\epsilon_r}{2 \theta_{34} m_{34}}}\right]- d_{f}^2 \tilde{n}_{12} \mu_{21} \exp \left[{-\frac{\epsilon_f}{2 \theta_{12} m_{12}}}\right] + \Delta^{r(2)0} \label{pmass2}
\end{align}
\begin{align}
\mcal P^{\alpha=3}_{R,0} =d_{f}^2 \tilde{n}_{12}\mu_{34} \exp\left[{-\frac{\epsilon_f}{2 \theta_{12} m_{12}}}\right]- d_{r}^2\tilde{n}_{34}\mu_{34}\exp \left[{-\frac{\epsilon_r}{2 \theta_{34} m_{34}}}\right] + \Delta^{r(3)0}	\label{pmass3}
\end{align}
\begin{align}
\mcal P^{\alpha=4}_{R,0}= d_{f}^2\tilde{n}_{12} \mu_{43}\exp \left[{-\frac{\epsilon_f}{2 \theta_{12} m_{12}}}\right]-d_{r}^2 \tilde{n}_{34} \mu_{43} \exp \left[{-\frac{\epsilon_r}{2 \theta_{34} m_{34}}}\right] + \Delta^{r(4)0} \label{pmass4}
\end{align}
\end{itemize}
where $\tilde{n}_{34} = 4\sqrt{\pi} \sqrt{\theta_{34}}M n_3 n_4 $, $\tilde{n}_{12} = 4\sqrt{\pi} \sqrt{\theta_{12}}  M n_1 n_2$ and $\Delta^{r(\alpha)0}$ is the contribution from 
$\Delta_{\alpha}$. Explicit expressions for $\Delta^{r(\alpha)0}$ have been given in 
\autoref{contri Delta energy} and \autoref{contri Delta mass}.

\subsection{Momentum Balance} \label{momentum chemical}
The generic form of the momentum production can be given as%
\begin{align}
\mcal P^{\alpha}_{R,i} = \sum_{\gamma = 1}^4\left(\tau^{\alpha}_{\gamma}u^{(\gamma)}_i + \omega^{\alpha}_{\gamma}q^{(\gamma)}_i \right)
\end{align}
The coefficients $\tau^{\alpha}_{\gamma}$ and $\omega^{\alpha}_{\gamma}$ can now be given as
\subsubsection{Component-1}
\begin{dgroup*}
 \begin{dmath*} 
  \tau^{(1)}_1 = \frac{1}{6} \beta_f \sqrt{\theta_1}\mu_{12}\left(\theta_{12}^2 m_{12}^2 \left(24 \theta_1^2+8 \theta_1 (5 \theta_2-7 \theta_{12})+3 \theta_2 (5 \theta_2-14 \theta_{12})\right)+2 \theta_1 \theta_{12} m_{12} \epsilon_f (4 \theta_1-7 \theta_{12}+5 \theta_2)+\theta_1^2 \epsilon_f^2\right) \\\\
  \end{dmath*}	
 \begin{dmath*}
  \tau^{(1)}_2 = -\frac{1}{6} \beta_f \theta_1 \sqrt{\theta_2}\mu_{12}\left(\theta_2 \left(4 \theta_{12}^2 m_{12}^2+3 \theta_{12} m_{12} \epsilon_f+\epsilon_f^2\right)+\theta_{12} m_{12} (5 \theta_1-14 \theta_{12}) (\theta_{12} m_{12}+\epsilon_f)\right)\\\\
 \end{dmath*}
 \begin{dmath*}
  \tau^{(1)}_3 = \frac{1}{6} \beta_r \sqrt{\theta_3}\mu_{12}\left(\theta_{34}^2 m_{34}^2 \left(-24 \theta_3^2+4\mu_{34}(2\theta_{34}) (6 \theta_3-14 \theta_{34}+5 \theta_4)+56 \theta_3 \theta_{34}-40 \theta_3 \theta_4+42 \theta_{34} \theta_4-15 \theta_4^2\right)+\theta_{34} m_{34} \epsilon_r (\mu_{34} (2\theta_{34}) (8 \theta_3-14 \theta_{34}+5 \theta_4)-2 \theta_3 (4 \theta_3-7 \theta_{34}+5 \theta_4))+\theta_3 \epsilon_r^2 (\theta_3 (\mu_{34}-1)+\theta_4 \mu_{34})\right)\\\\
 \end{dmath*}
 \begin{dmath*}
  \tau^{(1)}_4 = -\frac{1}{6} \beta_r \sqrt{\theta_4}\mu_{12}\left(\mu_{34} (2\theta_{34}) \left(\theta_4 \left(24 \theta_{34}^2 m_{34}^2+8 \theta_{34} m_{34} \epsilon_r+\epsilon_r^2\right)+\theta_{34} m_{34} (5 \theta_3-14 \theta_{34}) (4 \theta_{34} m_{34}+\epsilon_r)\right)-\theta_3 \theta_4 \left(4 \theta_{34}^2 m_{34}^2+3 \theta_{34} m_{34} \epsilon_r+\epsilon_r^2\right)+\theta_3 \theta_{34} (-m_{34}) (5 \theta_3-14 \theta_{34}) (\theta_{34} m_{34}+\epsilon_r)\right)\\\\
 \end{dmath*}
 \begin{dmath*}
  \omega^{(1)}_1 = -\frac{1}{15} \beta_f \sqrt{\theta_1}\mu_{12}\left(\theta_{12}^2 m_{12}^2 \left(24 \theta_1^2+40 \theta_1 (\theta_2-\theta_{12})+15 \theta_2 (\theta_2-2 \theta_{12})\right)+2 \theta_1 \theta_{12} m_{12} \epsilon_f (4 \theta_1-5 \theta_{12}+5 \theta_2)+\theta_1^2 \epsilon_f^2\right)\\\\
 \end{dmath*}
 \begin{dmath*}
  \omega^{(1)}_2 = \frac{1}{15} \beta_f \theta_1 \sqrt{\theta_2}\mu_{12}\left(\theta_2 \left(4 \theta_{12}^2 m_{12}^2+3 \theta_{12} m_{12} \epsilon_f+\epsilon_f^2\right)+5 \theta_{12} m_{12} (\theta_1-2 \theta_{12}) (\theta_{12} m_{12}+\epsilon_f)\right)\\\\
 \end{dmath*}
 \begin{dmath*}
  \omega^{(1)}_3 = -\frac{1}{15} \beta_r \sqrt{\theta_3}\mu_{12}\left(\theta_{34}^2 m_{34}^2 \left(-24 \theta_3^2+4\mu_{34}(2\theta_{34}) (6 \theta_3+5 (\theta_4-2 \theta_{34}))+40 \theta_3 \theta_{34}-40 \theta_3 \theta_4+30 \theta_{34} \theta_4-15 \theta_4^2\right)+\theta_{34} m_{34} \epsilon_r (\mu_{34} (2\theta_{34}) (8 \theta_3+5 (\theta_4-2 \theta_{34}))-2 \theta_3 (4 \theta_3-5 \theta_{34}+5 \theta_4))+\theta_3 \epsilon_r^2 (\theta_3 (\mu_{34}-1)+\theta_4 \mu_{34})\right)\\\\
 \end{dmath*}
 \begin{dmath*}
  \omega^{(1)}_4 = \frac{1}{15} \beta_r \sqrt{\theta_4}\mu_{12}\left(\mu_{34} (2\theta_{34}) \left(\theta_4 \left(24 \theta_{34}^2 m_{34}^2+8 \theta_{34} m_{34} \epsilon_r+\epsilon_r^2\right)+5 \theta_{34} m_{34} (\theta_3-2 \theta_{34}) (4 \theta_{34} m_{34}+\epsilon_r)\right)-\theta_3 \theta_4 \left(4 \theta_{34}^2 m_{34}^2+3 \theta_{34} m_{34} \epsilon_r+\epsilon_r^2\right)-5 \theta_3 \theta_{34} m_{34} (\theta_3-2 \theta_{34}) (\theta_{34} m_{34}+\epsilon_r)\right)\\\\
 \end{dmath*}
\end{dgroup*}

\subsubsection{Component-2}
\begin{dgroup*}
 \begin{dmath*}
  \tau^{(2)}_{1} = -\frac{1}{6} \beta_f \sqrt{\theta_1} \theta_2 \mu_{21} \left(\theta_{12}^2 m_{12}^2 (4 \theta_1-14 \theta_{12}+5 \theta_2)+\theta_{12} m_{12} \epsilon_f (3 \theta_1-14 \theta_{12}+5 \theta_2)+\theta_1 \epsilon_f^2\right)\\\\
 \end{dmath*}
 \begin{dmath*}
  \tau^{(2)}_2 = \frac{1}{6} \beta_f \sqrt{\theta_2} \mu_{21} \left(3 \theta_1 \theta_{12}^2 m_{12}^2 (5 \theta_1-14 \theta_{12})+\theta_2^2 \left(24 \theta_{12}^2 m_{12}^2+8 \theta_{12} m_{12} \epsilon_f+\epsilon_f^2\right)+2 \theta_{12} \theta_2 m_{12} (5 \theta_1-7 \theta_{12}) (4 \theta_{12} m_{12}+\epsilon_f)\right)\\\\
 \end{dmath*}
 \begin{dmath*}
  \tau^{(2)}_3 = -\frac{1}{6} \beta_r \sqrt{\theta_3} \mu_{21} \left(\theta_{34}^2 m_{34}^2 (4 \mu_{34} (2\theta_{34}) (6 \theta_3-14 \theta_{34}+5 \theta_4)+\theta_4 (-4 \theta_3+14 \theta_{34}-5 \theta_4))+\theta_{34} m_{34} \epsilon_r (\mu_{34} (2\theta_{34}) (8 \theta_3-14 \theta_{34}+5 \theta_4)+\theta_4 (-3 \theta_3+14 \theta_{34}-5 \theta_4))+\theta_3 \epsilon_r^2 (\theta_3 \mu_{34}+\theta_4 (\mu_{34}-1))\right)\\\\
 \end{dmath*}
 \begin{dmath*}
  \tau^{(2)}_4 = \frac{1}{6} \beta_r \sqrt{\theta_4} \mu_{21} \left(\theta_{34}^2 m_{34}^2 \left(-15 \theta_3^2+4 \mu_{34} (2\theta_{34}) (5 \theta_3-14 \theta_{34}+6 \theta_4)+42 \theta_3 \theta_{34}-40 \theta_3 \theta_4+56 \theta_{34} \theta_4-24 \theta_4^2\right)+\theta_{34} m_{34} \epsilon_r (\mu_{34} (2\theta_{34}) (5 \theta_3-14 \theta_{34}+8 \theta_4)-2 \theta_4 (5 \theta_3-7 \theta_{34}+4 \theta_4))+\theta_4 \epsilon_r^2 (\theta_3 \mu_{34}+\theta_4 (\mu_{34}-1))\right)\\\\
 \end{dmath*}
 \begin{dmath*}
  \omega^{(2)}_1 = \frac{1}{15} \beta_f \sqrt{\theta_1} \theta_2 \mu_{21} \left(\theta_{12}^2 m_{12}^2 (4 \theta_1+5 (\theta_2-2 \theta_{12}))+\theta_{12} m_{12} \epsilon_f (3 \theta_1+5 (\theta_2-2 \theta_{12}))+\theta_1 \epsilon_f^2\right)\\\\
 \end{dmath*}
 \begin{dmath*}
  \omega^{(2)}_2 = -\frac{1}{15} \beta_f \sqrt{\theta_2} \mu_{21} \left(15 \theta_1 \theta_{12}^2 m_{12}^2 (\theta_1-2 \theta_{12})+\theta_2^2 \left(24 \theta_{12}^2 m_{12}^2+8 \theta_{12} m_{12} \epsilon_f+\epsilon_f^2\right)+10 \theta_{12} \theta_2 m_{12} (\theta_1-\theta_{12}) (4 \theta_{12} m_{12}+\epsilon_f)\right)\\\\
 \end{dmath*}
  \begin{dmath*}
   \omega^{(2)}_3 = \frac{1}{15} \beta_r \sqrt{\theta_3} \mu_{21} \left(\theta_{34}^2 m_{34}^2 (4 \mu_{34} (2\theta_{34}) (6 \theta_3+5 (\theta_4-2 \theta_{34}))+\theta_4 (-4 \theta_3+10 \theta_{34}-5 \theta_4))+\theta_{34} m_{34} \epsilon_r (\mu_{34} (2\theta_{34}) (8 \theta_3+5 (\theta_4-2 \theta_{34}))+\theta_4 (-3 \theta_3+10 \theta_{34}-5 \theta_4))+\theta_3 \epsilon_r^2 (\theta_3 \mu_{34}+\theta_4 (\mu_{34}-1))\right)\\\\
  \end{dmath*}
  \begin{dmath*}
   \omega^{(2)}_4 = -\frac{1}{15} \beta_r \sqrt{\theta_4} \mu_{21} \left(\theta_{34}^2 m_{34}^2 \left(-15 \theta_3^2+4 \mu_{34} (2\theta_{34}) (5 \theta_3-10 \theta_{34}+6 \theta_4)+30 \theta_3 \theta_{34}-40 \theta_3 \theta_4+40 \theta_{34} \theta_4-24 \theta_4^2\right)+\theta_{34} m_{34} \epsilon_r (\mu_{34} (2\theta_{34}) (5 \theta_3-10 \theta_{34}+8 \theta_4)-2 \theta_4 (5 \theta_3-5 \theta_{34}+4 \theta_4))+\theta_4 \epsilon_r^2 (\theta_3 \mu_{34}+\theta_4 (\mu_{34}-1))\right)\\\\
  \end{dmath*}
\end{dgroup*}

\subsubsection{Component-3}
\begin{dgroup*}
 \begin{dmath*}
  \tau^{(3)}_1 = \frac{1}{6} \beta_f \sqrt{\theta_1} \mu_{34} \left(\theta_{12}^2 m_{12}^2 \left(-24 \theta_1^2+4 \mu_{21} (2\theta_{12}) (6 \theta_1-14 \theta_{12}+5 \theta_2)+56 \theta_1 \theta_{12}-40 \theta_1 \theta_2+42 \theta_{12} \theta_2-15 \theta_2^2\right)+\theta_{12} m_{12} \epsilon_f (\mu_{21} (2\theta_{12}) (8 \theta_1-14 \theta_{12}+5 \theta_2)-2 \theta_1 (4 \theta_1-7 \theta_{12}+5 \theta_2))+\theta_1 \epsilon_f^2 (\theta_1 (\mu_{21}-1)+\theta_2 \mu_{21})\right)\\\\
 \end{dmath*}
 \begin{dmath*}
  \tau^{(3)}_2 = -\frac{1}{6} \beta_f \sqrt{\theta_2} \mu_{34} \left(\mu_{21} (2\theta_{12}) \left(\theta_2 \left(24 \theta_{12}^2 m_{12}^2+8 \theta_{12} m_{12} \epsilon_f+\epsilon_f^2\right)+\theta_{12} m_{12} (5 \theta_1-14 \theta_{12}) (4 \theta_{12} m_{12}+\epsilon_f)\right)-\theta_1 \theta_2 \left(4 \theta_{12}^2 m_{12}^2+3 \theta_{12} m_{12} \epsilon_f+\epsilon_f^2\right)+\theta_1 \theta_{12} (-m_{12}) (5 \theta_1-14 \theta_{12}) (\theta_{12} m_{12}+\epsilon_f)\right)\\\\
 \end{dmath*}
  \begin{dmath*}
   \tau^{(3)}_3 = \frac{1}{6} \beta_r \sqrt{\theta_3} \mu_{34} \left(\theta_{34}^2 m_{34}^2 \left(24 \theta_3^2+8 \theta_3 (5 \theta_4-7 \theta_{34})+3 \theta_4 (5 \theta_4-14 \theta_{34})\right)+2 \theta_3 \theta_{34} m_{34} \epsilon_r (4 \theta_3-7 \theta_{34}+5 \theta_4)+\theta_3^2 \epsilon_r^2\right)\\\\
  \end{dmath*}
  \begin{dmath*}
   \tau^{(3)}_4 = -\frac{1}{6} \beta_r \theta_3 \sqrt{\theta_4} \mu_{34} \left(\theta_4 \left(4 \theta_{34}^2 m_{34}^2+3 \theta_{34} m_{34} \epsilon_r+\epsilon_r^2\right)+\theta_{34} m_{34} (5 \theta_3-14 \theta_{34}) (\theta_{34} m_{34}+\epsilon_r)\right)\\\\
  \end{dmath*}
  \begin{dmath*}
   \omega^{(3)}_1 = -\frac{1}{15} \beta_f \sqrt{\theta_1} \mu_{34} \left(\theta_{12}^2 m_{12}^2 \left(-24 \theta_1^2+4 \mu_{21} (2\theta_{12}) (6 \theta_1+5 (\theta_2-2 \theta_{12}))+40 \theta_1 \theta_{12}-40 \theta_1 \theta_2+30 \theta_{12} \theta_2-15 \theta_2^2\right)+\theta_{12} m_{12} \epsilon_f (\mu_{21} (2\theta_{12}) (8 \theta_1+5 (\theta_2-2 \theta_{12}))-2 \theta_1 (4 \theta_1-5 \theta_{12}+5 \theta_2))+\theta_1 \epsilon_f^2 (\theta_1 (\mu_{21}-1)+\theta_2 \mu_{21})\right)\\\\
  \end{dmath*}
  \begin{dmath*}
   \omega^{(3)}_2 = \frac{1}{15} \beta_f \sqrt{\theta_2} \mu_{34} \left(\mu_{21} (2\theta_{12}) \left(\theta_2 \left(24 \theta_{12}^2 m_{12}^2+8 \theta_{12} m_{12} \epsilon_f+\epsilon_f^2\right)+5 \theta_{12} m_{12} (\theta_1-2 \theta_{12}) (4 \theta_{12} m_{12}+\epsilon_f)\right)-\theta_1 \theta_2 \left(4 \theta_{12}^2 m_{12}^2+3 \theta_{12} m_{12} \epsilon_f+\epsilon_f^2\right)-5 \theta_1 \theta_{12} m_{12} (\theta_1-2 \theta_{12}) (\theta_{12} m_{12}+\epsilon_f)\right)\\\\
  \end{dmath*}
  \begin{dmath*}
   \omega^{(3)}_3 = -\frac{1}{15} \beta_r \sqrt{\theta_3} \mu_{34} \left(\theta_{34}^2 m_{34}^2 \left(24 \theta_3^2+40 \theta_3 (\theta_4-\theta_{34})+15 \theta_4 (\theta_4-2 \theta_{34})\right)+2 \theta_3 \theta_{34} m_{34} \epsilon_r (4 \theta_3-5 \theta_{34}+5 \theta_4)+\theta_3^2 \epsilon_r^2\right)\\\\
  \end{dmath*}
  \begin{dmath*}
   \omega^{(3)}_4 = \frac{1}{15} \beta_r \theta_3 \sqrt{\theta_4} \mu_{34} \left(\theta_4 \left(4 \theta_{34}^2 m_{34}^2+3 \theta_{34} m_{34} \epsilon_r+\epsilon_r^2\right)+5 \theta_{34} m_{34} (\theta_3-2 \theta_{34}) (\theta_{34} m_{34}+\epsilon_r)\right)\\\\
  \end{dmath*}
\end{dgroup*}
\subsubsection{Component-4}

\begin{dgroup*}
 \begin{dmath*}
  \tau^{(4)}_1 = -\frac{1}{6} \beta_f \sqrt{\theta_1} \mu_{43} \left(\theta_{12}^2 m_{12}^2 (4 \mu_{12} (2\theta_{12}) (6 \theta_1-14 \theta_{12}+5 \theta_2)+\theta_2 (-4 \theta_1+14 \theta_{12}-5 \theta_2))+\theta_{12} m_{12} \epsilon_f (\mu_{12} (2\theta_{12}) (8 \theta_1-14 \theta_{12}+5 \theta_2)+\theta_2 (-3 \theta_1+14 \theta_{12}-5 \theta_2))+\theta_1 \epsilon_f^2 (\theta_1 \mu_{12}+\theta_2 (\mu_{12}-1))\right)\\\\
 \end{dmath*}
  \begin{dmath*}
  \tau^{(4)}_2 = \frac{1}{6} \beta_f \sqrt{\theta_2} \mu_{43} \left(\theta_{12}^2 m_{12}^2 \left(-15 \theta_1^2+4 \mu_{12} (2\theta_{12}) (5 \theta_1-14 \theta_{12}+6 \theta_2)+42 \theta_1 \theta_{12}-40 \theta_1 \theta_2+56 \theta_{12} \theta_2-24 \theta_2^2\right)+\theta_{12} m_{12} \epsilon_f (\mu_{12} (2\theta_{12}) (5 \theta_1-14 \theta_{12}+8 \theta_2)-2 \theta_2 (5 \theta_1-7 \theta_{12}+4 \theta_2))+\theta_2 \epsilon_f^2 (\theta_1 \mu_{12}+\theta_2 (\mu_{12}-1))\right)\\\\
 \end{dmath*}
   \begin{dmath*}
  \tau^{(4)}_3 = -\frac{1}{6} \beta_r \sqrt{\theta_3} \theta_4 \mu_{43} \left(\theta_{34}^2 m_{34}^2 (4 \theta_3-14 \theta_{34}+5 \theta_4)+\theta_{34} m_{34} \epsilon_r (3 \theta_3-14 \theta_{34}+5 \theta_4)+\theta_3 \epsilon_r^2\right)\\\\
 \end{dmath*}
   \begin{dmath*}
  \tau^{(4)}_4 = \frac{1}{6} \beta_r \sqrt{\theta_4} \mu_{43} \left(3 \theta_3 \theta_{34}^2 m_{34}^2 (5 \theta_3-14 \theta_{34})+\theta_4^2 \left(24 \theta_{34}^2 m_{34}^2+8 \theta_{34} m_{34} \epsilon_r+\epsilon_r^2\right)+2 \theta_{34} \theta_4 m_{34} (5 \theta_3-7 \theta_{34}) (4 \theta_{34} m_{34}+\epsilon_r)\right)\\\\
 \end{dmath*}
  \begin{dmath*}
  \omega^{(4)}_1 = \frac{1}{15} \beta_f \sqrt{\theta_1} \mu_{43} \left(\theta_{12}^2 m_{12}^2 (4 \mu_{12} (2\theta_{12}) (6 \theta_1+5 (\theta_2-2 \theta_{12}))+\theta_2 (-4 \theta_1+10 \theta_{12}-5 \theta_2))+\theta_{12} m_{12} \epsilon_f (\mu_{12} (2\theta_{12}) (8 \theta_1+5 (\theta_2-2 \theta_{12}))+\theta_2 (-3 \theta_1+10 \theta_{12}-5 \theta_2))+\theta_1 \epsilon_f^2 (\theta_1 \mu_{12}+\theta_2 (\mu_{12}-1))\right)\\\\
 \end{dmath*}
  \begin{dmath*}
    \omega^{(4)}_2 = -\frac{1}{15} \beta_f \sqrt{\theta_2} \mu_{43} \left(\theta_{12}^2 m_{12}^2 \left(-15 \theta_1^2+4 \mu_{12} (2\theta_{12}) (5 \theta_1-10 \theta_{12}+6 \theta_2)+30 \theta_1 \theta_{12}-40 \theta_1 \theta_2+40 \theta_{12} \theta_2-24 \theta_2^2\right)+\theta_{12} m_{12} \epsilon_f (\mu_{12} (2\theta_{12}) (5 \theta_1-10 \theta_{12}+8 \theta_2)-2 \theta_2 (5 \theta_1-5 \theta_{12}+4 \theta_2))+\theta_2 \epsilon_f^2 (\theta_1 \mu_{12}+\theta_2 (\mu_{12}-1))\right)\\\\
  \end{dmath*}
  \begin{dmath*}
    \omega^{(4)}_3 = \frac{1}{15} \beta_r \sqrt{\theta_3} \theta_4 \mu_{43} \left(\theta_{34}^2 m_{34}^2 (4 \theta_3+5 (\theta_4-2 \theta_{34}))+\theta_{34} m_{34} \epsilon_r (3 \theta_3+5 (\theta_4-2 \theta_{34}))+\theta_3 \epsilon_r^2\right)\\\\
  \end{dmath*}
  \begin{dmath*}
    \omega^{(4)}_4 = -\frac{1}{15} \beta_r \sqrt{\theta_4} \mu_{43} \left(15 \theta_3 \theta_{34}^2 m_{34}^2 (\theta_3-2 \theta_{34})+\theta_4^2 \left(24 \theta_{34}^2 m_{34}^2+8 \theta_{34} m_{34} \epsilon_r+\epsilon_r^2\right)+10 \theta_{34} \theta_4 m_{34} (\theta_3-\theta_{34}) (4 \theta_{34} m_{34}+\epsilon_r)\right)\\\\
  \end{dmath*}
\end{dgroup*}
\begin{gather}
\beta_f = \left(\frac{\sqrt{\pi } d_{12}^2 M n_1 n_2 s_f^2 }{ \theta_{12}^{7/2} m_{12}^2} \right) \exp\left[{-\frac{\epsilon_f}{2 \theta_{12} m_{12}}}\right], \quad   \beta_r = \left(\frac{\sqrt{\pi } d_{34}^2  M n_3 n_4 s_r^2}{\theta_{34}^{7/2} m_{34}^2}\right)\exp\left[{-\frac{\epsilon_r}{2 \theta_{34} m_{34}}}\right]
\end{gather}
\subsection{Energy Balance} \label{energy chemical}
The following are the production terms corresponding to energy balance for all the components present in the mixture.
\subsubsection{Component-1}
\begin{dgroup*}
 \begin{dmath*}
\mcal P^{\alpha}_{R,1} = \frac{1}{8} \beta_r \theta_{34}^2 \mu_{12} \left(\theta_{34} (2\theta_{34}) e^{\frac{\epsilon_r}{2 \theta_{34} m_{34}}} (\theta_{34} m_{34} \Omega^{(5,2)}_{\epsilon_r}-2 \epsilon_r \Omega^{(3,2)}_{\epsilon_r})+8 \left(2 \theta_{34} \mu_{34}^2 (2\theta_{34}) (4 \theta_{34} m_{34}+\epsilon_r)-2 \theta_3 \mu_{34} (2\theta_{34}) (4 \theta_{34} m_{34}+\epsilon_r)+\theta_3 (\theta_{34} m_{34} (4 \theta_3+3 \theta_4)+\theta_3 \epsilon_r)\right)\right)-\beta_f \theta_1 \theta_{12}^2 \mu_{12} (\theta_{12} m_{12} (4 \theta_1+3 \theta_2)+\theta_1 \epsilon_f) + \Delta^{r(1)1}\\\\
 \end{dmath*}
\end{dgroup*}
\subsubsection{Component-2}
\begin{dgroup*}
 \begin{dmath*}
\mcal P^{\alpha}_{R,1} = \frac{\beta_r \theta_{34}^2}{8 \mu_{21}}\left(\theta_{34} \mu_{12}^2 (2\theta_{34}) e^{\frac{\epsilon_r}{2 \theta_{34} m_{34}}} (\theta_{34} m_{34} \Omega^{(5,2)}_{\epsilon_r}-2 \epsilon_r \Omega^{(3,2)}_{\epsilon_r})+8 \mu_{21}^2 \left(2 \theta_{34} \mu_{34}^2 (2\theta_{34}) (4 \theta_{34} m_{34}+\epsilon_r)-2 \theta_4 \mu_{34} (2\theta_{34}) (4 \theta_{34} m_{34}+\epsilon_r)+\theta_4 (\theta_{34} m_{34} (3 \theta_3+4 \theta_4)+\theta_4 \epsilon_r)\right)\right)-\beta_f \theta_{12}^2 \theta_2 \mu_{21} (\theta_{12} m_{12} (3 \theta_1+4 \theta_2)+\theta_2 \epsilon_f) + \Delta^{r(2)1}\\\\
 \end{dmath*}
 \end{dgroup*}
\subsubsection{Component-3}
\begin{dgroup*}
 \begin{dmath*}
\mcal P^{\alpha}_{R,1} = \frac{1}{8} \beta_f \theta_{12}^2 \mu_{34} \left(\theta_{12} (2\theta_{12}) e^{\frac{\epsilon_f}{2 \theta_{12} m_{12}}} (\theta_{12} m_{12} \Omega^{(5,2)}_{\epsilon_f}-2 \epsilon_f \Omega^{(3,2)}_{\epsilon_f})+8 \left(2 \theta_{12} \mu_{21}^2 (2\theta_{12}) (4 \theta_{12} m_{12}+\epsilon_f)-2 \theta_1 \mu_{21} (2\theta_{12}) (4 \theta_{12} m_{12}+\epsilon_f)+\theta_1 (\theta_{12} m_{12} (4 \theta_1+3 \theta_2)+\theta_1 \epsilon_f)\right)\right)-\beta_r \theta_3 \theta_{34}^2 \mu_{34} (\theta_{34} m_{34} (4 \theta_3+3 \theta_4)+\theta_3 \epsilon_r) + \Delta^{r(3)1}\\\\
 \end{dmath*}
\end{dgroup*}
\subsubsection{Component-4}
\begin{dgroup*}
 \begin{dmath*}
\mcal P^{\alpha}_{R,1} = \frac{\beta_f \theta_{12}^2}{8 \mu_{43}}\left(\theta_{12} \mu_{34}^2 (2\theta_{12}) e^{\frac{\epsilon_f}{2 \theta_{12} m_{12}}} (\theta_{12} m_{12} \Omega^{(5,2)}_{\epsilon_f}-2 \epsilon_f \Omega^{(3,2)}_{\epsilon_f})+8 \mu_{43}^2 \left(2 \theta_{12} \mu_{12}^2 (2\theta_{12}) (4 \theta_{12} m_{12}+\epsilon_f)-2 \theta_2 \mu_{12} (2\theta_{12}) (4 \theta_{12} m_{12}+\epsilon_f)+\theta_2 (\theta_{12} m_{12} (3 \theta_1+4 \theta_2)+\theta_2 \epsilon_f)\right)\right)-\beta_r \theta_{34}^2 \theta_4 \mu_{43} (\theta_{34} m_{34} (3 \theta_3+4 \theta_4)+\theta_4 \epsilon_r) + \Delta^{r(4)1}\\\\
 \end{dmath*}
\end{dgroup*}

\subsection{Stress Production} \label{stress chemical}
The production term for the stress tensor can be given in a generic form as
\begin{align}
\mcal P^{\alpha}_{R,ij} = \sum_{\gamma=1}^{4}\left(\xi^{(\alpha)}_{\gamma}\sigma^{(\alpha)}_{ij}\right)
\end{align}
The coefficients $\xi^{(\alpha)}_{\gamma}$ can now be given as 
\subsubsection{Component-1}
\begin{dgroup*}
 \begin{dmath*}
  \xi^{(1)}_1 = -\frac{1}{15} \beta_f \theta_1 \mu_{12} \left(\theta_{12}^2 m_{12}^2 \left(24 \theta_1^2+40 \theta_1 \theta_2+15 \theta_2^2\right)+2 \theta_1 \theta_{12} m_{12} \epsilon_f (4 \theta_1+5 \theta_2)+\theta_1^2 \epsilon_f^2\right)
 \end{dmath*}
  \begin{dmath*}
  \xi^{(1)}_2 = \frac{1}{15} \beta_f \theta_1^2 \theta_2 \mu_{12} \left(\theta_{12}^2 m_{12}^2+2 \theta_{12} m_{12} \epsilon_f-\epsilon_f^2\right)
 \end{dmath*}
  \begin{dmath*}
  \xi^{(1)}_3 = \frac{1}{15} \beta_r \theta_3 \mu_{12} \left(\theta_{34}^2 m_{34}^2 \left(24 \theta_3^2+48 \theta_{34} \mu_{34}^2 (2\theta_{34})-8 \mu_{34} (2\theta_{34}) (6 \theta_3+5 \theta_4)+40 \theta_3 \theta_4+15 \theta_4^2\right)+2 \theta_{34} m_{34} \epsilon_r \left(8 \theta_{34} \mu_{34}^2 (2\theta_{34})-\mu_{34} (2\theta_{34}) (8 \theta_3+5 \theta_4)+\theta_3 (4 \theta_3+5 \theta_4)\right)+\epsilon_r^2 \left(\theta_3^2+2 \theta_{34} \mu_{34}^2 (2\theta_{34})-2 \theta_3 \mu_{34} (2\theta_{34})\right)\right)\\\\
 \end{dmath*}
  \begin{dmath*}
  \xi^{(1)}_4 = \frac{1}{15} \beta_r \theta_4 \mu_{12} \left(\theta_3^2 \left(-\theta_{34}^2 m_{34}^2-2 \theta_{34} m_{34} \epsilon_r+\epsilon_r^2\right)+2 \theta_{34} \mu_{34}^2 (2\theta_{34}) \left(24 \theta_{34}^2 m_{34}^2+8 \theta_{34} m_{34} \epsilon_r+\epsilon_r^2\right)-2 \theta_3 \mu_{34} (2\theta_{34}) \left(4 \theta_{34}^2 m_{34}^2+3 \theta_{34} m_{34} \epsilon_r+\epsilon_r^2\right)\right)\\\\
 \end{dmath*}
\end{dgroup*}
\subsubsection{Component-2}
\begin{dgroup*}
 \begin{dmath*}
  \xi^{(2)}_1 = \frac{1}{15} \beta_f \theta_1 \theta_2^2 \mu_{21} \left(\theta_{12}^2 m_{12}^2+2 \theta_{12} m_{12} \epsilon_f-\epsilon_f^2\right)\\\\
 \end{dmath*}
 \begin{dmath*}
  \xi^{(2)}_2 = -\frac{1}{15} \beta_f \theta_2 \mu_{21} \left(\theta_{12}^2 m_{12}^2 \left(15 \theta_1^2+40 \theta_1 \theta_2+24 \theta_2^2\right)+2 \theta_{12} \theta_2 m_{12} \epsilon_f (5 \theta_1+4 \theta_2)+\theta_2^2 \epsilon_f^2\right)\\\\
 \end{dmath*}
 \begin{dmath*}
  \xi^{(2)}_3 = \frac{1}{15} \beta_r \theta_3 \mu_{21} \left(2 \theta_{34} \mu_{34}^2 (2\theta_{34}) \left(24 \theta_{34}^2 m_{34}^2+8 \theta_{34} m_{34} \epsilon_r+\epsilon_r^2\right)-2 \theta_4 \mu_{34} (2\theta_{34}) \left(4 \theta_{34}^2 m_{34}^2+3 \theta_{34} m_{34} \epsilon_r+\epsilon_r^2\right)+\theta_4^2 \left(-\theta_{34}^2 m_{34}^2-2 \theta_{34} m_{34} \epsilon_r+\epsilon_r^2\right)\right)\\\\
 \end{dmath*}
 \begin{dmath*}
  \xi^{(2)}_4 = \frac{1}{15} \beta_r \theta_4 \mu_{21} \left(\theta_{34}^2 m_{34}^2 \left(15 \theta_3^2+48 \theta_{34} \mu_{34}^2 (2\theta_{34})-8 \mu_{34} (2\theta_{34}) (5 \theta_3+6 \theta_4)+40 \theta_3 \theta_4+24 \theta_4^2\right)+2 \theta_{34} m_{34} \epsilon_r \left(8 \theta_{34} \mu_{34}^2 (2\theta_{34})-\mu_{34} (2\theta_{34}) (5 \theta_3+8 \theta_4)+\theta_4 (5 \theta_3+4 \theta_4)\right)+\epsilon_r^2 \left(2 \theta_{34} \mu_{34}^2 (2\theta_{34})-2 \theta_4 \mu_{34} (2\theta_{34})+\theta_4^2\right)\right)\\\\
 \end{dmath*}
\end{dgroup*}

\subsubsection{Component-3}
\begin{dgroup*}

 \begin{dmath*}
  \xi^{(3)}_1 = \frac{1}{15} \beta_f \theta_1 \mu_{34} \left(\theta_{12}^2 m_{12}^2 \left(24 \theta_1^2+48 \theta_{12} \mu_{21}^2 (2\theta_{12})-8 \mu_{21} (2\theta_{12}) (6 \theta_1+5 \theta_2)+40 \theta_1 \theta_2+15 \theta_2^2\right)+2 \theta_{12} m_{12} \epsilon_f \left(8 \theta_{12} \mu_{21}^2 (2\theta_{12})-\mu_{21} (2\theta_{12}) (8 \theta_1+5 \theta_2)+\theta_1 (4 \theta_1+5 \theta_2)\right)+\epsilon_f^2 \left(\theta_1^2+2 \theta_{12} \mu_{21}^2 (2\theta_{12})-2 \theta_1 \mu_{21} (2\theta_{12})\right)\right)\\\\
 \end{dmath*}

 \begin{dmath*}
  \xi^{(3)}_2 = \frac{1}{15} \beta_f \theta_2 \mu_{34} \left(\theta_1^2 \left(-\theta_{12}^2 m_{12}^2-2 \theta_{12} m_{12} \epsilon_f+\epsilon_f^2\right)+2 \theta_{12} \mu_{21}^2 (2\theta_{12}) \left(24 \theta_{12}^2 m_{12}^2+8 \theta_{12} m_{12} \epsilon_f+\epsilon_f^2\right)-2 \theta_1 \mu_{21} (2\theta_{12}) \left(4 \theta_{12}^2 m_{12}^2+3 \theta_{12} m_{12} \epsilon_f+\epsilon_f^2\right)\right)\\\\
 \end{dmath*}
 
 \begin{dmath*}
  \xi^{(3)}_3 = -\frac{1}{15} \beta_r \theta_3 \mu_{34} \left(\theta_{34}^2 m_{34}^2 \left(24 \theta_3^2+40 \theta_3 \theta_4+15 \theta_4^2\right)+2 \theta_3 \theta_{34} m_{34} \epsilon_r (4 \theta_3+5 \theta_4)+\theta_3^2 \epsilon_r^2\right)\\\\
 \end{dmath*}
 
 \begin{dmath*}
  \xi^{(3)}_4 = \frac{1}{15} \beta_r \theta_3^2 \theta_4 \mu_{34} \left(\theta_{34}^2 m_{34}^2+2 \theta_{34} m_{34} \epsilon_r-\epsilon_r^2\right)\\\\
 \end{dmath*}

\end{dgroup*}
\subsubsection{Component-4}
\begin{dgroup*}
 \begin{dmath*}
  \xi^{(4)}_1 = \frac{1}{15} \beta_f \theta_1 \mu_{43} \left(2 \theta_{12} \mu_{12}^2 (2\theta_{12}) \left(24 \theta_{12}^2 m_{12}^2+8 \theta_{12} m_{12} \epsilon_f+\epsilon_f^2\right)-2 \theta_2 \mu_{12} (2\theta_{12}) \left(4 \theta_{12}^2 m_{12}^2+3 \theta_{12} m_{12} \epsilon_f+\epsilon_f^2\right)+\theta_2^2 \left(-\theta_{12}^2 m_{12}^2-2 \theta_{12} m_{12} \epsilon_f+\epsilon_f^2\right)\right)\\\\
 \end{dmath*}
  \begin{dmath*}
   \xi^{(4)}_2 = \frac{1}{15} \beta_f \theta_2 \mu_{43} \left(\theta_{12}^2 m_{12}^2 \left(15 \theta_1^2+48 \theta_{12} \mu_{12}^2 (2\theta_{12})-8 \mu_{12} (2\theta_{12}) (5 \theta_1+6 \theta_2)+40 \theta_1 \theta_2+24 \theta_2^2\right)+2 \theta_{12} m_{12} \epsilon_f \left(8 \theta_{12} \mu_{12}^2 (2\theta_{12})-\mu_{12} (2\theta_{12}) (5 \theta_1+8 \theta_2)+\theta_2 (5 \theta_1+4 \theta_2)\right)+\epsilon_f^2 \left(2 \theta_{12} \mu_{12}^2 (2\theta_{12})-2 \theta_2 \mu_{12} (2\theta_{12})+\theta_2^2\right)\right)\\\\
  \end{dmath*}
  \begin{dmath*}
   \xi^{(4)}_3 = \frac{1}{15} \beta_r \theta_3 \theta_4^2 \mu_{43} \left(\theta_{34}^2 m_{34}^2+2 \theta_{34} m_{34} \epsilon_r-\epsilon_r^2\right)\\\\
  \end{dmath*}
  \begin{dmath*}
   \xi^{(4)}_4 = -\frac{1}{15} \beta_r \theta_4 \mu_{43} \left(\theta_{34}^2 m_{34}^2 \left(15 \theta_3^2+40 \theta_3 \theta_4+24 \theta_4^2\right)+2 \theta_{34} \theta_4 m_{34} \epsilon_r (5 \theta_3+4 \theta_4)+\theta_4^2 \epsilon_r^2\right)\\\\
  \end{dmath*}

\end{dgroup*}

\subsection{Heat Production} \label{heat chemical}
The production term for the heat flux can be given in a generic form as
\begin{align}
\mcal P^{\alpha}_{R,1,i}= \sum_{\gamma=1}^N\left(\kappa^{(\alpha)}_{\gamma}u^{(\alpha)}_i+\varsigma^{(\alpha)}_{\gamma}q^{(\alpha)}_i\right)
\end{align}
The coefficients $\kappa^{(\alpha)}_{\gamma}$ and $\varsigma^{(\alpha)}_{\gamma}$ can 
now be given as
\subsubsection{Component-1}
\begin{dgroup*}

 \begin{dmath*}
  \kappa^{(1)}_1 = 
 \end{dmath*}

 \begin{dmath*}
   \frac{1}{24} \beta^{(2)}_f \theta_1^{3/2} \left(\theta_{12}^3 m_{12}^3 \left(192 \theta_1^3+168 \theta_1^2 (3 \theta_2-2 \theta_{12})+140 \theta_1 \theta_2 (3 \theta_2-4 \theta_{12})+105 \theta_2^2 (\theta_2-2 \theta_{12})\right)+\theta_1 \theta_{12}^2 m_{12}^2 \epsilon_f \left(72 \theta_1^2+56 \theta_1 (3 \theta_2-2 \theta_{12})+35 \theta_2 (3 \theta_2-4 \theta_{12})\right)+\theta_1^2 \theta_{12} m_{12} \epsilon_f^2 (12 \theta_1-14 \theta_{12}+21 \theta_2)+\theta_1^3 \epsilon_f^3\right)\\\\
 \end{dmath*}
 
 \begin{dmath*}
  \kappa^{(1)}_2 = 
 \end{dmath*}

 \begin{dmath*}
   -\frac{1}{24} \beta^{(2)}_f \theta_1^2 \sqrt{\theta_2} \left(\theta_1 \theta_{12} m_{12} (5 \theta_1-14 \theta_{12}) \left(4 \theta_{12}^2 m_{12}^2+3 \theta_{12} m_{12} \epsilon_f+\epsilon_f^2\right)+5 \theta_{12} \theta_2^2 m_{12} \left(4 \theta_{12}^2 m_{12}^2+3 \theta_{12} m_{12} \epsilon_f+\epsilon_f^2\right)+\theta_2 \left(\theta_{12}^3 m_{12}^3 (43 \theta_1-70 \theta_{12})+\theta_{12}^2 m_{12}^2 \epsilon_f (39 \theta_1-70 \theta_{12})+\theta_1 \theta_{12} m_{12} \epsilon_f^2+\theta_1 \epsilon_f^3\right)\right)\\\\
 \end{dmath*}
 
 \begin{dmath*}
  \kappa^{(1)}_3 = 
 \end{dmath*}

 \begin{dmath*}
   \frac{\beta^{(2)}_r \sqrt{\theta_3}}{2304}\bigg(10 \theta_{34}^4 m_{34}^2 e^{\frac{\epsilon_r}{2 \theta_{34} m_{34}}} (4 \Omega^{(5,2)}_{\epsilon_r} (3 \theta_{34} \theta_4 m_{34} (14 \theta_{34}-5 \theta_4)+2 \theta_3 \epsilon_r (5 \theta_4-7 \theta_{34})+2 \theta_{34} \mu_{34} \epsilon_r (14 \theta_{34}-5 \theta_4))-2 \Omega^{(7,2)}_{\epsilon_r} (2 \theta_{34} \mu_{34} (\theta_{34} m_{34} (14 \theta_{34}-5 \theta_4)+\theta_3 \epsilon_r)-\theta_3 (2 \theta_{34} m_{34} (7 \theta_{34}-5 \theta_4)+\theta_3 \epsilon_r))+\theta_3 \theta_{34} m_{34} \Omega^{(9,2)}_{\epsilon_r} (\theta_3 (\mu_{34}-1)+\theta_4 \mu_{34})+24 \theta_4 \epsilon_r (5 \theta_4-14 \theta_{34}) \Omega^{(3,2)}_{\epsilon_r})+96 \left(\theta_{34}^3 m_{34}^3 \left(16 \theta_{34}^2 \mu_{34}^2 \left(-144 \theta_3^2+252 \theta_3 \theta_{34}-156 \theta_3 \theta_4+70 \theta_{34} \theta_4-25 \theta_4^2\right)+\theta_3 \left(48 \theta_3^2 (7 \theta_{34}-4 \theta_3)+210 \theta_4^2 (\theta_{34}-2 \theta_3)+56 \theta_3 \theta_4 (10 \theta_{34}-9 \
theta_3)-105 \theta_4^3\right)+24 \theta_3 \theta_{34} \mu_{34} \left(48 \theta_3^2+84 \theta_3 (\theta_4-\theta_{34})+35 \theta_4 (\theta_4-2 \theta_{34})\right)+192 \theta_{34}^3 \mu_{34}^3 (8 \theta_3-14 \theta_{34}+5 \theta_4)\right)+\theta_{34}^2 m_{34}^2 \epsilon_r \left(4 \theta_{34}^2 \mu_{34}^2 \left(-216 \theta_3^2+336 \theta_3 \theta_{34}-208 \theta_3 \theta_4+70 \theta_{34} \theta_4-25 \theta_4^2\right)+6 \theta_3 \theta_{34} \mu_{34} \left(72 \theta_3^2+112 \theta_3 (\theta_4-\theta_{34})+35 \theta_4 (\theta_4-2 \theta_{34})\right)+\theta_3^2 \left(-72 \theta_3^2+56 \theta_3 (2 \theta_{34}-3 \theta_4)+35 \theta_4 (4 \theta_{34}-3 \theta_4)\right)+64 \theta_{34}^3 \mu_{34}^3 (9 \theta_3-14 \theta_{34}+5 \theta_4)\right)+\theta_{34} m_{34} \epsilon_r^2 \left(\theta_3^3 (-12 \theta_3+14 \theta_{34}-21 \theta_4)+12 \theta_3^2 \theta_{34} \mu_{34} (6 \theta_3-7 \theta_{34}+7 \theta_4)+8 \theta_{34}^3 \mu_{34}^3 (12 \theta_3-14 \theta_{34}+5 \theta_4)-8 \theta_3 \theta_{34}^2 \mu_{34}^2 (18 \theta_3-
21 \theta_{34}+13 \theta_4)\right)+\theta_3 \epsilon_r^3 \left(-\theta_3^3+6 \theta_3^2 \theta_{34} \mu_{34}-12 \theta_3 \theta_{34}^2 \mu_{34}^2+8 \theta_{34}^3 \mu_{34}^3\right)\right)\bigg)\\\\
 \end{dmath*}
 
 \begin{dmath*}
  \kappa^{(1)}_4 = 
 \end{dmath*}

 \begin{dmath*}
   \frac{\beta^{(2)}_r \sqrt{\theta_4}}{2304}\bigg(5 \theta_{34}^3 m_{34}^2 (2\theta_{34}) e^{\frac{\epsilon_r}{2 \theta_{34} m_{34}}} (4 \Omega^{(5,2)}_{\epsilon_r} (\theta_3 (14 \theta_{34}-5 \theta_3) (3 \theta_{34} m_{34}+\epsilon_r)+\mu_{34} \epsilon_r (5 \theta_3-14 \theta_{34}) (2\theta_{34})+5 \theta_3 \theta_4 \epsilon_r)-2 \Omega^{(7,2)}_{\epsilon_r} (\mu_{34} (2\theta_{34}) (\theta_{34} m_{34} (5 \theta_3-14 \theta_{34})-\theta_4 \epsilon_r)+\theta_3 (\theta_{34} m_{34} (-5 \theta_3+14 \theta_{34}+5 \theta_4)+\theta_4 \epsilon_r))+\theta_{34} \theta_4 m_{34} \Omega^{(9,2)}_{\epsilon_r} (\theta_3-\mu_{34} (2\theta_{34}))+24 \theta_3 \epsilon_r (5 \theta_3-14 \theta_{34}) \Omega^{(3,2)}_{\epsilon_r})+96 \left(\theta_{34}^3 m_{34}^3 \left(-4 \theta_3 \mu_{34} (2\theta_{34}) \left(40 \theta_3^2-112 \theta_3 \theta_{34}+67 \theta_3 \theta_4-70 \theta_{34} \theta_4+30 \theta_4^2\right)+\theta_3^2 \left(20 \theta_3^2-56 \theta_3 \theta_{34}+43 \theta_3 \theta_4-70 \theta_{34} \theta_4+20 \theta_4^2\right)
+48 \theta_{34} \mu_{34}^3 (2\theta_{34})^2 (-5 \theta_3+14 \theta_{34}-8 \theta_4)+104 \theta_3 \theta_{34} \mu_{34}^2 (2\theta_{34}) (5 \theta_3-14 \theta_{34}+6 \theta_4)\right)+\theta_{34}^2 m_{34}^2 \epsilon_r \left(-\theta_3 \mu_{34} (2\theta_{34}) \left(70 \theta_3^2-196 \theta_3 \theta_{34}+95 \theta_3 \theta_4-70 \theta_{34} \theta_4+40 \theta_4^2\right)+\theta_3^2 \left(15 \theta_3^2-42 \theta_3 \theta_{34}+39 \theta_3 \theta_4-70 \theta_{34} \theta_4+15 \theta_4^2\right)+16 \theta_{34} \mu_{34}^3 (2\theta_{34})^2 (-5 \theta_3+14 \theta_{34}-9 \theta_4)+2 \theta_3 \theta_{34} \mu_{34}^2 (2\theta_{34}) (95 \theta_3-266 \theta_{34}+128 \theta_4)\right)+\theta_{34} m_{34} \epsilon_r^2 \left(-\theta_3 \mu_{34} (2\theta_{34}) \left(15 \theta_3^2+14 \theta_3 (\theta_4-3 \theta_{34})+5 \theta_4^2\right)+\theta_3^2 \left(5 \theta_3^2+\theta_3 (\theta_4-14 \theta_{34})+5 \theta_4^2\right)+2 \theta_{34} \mu_{34}^3 (2\theta_{34})^2 (-5 \theta_3+14 \theta_{34}-12 \theta_4)+2 \theta_3 \theta_{34} \mu_{34}^2 (2\
theta_{34}) (15 \theta_3-42 \theta_{34}+25 \theta_4)\right)+\theta_4 \epsilon_r^3 \left(\theta_3^3-3 \theta_3^2 \mu_{34} (2\theta_{34})-2 \theta_{34} \mu_{34}^3 (2\theta_{34})^2+6 \theta_3 \theta_{34} \mu_{34}^2 (2\theta_{34})\right)\right)\bigg)\\\\
 \end{dmath*}
 
 \begin{dmath*}
  \varsigma^{(1)}_1 = 
 \end{dmath*}

 \begin{dmath*}
   -\frac{1}{60} \beta^{(2)}_f \theta_1^{3/2} \left(\theta_{12}^3 m_{12}^3 \left(48 \theta_1^2 (4 \theta_1-5 \theta_{12})+30 \theta_2^2 (14 \theta_1-5 \theta_{12})+8 \theta_1 \theta_2 (63 \theta_1-50 \theta_{12})+105 \theta_2^3\right)+\theta_1 \theta_{12}^2 m_{12}^2 \epsilon_f \left(72 \theta_1^2+\theta_1 (168 \theta_2-80 \theta_{12})+5 \theta_2 (21 \theta_2-20 \theta_{12})\right)+\theta_1^2 \theta_{12} m_{12} \epsilon_f^2 (12 \theta_1-10 \theta_{12}+21 \theta_2)+\theta_1^3 \epsilon_f^3\right)\\\\
 \end{dmath*}

 \begin{dmath*}
  \varsigma^{(1)}_2 = 
 \end{dmath*}

  \begin{dmath*}
   \frac{1}{60} \beta^{(2)}_f \theta_1^2 \sqrt{\theta_2} \left(5 \theta_1 \theta_{12} m_{12} (\theta_1-2 \theta_{12}) \left(4 \theta_{12}^2 m_{12}^2+3 \theta_{12} m_{12} \epsilon_f+\epsilon_f^2\right)+5 \theta_{12} \theta_2^2 m_{12} \left(4 \theta_{12}^2 m_{12}^2+3 \theta_{12} m_{12} \epsilon_f+\epsilon_f^2\right)+\theta_2 \left(\theta_{12}^3 m_{12}^3 (43 \theta_1-50 \theta_{12})+\theta_{12}^2 m_{12}^2 \epsilon_f (39 \theta_1-50 \theta_{12})+\theta_1 \theta_{12} m_{12} \epsilon_f^2+\theta_1 \epsilon_f^3\right)\right)\\\\
  \end{dmath*}
 
  \begin{dmath*}
   \varsigma^{(1)}_3 = 
  \end{dmath*}

  \begin{dmath*}
   -\frac{\beta^{(2)}_r \sqrt{\theta_3}}{5760}\bigg(5 \theta_{34}^3 m_{34}^2 (2\theta_{34}) e^{\frac{\epsilon_r}{2 \theta_{34} m_{34}}} (20 \Omega^{(5,2)}_{\epsilon_r} (3 \theta_{34} \theta_4 m_{34} (2 \theta_{34}-\theta_4)+\mu_{34} \epsilon_r (2\theta_{34}) (2 \theta_{34}-\theta_4)+2 \theta_3 \epsilon_r (\theta_4-\theta_{34}))-2 \Omega^{(7,2)}_{\epsilon_r} (5 \theta_{34} m_{34} (2 \theta_3 (\theta_4-\theta_{34})-\mu_{34} (2\theta_{34}) (\theta_4-2 \theta_{34}))+\theta_3 \epsilon_r (\mu_{34} (2\theta_{34})-\theta_3))+\theta_3 \theta_{34} m_{34} \Omega^{(9,2)}_{\epsilon_r} (\theta_3 (\mu_{34}-1)+\theta_4 \mu_{34})+120 \theta_4 \epsilon_r (\theta_4-2 \theta_{34}) \Omega^{(3,2)}_{\epsilon_r})+96 \left(\theta_{34}^3 m_{34}^3 \left(8 \theta_{34} \mu_{34}^2 (2\theta_{34}) \left(-144 \theta_3^2+180 \theta_3 \theta_{34}-156 \theta_3 \theta_4+50 \theta_{34} \theta_4-25 \theta_4^2\right)+12 \theta_3 \mu_{34} (2\
theta_{34}) \left(48 \theta_3^2-60 \theta_3 \theta_{34}+84 \theta_3 \theta_4-50 \theta_{34} \theta_4+35 \theta_4^2\right)+\theta_3 \left(48 \theta_3^2 (5 \theta_{34}-4 \theta_3)+30 \theta_4^2 (5 \theta_{34}-14 \theta_3)+8 \theta_3 \theta_4 (50 \theta_{34}-63 \theta_3)-105 \theta_4^3\right)+48 \theta_{34} \mu_{34}^3 (2\theta_{34})^2 (8 \theta_3+5 (\theta_4-2 \theta_{34}))\right)+\theta_{34}^2 m_{34}^2 \epsilon_r \left(2 \theta_{34} \mu_{34}^2 (2\theta_{34}) \left(-216 \theta_3^2+240 \theta_3 \theta_{34}-208 \theta_3 \theta_4+50 \theta_{34} \theta_4-25 \theta_4^2\right)+3 \theta_3 \mu_{34} (2\theta_{34}) \left(72 \theta_3^2+16 \theta_3 (7 \theta_4-5 \theta_{34})+5 \theta_4 (7 \theta_4-10 \theta_{34})\right)+\theta_3^2 \left(-72 \theta_3^2+8 \theta_3 (10 \theta_{34}-21 \theta_4)+5 \theta_4 (20 \theta_{34}-21 \theta_4)\right)+16 \theta_{34} \mu_{34}^3 (2\theta_{34})^2 (9 \theta_3+5 (\theta_4-2 \theta_{34}))\right)+\theta_{34} m_{34} \epsilon_r^2 \left(\theta_3^3 (-12 \theta_3+10 \theta_{
34}-21 \theta_4)+6 \theta_3^2 \mu_{34} (2\theta_{34}) (6 \theta_3-5 \theta_{34}+7 \theta_4)+2 \theta_{34} \mu_{34}^3 (2\theta_{34})^2 (12 \theta_3+5 (\theta_4-2 \theta_{34}))-4 \theta_3 \theta_{34} \mu_{34}^2 (2\theta_{34}) (18 \theta_3-15 \theta_{34}+13 \theta_4)\right)+\theta_3 \epsilon_r^3 \left(-\theta_3^3+3 \theta_3^2 \mu_{34} (2\theta_{34})+2 \theta_{34} \mu_{34}^3 (2\theta_{34})^2-6 \theta_3 \theta_{34} \mu_{34}^2 (2\theta_{34})\right)\right)\bigg)  \\\\
  \end{dmath*}
 
  \begin{dmath*}
   \varsigma^{(1)}_4 = 
  \end{dmath*}

  \begin{dmath*}
  \frac{\beta^{(2)}_r \sqrt{\theta_4}}{5760}\bigg(5 \theta_{34}^3 m_{34}^2 (2\theta_{34}) e^{\frac{\epsilon_r}{2 \theta_{34} m_{34}}} (-20 \Omega^{(5,2)}_{\epsilon_r} (-\theta_3 (\theta_3-2 \theta_{34}) (3 \theta_{34} m_{34}+\epsilon_r)+\mu_{34} \epsilon_r (\theta_3-2 \theta_{34}) (2\theta_{34})+\theta_3 \theta_4 \epsilon_r)+2 \Omega^{(7,2)}_{\epsilon_r} (5 \theta_{34} m_{34} (\mu_{34} (\theta_3-2 \theta_{34}) (2\theta_{34})+\theta_3 (-\theta_3+2 \theta_{34}+\theta_4))+\theta_4 \epsilon_r (\theta_3-\mu_{34} (2\theta_{34})))+\theta_{34} \theta_4 m_{34} \Omega^{(9,2)}_{\epsilon_r} (\theta_3 (\mu_{34}-1)+\theta_4 \mu_{34})-120 \theta_3 \epsilon_r (\theta_3-2 \theta_{34}) \Omega^{(3,2)}_{\epsilon_r})+96 \left(\theta_{34}^3 m_{34}^3 \left(\theta_3^2 \left(-20 \theta_3^2+40 \theta_3 \theta_{34}-43 \theta_3 \theta_4+50 \theta_{34} \theta_4-20 \theta_4^2\right)+4 \theta_3 \mu_{34} (2\theta_{34}) \
left(\theta_4 (67 \theta_3-50 \theta_{34})+40 \theta_3 (\theta_3-2 \theta_{34})+30 \theta_4^2\right)+48 \theta_{34} \mu_{34}^3 (2\theta_{34})^2 (5 \theta_3-10 \theta_{34}+8 \theta_4)-104 \theta_3 \theta_{34} \mu_{34}^2 (2\theta_{34}) (5 \theta_3-10 \theta_{34}+6 \theta_4)\right)+\theta_{34}^2 m_{34}^2 \epsilon_r \left(\theta_3^2 \left(-15 \theta_3^2+30 \theta_3 \theta_{34}-39 \theta_3 \theta_4+50 \theta_{34} \theta_4-15 \theta_4^2\right)+5 \theta_3 \mu_{34} (2\theta_{34}) \left(\theta_4 (19 \theta_3-10 \theta_{34})+14 \theta_3 (\theta_3-2 \theta_{34})+8 \theta_4^2\right)+16 \theta_{34} \mu_{34}^3 (2\theta_{34})^2 (5 \theta_3-10 \theta_{34}+9 \theta_4)-2 \theta_3 \theta_{34} \mu_{34}^2 (2\theta_{34}) (95 (\theta_3-2 \theta_{34})+128 \theta_4)\right)+\theta_{34} m_{34} \epsilon_r^2 \left(-\theta_3^2 \left(5 \theta_3 (\theta_3-2 \theta_{34})+\theta_3 \theta_4+5 \theta_4^2\right)+\theta_3 \mu_{34} (2\theta_{34}) \left(15 \theta_3 (\theta_3-2 \theta_{34})+14 \theta_3 \theta_4+5 \theta_
4^2)+2 \theta_{34} \mu_{34}^3 (2\theta_{34})^2 (5 \theta_3-10 \theta_{34}+12 \theta_4)-10 \theta_3 \theta_{34} \mu_{34}^2 (2\theta_{34}) (3 \theta_3-6 \theta_{34}+5 \theta_4)\right)+\theta_4 \epsilon_r^3 \left(-\theta_3^3+3 \theta_3^2 \mu_{34} (2\theta_{34})+2 \theta_{34} \mu_{34}^3 (2\theta_{34})^2-6 \theta_3 \theta_{34} \mu_{34}^2 (2\theta_{34})\right)\right)\bigg)\\\\
 \end{dmath*}
\end{dgroup*}

\subsubsection{Component-2}
\begin{dgroup*}
 
 \begin{dmath*}
  \kappa^{(2)}_1 = 
 \end{dmath*}

 \begin{dmath*}
   -\frac{\mu_{21}\beta^{(2)}_f \sqrt{\theta_{1}} \theta_2^2}{24 \mu_{12}}\bigg(\theta_{12}^3 m_{12}^3 \left(20 \theta_{1}^2-70 \theta_{1} \theta_{12}+43 \theta_{1} \theta_2-56 \theta_{12} \theta_2+20 \theta_2^2\right)+\theta_{12}^2 m_{12}^2 \epsilon_f \left(15 \theta_{1}^2-70 \theta_{1} \theta_{12}+39 \theta_{1} \theta_2-42 \theta_{12} \theta_2+15 \theta_2^2\right)+\theta_{12} m_{12} \epsilon_f^2 \left(5 \theta_{1}^2+\theta_{1} \theta_2+\theta_2 (5 \theta_2-14 \theta_{12})\right)+\theta_{1} \theta_2 \epsilon_f^3\bigg)\theta_{34}\\\\
 \end{dmath*}

  \begin{dmath*}
   \kappa^{(2)}_2 = 
  \end{dmath*}

 \begin{dmath*}
   \frac{\mu_{21}\beta^{(2)}_f \theta_2^{3/2}}{24 \mu_{12}}\bigg(105 \theta_{1}^2 \theta_{12}^3 m_{12}^3 (\theta_{1}-2 \theta_{12})+7 \theta_{12} \theta_2^2 m_{12} (3 \theta_{1}-2 \theta_{12}) \left(24 \theta_{12}^2 m_{12}^2+8 \theta_{12} m_{12} \epsilon_f+\epsilon_f^2\right)+35 \theta_{1} \theta_{12}^2 \theta_2 m_{12}^2 (3 \theta_{1}-4 \theta_{12}) (4 \theta_{12} m_{12}+\epsilon_f)+\theta_2^3 \left(192 \theta_{12}^3 m_{12}^3+72 \theta_{12}^2 m_{12}^2 \epsilon_f+12 \theta_{12} m_{12} \epsilon_f^2+\epsilon_f^3\right)\bigg)\\\\
 \end{dmath*}
 
 \begin{dmath*}
  \kappa^{(2)}_3 = 
 \end{dmath*}

 \begin{dmath*}
   -\frac{\beta^{(2)}_r \sqrt{\theta_3}}{2304 \mu_{12} \mu_{21}}\bigg(10 \theta_{34}^4 \mu_{12}^2 m_{34}^2 e^{\frac{\epsilon_r}{2 \theta_{34} m_{34}}} (-4 \Omega^{(5,2)}_{\epsilon_r} (3 \theta_{34} \theta_4 m_{34} (14 \theta_{34}-5 \theta_4)+\theta_4 \epsilon_r (5 \theta_3+14 \theta_{34}-5 \theta_4)+2 \theta_{34} \mu_{34} \epsilon_r (5 \theta_4-14 \theta_{34}))-2 \Omega^{(7,2)}_{\epsilon_r} (2 \theta_{34} \mu_{34} (\theta_{34} m_{34} (14 \theta_{34}-5 \theta_4)+\theta_3 \epsilon_r)-\theta_4 (\theta_{34} m_{34} (5 \theta_3+14 \theta_{34}-5 \theta_4)+\theta_3 \epsilon_r))+\theta_3 \theta_{34} m_{34} \Omega^{(9,2)}_{\epsilon_r} (\theta_3 \mu_{34}+\theta_4 (\mu_{34}-1))+24 \theta_4 \epsilon_r (14 \theta_{34}-5 \theta_4) \Omega^{(3,2)}_{\epsilon_r})+96 \mu_{21}^2 \left(\theta_{34}^3 m_{34}^3 \left(\theta_4^2 \left(-20 \theta_3^2+70 \theta_3 \theta_{34}-43 \theta_3 \theta_4+56 \theta_{34} \theta_4-
20 \theta_4^2\right)+8 \theta_{34} \theta_4 \mu_{34} \left(30 \theta_3^2+\theta_3 (67 \theta_4-70 \theta_{34})+8 \theta_4 (5 \theta_4-14 \theta_{34})\right)+192 \theta_{34}^3 \mu_{34}^3 (8 \theta_3-14 \theta_{34}+5 \theta_4)+208 \theta_{34}^2 \theta_4 \mu_{34}^2 (-6 \theta_3+14 \theta_{34}-5 \theta_4)\right)+\theta_{34}^2 m_{34}^2 \epsilon_r \left(2 \theta_{34} \theta_4 \mu_{34} \left(40 \theta_3^2-70 \theta_3 \theta_{34}+95 \theta_3 \theta_4-196 \theta_{34} \theta_4+70 \theta_4^2\right)+\theta_4^2 \left(-15 \theta_3^2+70 \theta_3 \theta_{34}-39 \theta_3 \theta_4+42 \theta_{34} \theta_4-15 \theta_4^2\right)+64 \theta_{34}^3 \mu_{34}^3 (9 \theta_3-14 \theta_{34}+5 \theta_4)+4 \theta_{34}^2 \theta_4 \mu_{34}^2 (-128 \theta_3+266 \theta_{34}-95 \theta_4)\right)+\theta_{34} m_{34} \epsilon_r^2 \left(2 \theta_{34} \theta_4 \mu_{34} \left(5 \theta_3^2+14 \theta_4 (\theta_3-3 \theta_{34})+15 \theta_4^2\right)-\theta_4^2 \left(5 \theta_3^2+\theta_3 \theta_4+\theta_4 (5 \theta_4-14 \theta_{
34})\right)+8 \theta_{34}^3 \mu_{34}^3 (12 \theta_3-14 \theta_{34}+5 \theta_4)+4 \theta_{34}^2 \theta_4 \mu_{34}^2 (-25 \theta_3+42 \theta_{34}-15 \theta_4)\right)+\theta_3 \epsilon_r^3 \left(8 \theta_{34}^3 \mu_{34}^3-12 \theta_{34}^2 \theta_4 \mu_{34}^2+6 \theta_{34} \theta_4^2 \mu_{34}-\theta_4^3\right)\right)\bigg) \\\\
 \end{dmath*}
 
 \begin{dmath*}
  \kappa^{(2)}_4 = 
 \end{dmath*}

  \begin{dmath*}
   \frac{\beta^{(2)}_r \sqrt\mu_{21}{\theta_4}}{2304 \mu_{12} \mu_{21}}\bigg(10 \theta_{34}^4 \mu_{12}^2 m_{34}^2 e^{\frac{\epsilon_r}{2 \theta_{34} m_{34}}} (4 \Omega^{(5,2)}_{\epsilon_r} (3 \theta_3 \theta_{34} m_{34} (14 \theta_{34}-5 \theta_3)+2 \theta_4 \epsilon_r (5 \theta_3-7 \theta_{34})-2 \theta_{34} \mu_{34} \epsilon_r (5 \theta_3-14 \theta_{34}))+2 \Omega^{(7,2)}_{\epsilon_r} (\theta_{34} m_{34} (2 \theta_4 (7 \theta_{34}-5 \theta_3)+2 \theta_{34} \mu_{34} (5 \theta_3-14 \theta_{34}))+\theta_4 \epsilon_r (\theta_4-2 \theta_{34} \mu_{34}))+\theta_{34} \theta_4 m_{34} \Omega^{(9,2)}_{\epsilon_r} (\theta_3 \mu_{34}+\theta_4 (\mu_{34}-1))+24 \theta_3 \epsilon_r (5 \theta_3-14 \theta_{34}) \Omega^{(3,2)}_{\epsilon_r})+96 \mu_{21}^2 \left(\theta_{34}^3 m_{34}^3 \left(16 \theta_{34}^2 \mu_{34}^2 \left(-25 \theta_3^2+2 \theta_3 (35 \theta_{34}-78 \theta_4)+36 \theta_4 (7 \theta_{34}-4 \theta_4)\right)+\
\theta_4 \left(-105 \theta_3^3+210 \theta_3^2 (\theta_{34}-2 \theta_4)+56 \theta_3 \theta_4 (10 \theta_{34}-9 \theta_4)+48 \theta_4^2 (7 \theta_{34}-4 \theta_4)\right)+192 \theta_{34}^3 \mu_{34}^3 (5 \theta_3-14 \theta_{34}+8 \theta_4)+24 \theta_{34} \theta_4 \mu_{34} \left(84 \theta_4 (\theta_3-\theta_{34})+35 \theta_3 (\theta_3-2 \theta_{34})+48 \theta_4^2\right)\right)+\theta_{34}^2 m_{34}^2 \epsilon_r \left(4 \theta_{34}^2 \mu_{34}^2 \left(-25 \theta_3^2+\theta_3 (70 \theta_{34}-208 \theta_4)+24 \theta_4 (14 \theta_{34}-9 \theta_4)\right)+64 \theta_{34}^3 \mu_{34}^3 (5 \theta_3-14 \theta_{34}+9 \theta_4)+6 \theta_{34} \theta_4 \mu_{34} \left(112 \theta_4 (\theta_3-\theta_{34})+35 \theta_3 (\theta_3-2 \theta_{34})+72 \theta_4^2\right)+\theta_4^2 \left(56 \theta_4 (2 \theta_{34}-3 \theta_3)+35 \theta_3 (4 \theta_{34}-3 \theta_3)-72 \theta_4^2\right)\right)+\theta_{34} m_{34} \epsilon_r^2 \left(8 \theta_{34}^3 \mu_{34}^3 (5 \theta_3-14 \theta_{34}+12 \theta_4)+8 \theta_{34}^2 \theta_
4 \mu_{34}^2 (-13 \theta_3+21 \theta_{34}-18 \theta_4)+\theta_4^3 (-21 \theta_3+14 \theta_{34}-12 \theta_4)+12 \theta_{34} \theta_4^2 \mu_{34} (7 \theta_3-7 \theta_{34}+6 \theta_4)\right)+\theta_4 \epsilon_r^3 \left(8 \theta_{34}^3 \mu_{34}^3-12 \theta_{34}^2 \theta_4 \mu_{34}^2+6 \theta_{34} \theta_4^2 \mu_{34}-\theta_4^3\right)\right)\bigg)\\\\
  \end{dmath*}
 
\end{dgroup*}

The coefficients $\beta^{(2)}_r$ and $\beta^{(2)}_f$ appearing the above expressions are given as
\begin{gather}
  \beta^{(2)}_f = \left(\frac{\mu_{12}\sqrt{\pi } d_{12}^2 M n_1 n_2 s_f^2 }{\theta_{12}^{11/2} m_{12}^3}\right) \exp\left[{-\frac{\epsilon_f}{2 \theta_{12} m_{12}}}\right], \quad  \beta^{(2)}_r = \left(\frac{\mu_{12}\sqrt{\pi } d_{34}^2 M n_3 n_4 s_r^2 }{\theta_{34}^{11/2} m_{34}^3}\right)\exp\left[{-\frac{\epsilon_r}{2 \theta_{34} m_{34}}}\right]
\end{gather}
\section{Supplementary material: Contribution from $\Delta_{\alpha}$}
We can now look into the contribution arising from $\Delta_{\alpha}$
into the production terms mentioned above. Since $\Delta_{\alpha}$ is a non-equilibrium quantity and in the present work we are only dealing with
linearised production terms so the production terms corresponding to 
$u_i^{(\alpha)}$, $q_i^{(\alpha)}$ and $\sigma_{ij}^{(\alpha)}$ will not see any contribution from $\Delta_{\alpha}$. Only the reactions rates, the mass balance and the energy balance will be influenced by $\Delta_{\alpha}$.

\subsection{Mass balance} \label{contri Delta mass}
The generic form of the contribution from $\Delta_{\alpha}$ into the mass production is given as:
\begin{align}
 \Delta^{r(\alpha)0} = \sum_{\gamma=1}^4\left(\phi^{(\alpha)}_{\gamma}\Delta_{\gamma}\right)
\end{align}
\subsubsection{Component-1}
\begin{gather*}
  \phi^{(1)}_{1} = \frac{\mu_{12}\beta_f \theta_{1}^2}{120}\bigg(\theta_{12}^2 m_{12}^2+2 \theta_{12} m_{12} \epsilon_f-\epsilon_f^2\bigg),
  \quad   \phi^{(1)}_2 = \frac{\theta_{2}^2 \mu_{12}\beta_f}{120}\bigg(\theta_{12}^2 m_{12}^2+2 \theta_{12} m_{12} \epsilon_f-\epsilon_f^2\bigg)\\
    \phi^{(1)}_3 = \frac{\mu_{12}\beta_r \theta_{3}^2}{120}\bigg(-\theta_{34}^2 m_{34}^2-2 \theta_{34} m_{34} \epsilon_r+\epsilon_r^2\bigg), \quad
   \phi^{(1)}_4 = \frac{\theta_{4}^2 \mu_{12}\beta_r}{120}\bigg(-\theta_{34}^2 m_{34}^2-2 \theta_{34} m_{34} \epsilon_r+\epsilon_r^2\bigg)
 \end{gather*}
 
\subsubsection{Component-2}
\begin{gather*}
 \phi^{(2)}_1 = \frac{\mu_{21}\beta_f \theta_{1}^2}{120}\bigg(\theta_{12}^2 m_{12}^2+2 \theta_{12} m_{12} \epsilon_f-\epsilon_f^2\bigg), \quad
 \phi^{(2)}_2 = \frac{\theta_{2}^2 \mu_{21}\beta_f}{120}\bigg(\theta_{12}^2 m_{12}^2+2 \theta_{12} m_{12} \epsilon_f-\epsilon_f^2\bigg)\\
 \phi^{(2)}_3 = \frac{\mu_{21}\beta_r \theta_{3}^2}{120}\bigg(-\theta_{34}^2 m_{34}^2-2 \theta_{34} m_{34} \epsilon_r+\epsilon_r^2\bigg), \quad
  \phi^{(2)}_4 = \frac{\theta_{4}^2 \mu_{21}\beta_r}{120}\bigg(-\theta_{34}^2 m_{34}^2-2 \theta_{34} m_{34} \epsilon_r+\epsilon_r^2\bigg) 
\end{gather*}

\subsubsection{Component-3}
\begin{gather*}
   \phi^{(3)}_1 = \frac{1}{120} \beta_f \theta_{1}^2 \mu_{34} \left(-\theta_{12}^2 m_{12}^2-2 \theta_{12} m_{12} \epsilon_f+\epsilon_f^2\right), \quad
  \phi^{(3)}_2 = \frac{1}{120} \beta_f \theta_{2}^2 \mu_{34} \left(-\theta_{12}^2 m_{12}^2-2 \theta_{12} m_{12} \epsilon_f+\epsilon_f^2\right)\\
   \phi^{(3)}_3 = \frac{1}{120} \beta_r \theta_{3}^2 \mu_{34} \left(\theta_{34}^2 m_{34}^2+2 \theta_{34} m_{34} \epsilon_r-\epsilon_r^2\right), \quad
   \phi^{(3)}_4 = \frac{1}{120} \beta_r \theta_{4}^2 \mu_{34} \left(\theta_{34}^2 m_{34}^2+2 \theta_{34} m_{34} \epsilon_r-\epsilon_r^2\right)
\end{gather*}
\subsubsection{Component-4}
\begin{gather*}
  \phi^{(4)}_1 = \frac{1}{120} \beta_f \theta_{1}^2 \mu_{43} \left(-\theta_{12}^2 m_{12}^2-2 \theta_{12} m_{12} \epsilon_f+\epsilon_f^2\right), \quad
  \phi^{(4)}_2 = \frac{1}{120} \beta_f \theta_{2}^2 \mu_{43} \left(-\theta_{12}^2 m_{12}^2-2 \theta_{12} m_{12} \epsilon_f+\epsilon_f^2\right)\\
  \phi^{(4)}_3 = \frac{1}{120} \beta_r \theta_{3}^2 \mu_{43} \left(\theta_{34}^2 m_{34}^2+2 \theta_{34} m_{34} \epsilon_r-\epsilon_r^2\right),\quad
  \phi^{(4)}_4 = \frac{1}{120} \beta_r \theta_{4}^2 \mu_{43} \left(\theta_{34}^2 m_{34}^2+2 \theta_{34} m_{34} \epsilon_r-\epsilon_r^2\right)
\end{gather*}
\subsection{Energy Balance} \label{contri Delta energy}
The contribution from the $\nth{14}$-moment into the energy production rate, given by 
$\Delta^{r(\alpha)1}$ in the above sections, can be written in a generic form as
\begin{align}
 \Delta^{r(\alpha)1} = \sum_{\gamma=1}^4\left(\psi^{(\alpha)}_{\gamma}\Delta_{\gamma}\right)
\end{align}
\subsubsection{Component-1}

\begin{dgroup*}

  \begin{dmath*}
   \psi^{(1)}_1 = 
  \end{dmath*}

  \begin{dmath*}
    -\frac{\mu_{12}\beta_f \theta_{1}^2}{480 \theta_{12}^2 m_{12}}\bigg(\theta_{12}^3 m_{12}^3 \left(12 \theta_{1}^2+29 \theta_{1} \theta_{2}+20 \theta_{2}^2\right)+\theta_{12}^2 m_{12}^2 \epsilon_f \left(7 \theta_{1}^2+18 \theta_{1} \theta_{2}+20 \theta_{2}^2\right)+\theta_{1} \theta_{12} m_{12} \epsilon_f^2 (2 \theta_{1}+11 \theta_{2})+\theta_{1}^2 \epsilon_f^3\bigg)\\\\
  \end{dmath*}

  \begin{dmath*}
   \psi^{(1)}_2 = 
  \end{dmath*}

  \begin{dmath*}
   \frac{\mu_{12}\beta_f \theta_{1} \theta_{2}^2}{480 \theta_{12}^2 m_{12}}\bigg(3 \theta_{12}^3 \theta_{2} m_{12}^3-3 \theta_{12}^2 m_{12}^2 \epsilon_f (\theta_{1}-2 \theta_{2})+3 \theta_{12} m_{12} \epsilon_f^2 (2 \theta_{1}-\theta_{2})-\theta_{1} \epsilon_f^3\bigg)\\\\
  \end{dmath*}
  
  \begin{dmath*}
   \psi^{(1)}_3 = 
  \end{dmath*}

  \begin{dmath*}
  \frac{\beta_r \theta_{3}^2 \mu_{12}}{30720 \theta_{34}^2 m_{12}}\bigg(4 \theta_{34}^4 m_{34}^2 e^{\frac{\epsilon_r}{2 \theta_{34} m_{34}}} (20 \Omega^{(5,2)}_{\epsilon_r} (3 \theta_{34} m_{34}+2 \epsilon_r)-2 \Omega^{(7,2)}_{\epsilon_r} (10 \theta_{34} m_{34}+\epsilon_r)+\theta_{34} m_{34} \Omega^{(9,2)}_{\epsilon_r}-120 \epsilon_r \Omega^{(3,2)}_{\epsilon_r})+64 \left(\theta_{34}^3 m_{34}^3 \left(12 \theta_{3}^2 (\mu_{34}-1)^2+\theta_{3} \theta_{4} (8 \mu_{34} (3 \mu_{34}-7)+29)+4 \theta_{4}^2 (\mu_{34}-1) (3 \mu_{34}-5)\right)+\theta_{34}^2 m_{34}^2 \epsilon_r \left(7 \theta_{3}^2 (\mu_{34}-1)^2+2 \theta_{3} \theta_{4} (\mu_{34} (7 \mu_{34}-19)+9)+\theta_{4}^2 (\mu_{34}-2) (7 \mu_{34}-10)\right)+\theta_{34} m_{34} \epsilon_r^2 \left(-8 \theta_{34} \mu_{34} (\theta_{3}+2 \theta_{4})+\theta_{3} (2 \theta_{3}+11 \theta_{4})+8 \theta_{34}^2 \mu_{34}^2\right)+\epsilon_r^3 (\theta_{3} (\mu_{34}-1)+\theta_{4} \mu_{34})^2\right)\bigg)\\\\
  \end{dmath*}
  
  \begin{dmath*}
   \psi^{(1)}_4 = 
  \end{dmath*}

  \begin{dmath*}
  \frac{\beta_r \theta_{4}^2 \mu_{12}}{30720 \theta_{34}^2 m_{12}}\bigg(4 \theta_{34}^4 m_{34}^2 e^{\frac{\epsilon_r}{2 \theta_{34} m_{34}}} (20 \Omega^{(5,2)}_{\epsilon_r} (3 \theta_{34} m_{34}+2 \epsilon_r)-2 \Omega^{(7,2)}_{\epsilon_r} (10 \theta_{34} m_{34}+\epsilon_r)+\theta_{34} m_{34} \Omega^{(9,2)}_{\epsilon_r}-120 \epsilon_r \Omega^{(3,2)}_{\epsilon_r})+64 \left(\theta_{34}^3 m_{34}^3 \left(16 \theta_{3} \theta_{34} \mu_{34}-3 \theta_{3} \theta_{4}+48 \theta_{34}^2 \mu_{34}^2\right)+\theta_{34}^2 m_{34}^2 \epsilon_r \left(20 \theta_{3} \theta_{34} \mu_{34}+3 \theta_{3} (\theta_{3}-2 \theta_{4})+28 \theta_{34}^2 \mu_{34}^2\right)+\theta_{34} m_{34} \epsilon_r^2 \left(8 \theta_{3} \theta_{34} \mu_{34}+3 \theta_{3} (\theta_{4}-2 \theta_{3})+8 \theta_{34}^2 \mu_{34}^2\right)+\epsilon_r^3 (\theta_{3} (\mu_{34}-1)+\theta_{4} \mu_{34})^2\right)\bigg)\\\\
  \end{dmath*}
 
\end{dgroup*}

\subsubsection{Component-2}

\begin{dgroup*}
 \begin{dmath*}
  \psi^{(2)}_1 = 
 \end{dmath*}

 \begin{dmath*}
   \frac{\beta_f \theta_{1}^2 \theta_{2} \mu_{21} \left(3 \theta_{1} \theta_{12}^3 m_{12}^3+3 \theta_{12}^2 m_{12}^2 \epsilon_f (2 \theta_{1}-\theta_{2})-3 \theta_{12} m_{12} \epsilon_f^2 (\theta_{1}-2 \theta_{2})-\theta_{2} \epsilon_f^3\right)}{480 \theta_{12}^2 m_{12}}\\\\
 \end{dmath*}

 \begin{dmath*}
  \psi^{(2)}_2 = 
 \end{dmath*}

 \begin{dmath*}
   -\frac{\mu_{21}\beta_f \theta_{2}^2}{480 \theta_{12}^2 m_{12}}\bigg(\theta_{12}^3 m_{12}^3 \left(20 \theta_{1}^2+29 \theta_{1} \theta_{2}+12 \theta_{2}^2\right)+\theta_{12}^2 m_{12}^2 \epsilon_f \left(20 \theta_{1}^2+18 \theta_{1} \theta_{2}+7 \theta_{2}^2\right)+\theta_{12} \theta_{2} m_{12} \epsilon_f^2 (11 \theta_{1}+2 \theta_{2})+\theta_{2}^2 \epsilon_f^3\bigg)\\\\
 \end{dmath*}
 
 \begin{dmath*}
  \psi^{(2)}_3 = 
 \end{dmath*}

 \begin{dmath*}
  \frac{\beta_r \theta_{3}^2}{30720 \theta_{34}^2 \mu_{21} m_{12}}\bigg(4 \theta_{34}^4 \mu_{12}^2 m_{34}^2 e^{\frac{\epsilon_r}{2 \theta_{34} m_{34}}} (20 \Omega^{(5,2)}_{\epsilon_r} (3 \theta_{34} m_{34}+2 \epsilon_r)-2 \Omega^{(7,2)}_{\epsilon_r} (10 \theta_{34} m_{34}+\epsilon_r)+\theta_{34} m_{34} \Omega^{(9,2)}_{\epsilon_r}-120 \epsilon_r \Omega^{(3,2)}_{\epsilon_r})+64 \mu_{21}^2 \left(\theta_{34}^3 m_{34}^3 \left(-3 \theta_{3} \theta_{4}+48 \theta_{34}^2 \mu_{34}^2+16 \theta_{34} \theta_{4} \mu_{34}\right)+\theta_{34}^2 m_{34}^2 \epsilon_r \left(3 \theta_{4} (\theta_{4}-2 \theta_{3})+28 \theta_{34}^2 \mu_{34}^2+20 \theta_{34} \theta_{4} \mu_{34}\right)+\theta_{34} m_{34} \epsilon_r^2 \left(3 \theta_{4} (\theta_{3}-2 \theta_{4})+8 \theta_{34}^2 \mu_{34}^2+8 \theta_{34} \theta_{4} \mu_{34}\right)+\epsilon_r^3 (\theta_{3} \mu_{34}+\theta_{4} (\mu_{34}-1))^2\right)\bigg)\\\\
 \end{dmath*}
 
 \begin{dmath*}
  \psi^{(2)}_4 = 
 \end{dmath*}

 \begin{dmath*}
   \frac{\beta_r \theta_{4}^2}{30720 \theta_{34}^2 \mu_{21} m_{12}}\bigg(4 \theta_{34}^4 \mu_{12}^2 m_{34}^2 e^{\frac{\epsilon_r}{2 \theta_{34} m_{34}}} (20 \Omega^{(5,2)}_{\epsilon_r} (3 \theta_{34} m_{34}+2 \epsilon_r)-2 \Omega^{(7,2)}_{\epsilon_r} (10 \theta_{34} m_{34}+\epsilon_r)+\theta_{34} m_{34} \Omega^{(9,2)}_{\epsilon_r}-120 \epsilon_r \Omega^{(3,2)}_{\epsilon_r})+64 \mu_{21}^2 \left(\theta_{34}^3 m_{34}^3 \left(4 \theta_{3}^2 (\mu_{34}-1) (3 \mu_{34}-5)+\theta_{3} \theta_{4} (8 \mu_{34} (3 \mu_{34}-7)+29)+12 \theta_{4}^2 (\mu_{34}-1)^2\right)+\theta_{34}^2 m_{34}^2 \epsilon_r \left(\theta_{3}^2 (\mu_{34}-2) (7 \mu_{34}-10)+2 \theta_{3} \theta_{4} (\mu_{34} (7 \mu_{34}-19)+9)+7 \theta_{4}^2 (\mu_{34}-1)^2\right)+\theta_{34} m_{34} \epsilon_r^2 \left(-8 \theta_{34} \mu_{34} (2 \theta_{3}+\theta_{4})+\theta_{4} (11 \theta_{3}+2 \theta_{4})+8 \theta_{34}^2 \mu_{34}^2\right)+\epsilon_r^3 (\theta_{3} \mu_{34}+\theta_{4} (\mu_{34}-1))^2\right)\bigg)\\\\
 \end{dmath*}

\end{dgroup*}

\subsubsection{Component-3}
\begin{dgroup*}
 
 \begin{dmath*}
  \psi^{(3)}_1 = 
 \end{dmath*}

 \begin{dmath*}
     \frac{\beta_f \theta_{1}^2 \mu_{34}}{30720 \theta_{12}^2 m_{12}}\bigg(4 \theta_{12}^4 m_{12}^2 e^{\frac{\epsilon_f}{2 \theta_{12} m_{12}}} (20 \Omega^{(5,2)}_{\epsilon_f} (3 \theta_{12} m_{12}+2 \epsilon_f)-2 \Omega^{(7,2)}_{\epsilon_f} (10 \theta_{12} m_{12}+\epsilon_f)+\theta_{12} m_{12} \Omega^{(9,2)}_{\epsilon_f}-120 \epsilon_f \Omega^{(3,2)}_{\epsilon_f})+64 \left(\theta_{12}^3 m_{12}^3 \left(12 \theta_{1}^2 (\mu_{21}-1)^2+\theta_{1} \theta_{2} (8 \mu_{21} (3 \mu_{21}-7)+29)+4 \theta_{2}^2 (\mu_{21}-1) (3 \mu_{21}-5)\right)+\theta_{12}^2 m_{12}^2 \epsilon_f \left(7 \theta_{1}^2 (\mu_{21}-1)^2+2 \theta_{1} \theta_{2} (\mu_{21} (7 \mu_{21}-19)+9)+\theta_{2}^2 (\mu_{21}-2) (7 \mu_{21}-10)\right)+\theta_{12} m_{12} \epsilon_f^2 \left(-8 \theta_{12} \mu_{21} (\theta_{1}+2 \theta_{2})+\theta_{1} (2 \theta_{1}+11 \theta_{2})+8 \theta_{12}^2 \mu_{21}^2\right)+\epsilon_f^3 (\theta_{1} (\mu_{21}-1)+\theta_{2} \mu_{21})^2\right)\bigg)\\\\
 \end{dmath*}

 \begin{dmath*}
  \psi^{(3)}_2 = 
 \end{dmath*}

 \begin{dmath*}
   \frac{\beta_f \theta_{2}^2 \mu_{34}}{30720 \theta_{12}^2 m_{12}}\bigg(4 \theta_{12}^4 m_{12}^2 e^{\frac{\epsilon_f}{2 \theta_{12} m_{12}}} (20 \Omega^{(5,2)}_{\epsilon_f} (3 \theta_{12} m_{12}+2 \epsilon_f)-2 \Omega^{(7,2)}_{\epsilon_f} (10 \theta_{12} m_{12}+\epsilon_f)+\theta_{12} m_{12} \Omega^{(9,2)}_{\epsilon_f}-120 \epsilon_f \Omega^{(3,2)}_{\epsilon_f})+64 \left(\theta_{12}^3 m_{12}^3 \left(16 \theta_{1} \theta_{12} \mu_{21}-3 \theta_{1} \theta_{2}+48 \theta_{12}^2 \mu_{21}^2\right)+\theta_{12}^2 m_{12}^2 \epsilon_f \left(20 \theta_{1} \theta_{12} \mu_{21}+3 \theta_{1} (\theta_{1}-2 \theta_{2})+28 \theta_{12}^2 \mu_{21}^2\right)+\theta_{12} m_{12} \epsilon_f^2 \left(8 \theta_{1} \theta_{12} \mu_{21}+3 \theta_{1} (\theta_{2}-2 \theta_{1})+8 \theta_{12}^2 \mu_{21}^2\right)+\epsilon_f^3 (\theta_{1} (\mu_{21}-1)+\theta_{2} \mu_{21})^2\right)\bigg)\\\\
 \end{dmath*}
 
 \begin{dmath*}
  \psi^{(3)}_3 = 
 \end{dmath*}

 \begin{dmath*}
    -\frac{\mu_{34}\beta_r \theta_{3}^2}{480 \theta_{34}^2 m_{12}}\bigg(\theta_{34}^3 m_{34}^3 \left(12 \theta_{3}^2+29 \theta_{3} \theta_{4}+20 \theta_{4}^2\right)+\theta_{34}^2 m_{34}^2 \epsilon_r \left(7 \theta_{3}^2+18 \theta_{3} \theta_{4}+20 \theta_{4}^2\right)+\theta_{3} \theta_{34} m_{34} \epsilon_r^2 (2 \theta_{3}+11 \theta_{4})+\theta_{3}^2 \epsilon_r^3\bigg)
 \end{dmath*}

 \begin{dmath*}
  \psi^{(3)}_4 = 
 \end{dmath*}

 \begin{dmath*}
    \frac{\mu_{34}\beta_r \theta_{3} \theta_{4}^2}{480 \theta_{34}^2 m_{12}}\bigg(3 \theta_{34}^3 \theta_{4} m_{34}^3-3 \theta_{34}^2 m_{34}^2 \epsilon_r (\theta_{3}-2 \theta_{4})+3 \theta_{34} m_{34} \epsilon_r^2 (2 \theta_{3}-\theta_{4})-\theta_{3} \epsilon_r^3\bigg)\\\\
 \end{dmath*}

\end{dgroup*}

\subsubsection{Component-4}
\begin{dgroup*}

 \begin{dmath*}
  \psi^{(4)}_1 = 
 \end{dmath*}

 \begin{dmath*}
  \frac{\beta_f \theta_{1}^2}{30720 \theta_{12}^2 \mu_{43} m_{12}}\bigg(4 \theta_{12}^4 \mu_{34}^2 m_{12}^2 e^{\frac{\epsilon_f}{2 \theta_{12} m_{12}}} (20 \Omega^{(5,2)}_{\epsilon_f} (3 \theta_{12} m_{12}+2 \epsilon_f)-2 \Omega^{(7,2)}_{\epsilon_f} (10 \theta_{12} m_{12}+\epsilon_f)+\theta_{12} m_{12} \Omega^{(9,2)}_{\epsilon_f}-120 \epsilon_f \Omega^{(3,2)}_{\epsilon_f})+64 \mu_{43}^2 \left(\theta_{12}^3 m_{12}^3 \left(-3 \theta_{1} \theta_{2}+48 \theta_{12}^2 \mu_{12}^2+16 \theta_{12} \theta_{2} \mu_{12}\right)+\theta_{12}^2 m_{12}^2 \epsilon_f \left(3 \theta_{2} (\theta_{2}-2 \theta_{1})+28 \theta_{12}^2 \mu_{12}^2+20 \theta_{12} \theta_{2} \mu_{12}\right)+\theta_{12} m_{12} \epsilon_f^2 \left(3 \theta_{2} (\theta_{1}-2 \theta_{2})+8 \theta_{12}^2 \mu_{12}^2+8 \theta_{12} \theta_{2} \mu_{12}\right)+\epsilon_f^3 (\theta_{1} \mu_{12}+\theta_{2} (\mu_{12}-1))^2\right)\bigg)\\\\
 \end{dmath*}

  \begin{dmath*}
   \psi^{(4)}_2 = 
  \end{dmath*}

  \begin{dmath*}
    \frac{\beta_f \theta_{2}^2}{30720 \theta_{12}^2 \mu_{43} m_{12}}\bigg(4 \theta_{12}^4 \mu_{34}^2 m_{12}^2 e^{\frac{\epsilon_f}{2 \theta_{12} m_{12}}} (20 \Omega^{(5,2)}_{\epsilon_f} (3 \theta_{12} m_{12}+2 \epsilon_f)-2 \Omega^{(7,2)}_{\epsilon_f} (10 \theta_{12} m_{12}+\epsilon_f)+\theta_{12} m_{12} \Omega^{(9,2)}_{\epsilon_f}-120 \epsilon_f \Omega^{(3,2)}_{\epsilon_f})+64 \mu_{43}^2 \left(\theta_{12}^3 m_{12}^3 \left(4 \theta_{1}^2 (\mu_{12}-1) (3 \mu_{12}-5)+\theta_{1} \theta_{2} (8 \mu_{12} (3 \mu_{12}-7)+29)+12 \theta_{2}^2 (\mu_{12}-1)^2\right)+\theta_{12}^2 m_{12}^2 \epsilon_f \left(\theta_{1}^2 (\mu_{12}-2) (7 \mu_{12}-10)+2 \theta_{1} \theta_{2} (\mu_{12} (7 \mu_{12}-19)+9)+7 \theta_{2}^2 (\mu_{12}-1)^2\right)+\theta_{12} m_{12} \epsilon_f^2 \left(-8 \theta_{12} \mu_{12} (2 \theta_{1}+\theta_{2})+\theta_{2} (11 \theta_{1}+2 \theta_{2})+8 \theta_{12}^2 \mu_{12}^2\right)+\epsilon_f^3 (\theta_{1} \mu_{12}+\theta_{2} (\mu_{12}-1))^2\right)\bigg)\\\\
 \end{dmath*}
 
 \begin{dmath*}
  \psi^{(4)}_3 = 
 \end{dmath*}

 \begin{dmath*}
   \frac{\mu_{43}\beta_r \theta_{3}^2 \theta_{4}}{480 \theta_{34}^2 m_{12}}\bigg(3 \theta_{3} \theta_{34}^3 m_{34}^3+3 \theta_{34}^2 m_{34}^2 \epsilon_r (2 \theta_{3}-\theta_{4})-3 \theta_{34} m_{34} \epsilon_r^2 (\theta_{3}-2 \theta_{4})-\theta_{4} \epsilon_r^3\bigg)\\\\
 \end{dmath*}
 
 \begin{dmath*}
  \psi^{(4)}_4 = 
 \end{dmath*}

 \begin{dmath*}
  -\frac{\mu_{43}\beta_r \theta_{4}^2}{480 \theta_{34}^2 m_{12}}\bigg(\theta_{34}^3 m_{34}^3 \left(20 \theta_{3}^2+29 \theta_{3} \theta_{4}+12 \theta_{4}^2\right)+\theta_{34}^2 m_{34}^2 \epsilon_r \left(20 \theta_{3}^2+18 \theta_{3} \theta_{4}+7 \theta_{4}^2\right)+\theta_{34} \theta_{4} m_{34} \epsilon_r^2 (11 \theta_{3}+2 \theta_{4})+\theta_{4}^2 \epsilon_r^3\bigg)\\\\
 \end{dmath*}
\end{dgroup*}
\subsection{$u^{2}_{\alpha}$ Production(Elastic)} \label{Delta elastic}
The generic form of this production term is given as:
\begin{align}
P^{e(\alpha)1} = \displaystyle \sum_{\beta = 1}^N \bar{\nu}_{\alpha\beta} \left( \lambda_{\alpha\beta} + \varsigma_{\alpha}\Delta_{\alpha} + \varsigma_{\beta}\Delta_{\beta}\right)
\end{align}
The coefficients $\lambda_{\alpha\beta}$ and $\varsigma_{\gamma}$ can now be given as
\begin{dgroup*}
  \begin{dmath*}
   \lambda_{\alpha\beta} = 
  \end{dmath*}

 \begin{dmath*}
   -\frac{5 \sqrt{2}}{\sqrt{2\theta_{\alpha\beta}}}\bigg(\theta_{\alpha} \theta_{\beta} ^2 \mu_{\beta}  \left(-5 \mu_{\alpha} ^2+6 \mu_{\alpha}  \mu_{\beta} -13 \mu_{\beta} ^2\right)-8 \theta_{\beta} ^3 \mu_{\beta} ^3+2 \theta_{\alpha}^3 \mu_{\alpha}  \left(3 \mu_{\alpha} ^2+2 \mu_{\alpha}  \mu_{\beta} +3 \mu_{\beta} ^2\right)+\theta_{\alpha}^2 \theta_{\beta}  \left(5 \mu_{\alpha} ^3-4 \mu_{\alpha} ^2 \mu_{\beta} +9 \mu_{\alpha}  \mu_{\beta} ^2-6 \mu_{\beta} ^3\right)\bigg)\\\\
 \end{dmath*}

 \begin{dmath*}
  \varsigma_{\alpha} = 
 \end{dmath*}

 \begin{dmath*}
  -\frac{\theta_{\alpha}^2}{12 \sqrt{2} (2\theta_{\alpha\beta})^{5/2}}\bigg(30 \theta_{\alpha}^3 \mu_{\alpha}  \left(3 \mu_{\alpha} ^2+2 \mu_{\alpha}  \mu_{\beta} +3 \mu_{\beta} ^2\right)+\theta_{\alpha} \theta_{\beta} ^2 \left(184 \mu_{\alpha} ^3+69 \mu_{\alpha} ^2 \mu_{\beta} +210 \mu_{\alpha}  \mu_{\beta} ^2-35 \mu_{\beta} ^3\right)+3 \theta_{\alpha}^2 \theta_{\beta}  \left(77 \mu_{\alpha} ^3+44 \mu_{\alpha} ^2 \mu_{\beta} +81 \mu_{\alpha}  \mu_{\beta} ^2-6 \mu_{\beta} ^3\right)+4 \theta_{\beta} ^3 \left(10 \mu_{\alpha} ^3-3 \mu_{\alpha} ^2 \mu_{\beta} +12 \mu_{\alpha}  \mu_{\beta} ^2-5 \mu_{\beta} ^3\right)\bigg) \\\\
 \end{dmath*}

 \begin{dmath*}
  \varsigma_{\beta} = 
 \end{dmath*}

 \begin{dmath*}
 \frac{\theta_{\beta} ^2}{12 \sqrt{2} (2\theta_{\alpha\beta})^{5/2}}\bigg(3 \theta_{\alpha} \theta_{\beta} ^2 \mu_{\beta}  \left(5 \mu_{\alpha} ^2-6 \mu_{\alpha}  \mu_{\beta} +109 \mu_{\beta} ^2\right)+120 \theta_{\beta} ^3 \mu_{\beta} ^3 +2 \theta_{\alpha}^3 \left(\mu_{\alpha} ^3+6 \mu_{\alpha} ^2 \mu_{\beta} -15 \mu_{\alpha}  \mu_{\beta} ^2+40 \mu_{\beta} ^3\right)+\theta_{\alpha}^2 \theta_{\beta}  \left(5 \mu_{\alpha} ^3+36 \mu_{\alpha} ^2 \mu_{\beta} -39 \mu_{\alpha}  \mu_{\beta} ^2+290 \mu_{\beta} ^3\right)\bigg)\\\\
 \end{dmath*}

\end{dgroup*}

\subsection{$u^{2}_{\alpha}$ Production(Chemical)} \label{Delta chemical}
The generic form of the production term of the fourteenth moment can be given as
\begin{align}
  P^{r(\alpha)2} = \sum_{\gamma=1}^4\left(\varepsilon^{(\alpha)}_{\gamma}\Delta_{\alpha}\right) + \chi^{(\alpha)}
\end{align}
The coefficients $\varepsilon^{(\alpha)}_{\gamma}$ and $\chi^{(\alpha)}$ can now be given as
\subsubsection{Component-1}
\begin{dgroup*}
 \begin{dmath*}
  \chi^{(1)} = 
 \end{dmath*}
 
 \begin{dmath*}
  \frac{\beta_r \mu_{12}}{24}\bigg(\theta_{34}^2 m_{34} 2\theta_{34} e^{\frac{\epsilon_r}{2 \theta_{34} m_{34}}} \left(10 \Omega^{(5,2)}_{\epsilon_r} \left(3 \theta_{3} \theta_{34} \theta_{4} m_{34}+\epsilon_r \left(-\theta_{3}^2-2 \theta_{34} \mu_{34}^2 2\theta_{34}+2 \theta_{3} \mu_{34} 2\theta_{34}\right)\right)+\theta_{34} \left(5 m_{34} \Omega^{(7,2)}_{\epsilon_r} \left(\theta_{3}^2+2 \theta_{34} \mu_{34}^2 2\theta_{34}-2 \theta_{3} \mu_{34} 2\theta_{34}\right)+3 \theta_{34} m_{34} 2\theta_{34} \Omega^{(7,4)}_{\epsilon_r}-6 \epsilon_r 2\theta_{34} \Omega^{(5,4)}_{\epsilon_r}\right)-60 \theta_{3} \theta_{4} \epsilon_r \Omega^{(3,2)}_{\epsilon_r}\right)+24 \left(\theta_{3}^2 \left(\theta_{34}^2 m_{34}^2 \left(24 \theta_{3}^2+40 \theta_{3} \theta_{4}+15 \theta_{4}^2\right)+2 \theta_{3} \theta_{34} m_{34} \epsilon_r (4 \theta_{3}+5 \theta_{4})+\theta_{3}^2 \epsilon_r^2\right)-4 \theta_{3}^2 \mu_{34} 2\theta_{34} \left(4 \theta_{34}^2 m_{34}^2 (6 \theta_{3}+5 \theta_{4})+\theta_{34} m_{34} \epsilon_r (8 \
theta_{3}+5 \theta_{4})+\theta_{3} \epsilon_r^2\right)+4 \theta_{34}^2 \mu_{34}^4 2\theta_{34}^2 \left(24 \theta_{34}^2 m_{34}^2+8 \theta_{34} m_{34} \epsilon_r+\epsilon_r^2\right)-8 \theta_{3} \theta_{34} \mu_{34}^3 2\theta_{34}^2 \left(24 \theta_{34}^2 m_{34}^2+8 \theta_{34} m_{34} \epsilon_r+\epsilon_r^2\right)+4 \theta_{3} \theta_{34} \mu_{34}^2 2\theta_{34} \left(3 \theta_{3} \left(24 \theta_{34}^2 m_{34}^2+8 \theta_{34} m_{34} \epsilon_r+\epsilon_r^2\right)+5 \theta_{34} \theta_{4} m_{34} (4 \theta_{34} m_{34}+\epsilon_r)\right)\right)\bigg)
 \end{dmath*}

   \begin{dmath*}
    -\beta_f \theta_{1}^2 \mu_{12} \left(\theta_{12}^2 m_{12}^2 \left(24 \theta_{1}^2+40 \theta_{1} \theta_{2}+15 \theta_{2}^2\right)+2 \theta_{1} \theta_{12} m_{12} \epsilon_f (4 \theta_{1}+5 \theta_{2})+\theta_{1}^2 \epsilon_f^2\right)\\\\
   \end{dmath*}

  \begin{dmath*}
   \varepsilon^{(1)}_1 = 
  \end{dmath*}

  \begin{dmath*}
    -\frac{1}{480} \beta^{(3)}_f \theta_{1}^2 \mu_{12} \left(3 \theta_{12}^4 m_{12}^4 \left(120 \theta_{1}^4+424 \theta_{1}^3 \theta_{2}+539 \theta_{1}^2 \theta_{2}^2+280 \theta_{1} \theta_{2}^3+40 \theta_{2}^4\right)+6 \theta_{1} \theta_{12}^3 m_{12}^3 \epsilon_f (7 \theta_{1}+12 \theta_{2}) \left(4 \theta_{1}^2+7 \theta_{1} \theta_{2}+5 \theta_{2}^2\right)+3 \theta_{1}^2 \theta_{12}^2 m_{12}^2 \epsilon_f^2 \left(13 \theta_{1}^2+44 \theta_{1} \theta_{2}+61 \theta_{2}^2\right)+2 \theta_{1}^3 \theta_{12} m_{12} \epsilon_f^3 (3 \theta_{1}+13 \theta_{2})+\theta_{1}^4 \epsilon_f^4\right)\\\\
  \end{dmath*}
  
  \begin{dmath*}
   \varepsilon^{(1)}_2 = 
  \end{dmath*}

  \begin{dmath*}
   -\frac{1}{480} \beta^{(3)}_f \theta_{1}^2 \theta_{2}^2 \mu_{12} \left(-15 \theta_{12}^4 \theta_{2}^2 m_{12}^4+30 \theta_{12}^3 \theta_{2} m_{12}^3 \epsilon_f (\theta_{1}-\theta_{2})+15 \theta_{12}^2 m_{12}^2 \epsilon_f^2 \left(\theta_{1}^2-4 \theta_{1} \theta_{2}+\theta_{2}^2\right)+10 \theta_{1} \theta_{12} m_{12} \epsilon_f^3 (\theta_{2}-\theta_{1})+\theta_{1}^2 \epsilon_f^4\right)\\\\
  \end{dmath*}
  
  \begin{dmath*}
   \varepsilon^{(1)}_3=
  \end{dmath*}

  \begin{dmath*}
   \frac{\beta^{(3)}_r \theta_{3}^2 \mu_{12}}{184320}\bigg(384 \left(\theta_{34}^4 m_{34}^4 \left(360 \theta_{3}^4 (\mu_{34}-1)^4+24 \theta_{3}^3 \theta_{4} (\mu_{34}-1)^2 (12 \mu_{34} (5 \mu_{34}-9)+53)+\theta_{3}^2 \theta_{4}^2 (16 \mu_{34} (\mu_{34} (27 \mu_{34} (5 \mu_{34}-18)+664)-415)+1617)+8 \theta_{3} \theta_{4}^3 (\mu_{34}-1) (\mu_{34} (36 \mu_{34} (5 \mu_{34}-12)+343)-105)+8 \theta_{4}^4 (\mu_{34}-1) (\mu_{34} (9 \mu_{34} (5 \mu_{34}-11)+65)-15)\right)+2 \theta_{34}^3 m_{34}^3 \epsilon_r \left(4 \theta_{34}^2 \mu_{34}^2 \left(504 \theta_{3}^2+803 \theta_{3} \theta_{4}+252 \theta_{4}^2\right)+3 \theta_{3} (7 \theta_{3}+12 \theta_{4}) \left(4 \theta_{3}^2+7 \theta_{3} \theta_{4}+5 \theta_{4}^2\right)-4 \theta_{34} \mu_{34} \left(168 \theta_{3}^3+419 \theta_{3}^2 \theta_{4}+312 \theta_{3} \theta_{4}^2+40 \theta_{4}^3\right)-128 \theta_{34}^3 \mu_{34}^3 (21 \theta_{3}+16 \theta_{4})+1344 \theta_{34}^4 \mu_{34}^4\right)+\theta_{34}^2 m_{34}^2 \epsilon_r^2 \left(8 \theta_{34}^2 \mu_{34}^2 \left(117 \
theta_{3}^2+178 \theta_{3} \theta_{4}+44 \theta_{4}^2\right)-8 \theta_{3} \theta_{34} \mu_{34} \left(39 \theta_{3}^2+94 \theta_{3} \theta_{4}+64 \theta_{4}^2\right)+3 \theta_{3}^2 \left(13 \theta_{3}^2+44 \theta_{3} \theta_{4}+61 \theta_{4}^2\right)-32 \theta_{34}^3 \mu_{34}^3 (39 \theta_{3}+28 \theta_{4})+624 \theta_{34}^4 \mu_{34}^4\right)+2 \theta_{34} m_{34} \epsilon_r^3 (\theta_{3} (\mu_{34}-1)+\theta_{4} \mu_{34})^2 \left(-4 \theta_{34} \mu_{34} (3 \theta_{3}+4 \theta_{4})+\theta_{3} (3 \theta_{3}+13 \theta_{4})+12 \theta_{34}^2 \mu_{34}^2\right)+(\theta_{3} \epsilon_r-2 \theta_{34} \mu_{34} \epsilon_r)^4\right)
  \end{dmath*}
  
  \begin{dmath*}
   +4 \theta_{34}^5 m_{34}^3 e^{\frac{\epsilon_r}{2 \theta_{34} m_{34}}} \left(-400 \Omega^{(5,2)}_{\epsilon_r} \left(3 \theta_{34} \theta_{4} m_{34} (4 \theta_{4}-3 \theta_{3})+\epsilon_r \left(3 \theta_{3}^2 (\mu_{34}-1)^2+2 \theta_{3} \theta_{4} \left(3 \mu_{34}^2+\mu_{34}-7\right)+\theta_{4}^2 (\mu_{34}+2) (3 \mu_{34}+2)\right)\right)+600 \theta_{3}^2 \theta_{34} \mu_{34}^2 m_{34} \Omega^{(7,2)}_{\epsilon_r}-200 \theta_{3}^2 \theta_{34} \mu_{34}^2 m_{34} \Omega^{(9,2)}_{\epsilon_r}+10 \theta_{3}^2 \theta_{34} \mu_{34}^2 m_{34} \Omega^{(11,2)}_{\epsilon_r}-1200 \theta_{3}^2 \theta_{34} \mu_{34} m_{34} \Omega^{(7,2)}_{\epsilon_r}+400 \theta_{3}^2 \theta_{34} \mu_{34} m_{34} \Omega^{(9,2)}_{\epsilon_r}-20 \theta_{3}^2 \theta_{34} \mu_{34} m_{34} \Omega^{(11,2)}_{\epsilon_r}+600 \theta_{3}^2 \theta_{34} m_{34} \Omega^{(7,2)}_{\epsilon_r}+180 \theta_{3}^2 \theta_{34} m_{34} \Omega^{(7,4)}_{\epsilon_r}-200 \theta_{3}^2 \theta_{34} m_{34} \Omega^{(9,2)}_{\epsilon_r}-60 \theta_{3}^2 \theta_{34} m_{34} \Omega^{(9,
4)}_{\epsilon_r}+10 \theta_{3}^2 \theta_{34} m_{34} \Omega^{(11,2)}_{\epsilon_r}+1200 \theta_{3} \theta_{34} \theta_{4} \mu_{34}^2 m_{34} \Omega^{(7,2)}_{\epsilon_r}-400 \theta_{3} \theta_{34} \theta_{4} \mu_{34}^2 m_{34} \Omega^{(9,2)}_{\epsilon_r}+20 \theta_{3} \theta_{34} \theta_{4} \mu_{34}^2 m_{34} \Omega^{(11,2)}_{\epsilon_r}+400 \theta_{3} \theta_{34} \theta_{4} \mu_{34} m_{34} \Omega^{(7,2)}_{\epsilon_r}+240 \theta_{3} \theta_{34} \theta_{4} \mu_{34} m_{34} \Omega^{(9,2)}_{\epsilon_r}-20 \theta_{3} \theta_{34} \theta_{4} \mu_{34} m_{34} \Omega^{(11,2)}_{\epsilon_r}-2800 \theta_{3} \theta_{34} \theta_{4} m_{34} \Omega^{(7,2)}_{\epsilon_r}+360 \theta_{3} \theta_{34} \theta_{4} m_{34} \Omega^{(7,4)}_{\epsilon_r}+220 \theta_{3} \theta_{34} \theta_{4} m_{34} \Omega^{(9,2)}_{\epsilon_r}-120 \theta_{3} \theta_{34} \theta_{4} m_{34} \Omega^{(9,4)}_{\epsilon_r}+12 \theta_{34}^3 m_{34} \Omega^{(11,4)}_{\epsilon_r}+600 \theta_{34} \theta_{4}^2 \mu_{34}^2 m_{34} \Omega^{(7,2)}_{\epsilon_r}-200 \theta_{34} \theta_
{4}^2 \mu_{34}^2 m_{34} \Omega^{(9,2)}_{\epsilon_r}+10 \theta_{34} \theta_{4}^2 \mu_{34}^2 m_{34} \Omega^{(11,2)}_{\epsilon_r}+1600 \theta_{34} \theta_{4}^2 \mu_{34} m_{34} \Omega^{(7,2)}_{\epsilon_r}-160 \theta_{34} \theta_{4}^2 \mu_{34} m_{34} \Omega^{(9,2)}_{\epsilon_r}+800 \theta_{34} \theta_{4}^2 m_{34} \Omega^{(7,2)}_{\epsilon_r}+180 \theta_{34} \theta_{4}^2 m_{34} \Omega^{(7,4)}_{\epsilon_r}-60 \theta_{34} \theta_{4}^2 m_{34} \Omega^{(9,4)}_{\epsilon_r}+400 \theta_{3}^2 \mu_{34}^2 \epsilon_r \Omega^{(7,2)}_{\epsilon_r}-20 \theta_{3}^2 \mu_{34}^2 \epsilon_r \Omega^{(9,2)}_{\epsilon_r}-800 \theta_{3}^2 \mu_{34} \epsilon_r \Omega^{(7,2)}_{\epsilon_r}+40 \theta_{3}^2 \mu_{34} \epsilon_r \Omega^{(9,2)}_{\epsilon_r}-360 \theta_{3}^2 \epsilon_r \Omega^{(5,4)}_{\epsilon_r}+400 \theta_{3}^2 \epsilon_r \Omega^{(7,2)}_{\epsilon_r}+120 \theta_{3}^2 \epsilon_r \Omega^{(7,4)}_{\epsilon_r}-20 \theta_{3}^2 \epsilon_r \Omega^{(9,2)}_{\epsilon_r}-6 \theta_{3}^2 \epsilon_r \Omega^{(9,4)}_{\epsilon_r}+800 \theta_{3} \
theta_{4} \mu_{34}^2 \epsilon_r \Omega^{(7,2)}_{\epsilon_r}-40 \theta_{3} \theta_{4} \mu_{34}^2 \epsilon_r \Omega^{(9,2)}_{\epsilon_r}-480 \theta_{3} \theta_{4} \mu_{34} \epsilon_r \Omega^{(7,2)}_{\epsilon_r}+40 \theta_{3} \theta_{4} \mu_{34} \epsilon_r \Omega^{(9,2)}_{\epsilon_r}+2400 \theta_{4} \epsilon_r (4 \theta_{4}-3 \theta_{3}) \Omega^{(3,2)}_{\epsilon_r}-720 \theta_{3} \theta_{4} \epsilon_r \Omega^{(5,4)}_{\epsilon_r}-440 \theta_{3} \theta_{4} \epsilon_r \Omega^{(7,2)}_{\epsilon_r}+240 \theta_{3} \theta_{4} \epsilon_r \Omega^{(7,4)}_{\epsilon_r}-12 \theta_{3} \theta_{4} \epsilon_r \Omega^{(9,4)}_{\epsilon_r}+400 \theta_{4}^2 \mu_{34}^2 \epsilon_r \Omega^{(7,2)}_{\epsilon_r}-20 \theta_{4}^2 \mu_{34}^2 \epsilon_r \Omega^{(9,2)}_{\epsilon_r}+320 \theta_{4}^2 \mu_{34} \epsilon_r \Omega^{(7,2)}_{\epsilon_r}-360 \theta_{4}^2 \epsilon_r \Omega^{(5,4)}_{\epsilon_r}+120 \theta_{4}^2 \epsilon_r \Omega^{(7,4)}_{\epsilon_r}-6 \theta_{4}^2 \epsilon_r \Omega^{(9,4)}_{\epsilon_r}\right)\bigg)\\\\
  \end{dmath*}

  \begin{dmath*}
   \varepsilon^{(1)}_{4} = 
  \end{dmath*}

   \begin{dmath*}
    \frac{\beta^{(3)}_r \theta_{4}^2 \mu_{12}}{184320}\bigg(384 \left(\theta_{34}^4 m_{34}^4 \left(32 \theta_{3}^2 \theta_{34} \mu_{34} (2 \theta_{3}+5 \theta_{4})-15 \theta_{3}^2 \theta_{4}^2-2304 \theta_{3} \theta_{34}^3 \mu_{34}^3+32 \theta_{3} \theta_{34}^2 \mu_{34}^2 (2 \theta_{3}+15 \theta_{4})+5760 \theta_{34}^4 \mu_{34}^4\right)+2 \theta_{34}^3 m_{34}^3 \epsilon_r \left(4 \theta_{3}^2 \theta_{34} \mu_{34} (4 \theta_{3}+25 \theta_{4})+15 \theta_{3}^2 \theta_{4} (\theta_{3}-\theta_{4})-640 \theta_{3} \theta_{34}^3 \mu_{34}^3-4 \theta_{3} \theta_{34}^2 \mu_{34}^2 (12 \theta_{3}-35 \theta_{4})+1344 \theta_{34}^4 \mu_{34}^4\right)+\theta_{34}^2 m_{34}^2 \epsilon_r^2 \left(8 \theta_{3}^2 \theta_{34} \mu_{34} (\theta_{3}+10 \theta_{4})+15 \theta_{3}^2 \left(\theta_{3}^2-4 \theta_{3} \theta_{4}+\theta_{4}^2\right)-352 \theta_{3} \theta_{34}^3 \mu_{34}^3-8 \theta_{3} \theta_{34}^2 \mu_{34}^2 (7 \theta_{3}-10 \theta_{4})+624 \theta_{34}^4 \mu_{34}^4\right)+2 \theta_{34} m_{34} \epsilon_r^3 (\theta_{3} (\mu_{34}
-1)+\theta_{4} \mu_{34})^2 \left(4 \theta_{3} \theta_{34} \mu_{34}+5 \theta_{3} (\theta_{4}-\theta_{3})+12 \theta_{34}^2 \mu_{34}^2\right)+(\theta_{3} \epsilon_r-2 \theta_{34} \mu_{34} \epsilon_r)^4\right)
   \end{dmath*}
   
   \begin{dmath*}
    +4 \theta_{34}^5 m_{34}^3 e^{\frac{\epsilon_r}{2 \theta_{34} m_{34}}} \left(600 \theta_{3}^2 \theta_{34} \mu_{34}^2 m_{34} \Omega^{(7,2)}_{\epsilon_r}-200 \theta_{3}^2 \theta_{34} \mu_{34}^2 m_{34} \Omega^{(9,2)}_{\epsilon_r}+10 \theta_{3}^2 \theta_{34} \mu_{34}^2 m_{34} \Omega^{(11,2)}_{\epsilon_r}-2800 \theta_{3}^2 \theta_{34} \mu_{34} m_{34} \Omega^{(7,2)}_{\epsilon_r}+560 \theta_{3}^2 \theta_{34} \mu_{34} m_{34} \Omega^{(9,2)}_{\epsilon_r}-20 \theta_{3}^2 \theta_{34} \mu_{34} m_{34} \Omega^{(11,2)}_{\epsilon_r}+3000 \theta_{3}^2 \theta_{34} m_{34} \Omega^{(7,2)}_{\epsilon_r}+180 \theta_{3}^2 \theta_{34} m_{34} \Omega^{(7,4)}_{\epsilon_r}-360 \theta_{3}^2 \theta_{34} m_{34} \Omega^{(9,2)}_{\epsilon_r}-60 \theta_{3}^2 \theta_{34} m_{34} \Omega^{(9,4)}_{\epsilon_r}+10 \theta_{3}^2 \theta_{34} m_{34} \Omega^{(11,2)}_{\epsilon_r}-400 \Omega^{(5,2)}_{\epsilon_r} \left(3 \theta_{3} (4 \theta_{3} \theta_{34} m_{34}-3 \theta_{34} \theta_{4} m_{34}+5 \theta_{3} \epsilon_r-2 \theta_{4} \epsilon_r)-28 \theta_{3} 
\theta_{34} \mu_{34} \epsilon_r+12 \theta_{34}^2 \mu_{34}^2 \epsilon_r\right)+1200 \theta_{3} \theta_{34} \theta_{4} \mu_{34}^2 m_{34} \Omega^{(7,2)}_{\epsilon_r}-400 \theta_{3} \theta_{34} \theta_{4} \mu_{34}^2 m_{34} \Omega^{(9,2)}_{\epsilon_r}+20 \theta_{3} \theta_{34} \theta_{4} \mu_{34}^2 m_{34} \Omega^{(11,2)}_{\epsilon_r}-2800 \theta_{3} \theta_{34} \theta_{4} \mu_{34} m_{34} \Omega^{(7,2)}_{\epsilon_r}+560 \theta_{3} \theta_{34} \theta_{4} \mu_{34} m_{34} \Omega^{(9,2)}_{\epsilon_r}-20 \theta_{3} \theta_{34} \theta_{4} \mu_{34} m_{34} \Omega^{(11,2)}_{\epsilon_r}-1200 \theta_{3} \theta_{34} \theta_{4} m_{34} \Omega^{(7,2)}_{\epsilon_r}+360 \theta_{3} \theta_{34} \theta_{4} m_{34} \Omega^{(7,4)}_{\epsilon_r}+60 \theta_{3} \theta_{34} \theta_{4} m_{34} \Omega^{(9,2)}_{\epsilon_r}-120 \theta_{3} \theta_{34} \theta_{4} m_{34} \Omega^{(9,4)}_{\epsilon_r}+12 \theta_{34}^3 m_{34} \Omega^{(11,4)}_{\epsilon_r}+600 \theta_{34} \theta_{4}^2 \mu_{34}^2 m_{34} \Omega^{(7,2)}_{\epsilon_r}-200 \theta_{34} \theta_{4}
^2 \mu_{34}^2 m_{34} \Omega^{(9,2)}_{\epsilon_r}+10 \theta_{34} \theta_{4}^2 \mu_{34}^2 m_{34} \Omega^{(11,2)}_{\epsilon_r}+180 \theta_{34} \theta_{4}^2 m_{34} \Omega^{(7,4)}_{\epsilon_r}-60 \theta_{34} \theta_{4}^2 m_{34} \Omega^{(9,4)}_{\epsilon_r}+400 \theta_{3}^2 \mu_{34}^2 \epsilon_r \Omega^{(7,2)}_{\epsilon_r}-20 \theta_{3}^2 \mu_{34}^2 \epsilon_r \Omega^{(9,2)}_{\epsilon_r}-1120 \theta_{3}^2 \mu_{34} \epsilon_r \Omega^{(7,2)}_{\epsilon_r}+40 \theta_{3}^2 \mu_{34} \epsilon_r \Omega^{(9,2)}_{\epsilon_r}-360 \theta_{3}^2 \epsilon_r \Omega^{(5,4)}_{\epsilon_r}+720 \theta_{3}^2 \epsilon_r \Omega^{(7,2)}_{\epsilon_r}+120 \theta_{3}^2 \epsilon_r \Omega^{(7,4)}_{\epsilon_r}-20 \theta_{3}^2 \epsilon_r \Omega^{(9,2)}_{\epsilon_r}-6 \theta_{3}^2 \epsilon_r \Omega^{(9,4)}_{\epsilon_r}+800 \theta_{3} \theta_{4} \mu_{34}^2 \epsilon_r \Omega^{(7,2)}_{\epsilon_r}-40 \theta_{3} \theta_{4} \mu_{34}^2 \epsilon_r \Omega^{(9,2)}_{\epsilon_r}-1120 \theta_{3} \theta_{4} \mu_{34} \epsilon_r \Omega^{(7,2)}_{\epsilon_r}+40 \
theta_{3} \theta_{4} \mu_{34} \epsilon_r \Omega^{(9,2)}_{\epsilon_r}+2400 \theta_{3} \epsilon_r (4 \theta_{3}-3 \theta_{4}) \Omega^{(3,2)}_{\epsilon_r}-720 \theta_{3} \theta_{4} \epsilon_r \Omega^{(5,4)}_{\epsilon_r}-120 \theta_{3} \theta_{4} \epsilon_r \Omega^{(7,2)}_{\epsilon_r}+240 \theta_{3} \theta_{4} \epsilon_r \Omega^{(7,4)}_{\epsilon_r}-12 \theta_{3} \theta_{4} \epsilon_r \Omega^{(9,4)}_{\epsilon_r}+400 \theta_{4}^2 \mu_{34}^2 \epsilon_r \Omega^{(7,2)}_{\epsilon_r}-20 \theta_{4}^2 \mu_{34}^2 \epsilon_r \Omega^{(9,2)}_{\epsilon_r}-360 \theta_{4}^2 \epsilon_r \Omega^{(5,4)}_{\epsilon_r}+120 \theta_{4}^2 \epsilon_r \Omega^{(7,4)}_{\epsilon_r}-6 \theta_{4}^2 \epsilon_r \Omega^{(9,4)}_{\epsilon_r}\right)\bigg)\\\\
   \end{dmath*}

\end{dgroup*}

\subsubsection{Component-2}
\begin{dgroup*}
 \begin{dmath*}
  \chi^{(2)} = 
 \end{dmath*}
  
  \begin{dmath*}
-\beta_f \theta_{2}^2 \mu_{21} \left(\theta_{12}^2 m_{12}^2 \left(15 \theta_{1}^2+40 \theta_{1} \theta_{2}+24 \theta_{2}^2\right)+2 \theta_{12} \theta_{2} m_{12} \epsilon_f (5 \theta_{1}+4 \theta_{2})+\theta_{2}^2 \epsilon_f^2\right)  
 \end{dmath*}
 
 \begin{dmath*}
  +\frac{\beta_r}{96 \mu_{21}^3}\bigg(4\theta_{34}^3 \mu_{12}^2 m_{34} e^{\frac{\epsilon_r}{2 \theta_{34} m_{34}}} \left(20 \mu_{21}^2 \Omega^{(5,2)}_{\epsilon_r} \left(3 \theta_{3} \theta_{34} \theta_{4} m_{34}-\epsilon_r (\theta_{4}-2 \theta_{34} \mu_{34})^2\right)+12 \theta_{34}^3 \mu_{12}^2 m_{34} \Omega^{(7,4)}_{\epsilon_r}+10 \theta_{34} \mu_{21}^2 m_{34} \Omega^{(7,2)}_{\epsilon_r} (\theta_{4}-2 \theta_{34} \mu_{34})^2-120 \theta_{3} \theta_{4} \mu_{21}^2 \epsilon_r \Omega^{(3,2)}_{\epsilon_r}-24 \theta_{34}^2 \mu_{12}^2 \epsilon_r \Omega^{(5,4)}_{\epsilon_r}\right)+96 \mu_{21}^4 \left(\theta_{34}^2 m_{34}^2 \left(\theta_{4}^2 \left(15 \theta_{3}^2+40 \theta_{3} \theta_{4}+24 \theta_{4}^2\right)+32 \theta_{34}^2 \theta_{4} \mu_{34}^2 (5 \theta_{3}+18 \theta_{4})-32 \theta_{34} \theta_{4}^2 \mu_{34} (5 \theta_{3}+6 \theta_{4})+384 \theta_{34}^4 \mu_{34}^4-768 \theta_{34}^3 \theta_{4} \mu_{34}^3\right)+2 \theta_{34} m_{34} \epsilon_r (\theta_{3} \mu_{34}+\theta_{4} (\mu_{34}-1))^2 \left(\theta_{4} (5 \
theta_{3}+4 \theta_{4})+16 \theta_{34}^2 \mu_{34}^2-16 \theta_{34} \theta_{4} \mu_{34}\right)+\epsilon_r^2 (\theta_{3} \mu_{34}+\theta_{4} (\mu_{34}-1))^4\right)\bigg)\\\\
 \end{dmath*}

 \begin{dmath*}
   \varepsilon^{(2)}_{1} = 
  \end{dmath*}
  
  \begin{dmath*}
   -\frac{\mu_{21}}{480} \beta^{(3)}_f \theta_{1}^2 \theta_{2}^2\bigg(-15 \theta_{1}^2 \theta_{12}^4 m_{12}^4+30 \theta_{1} \theta_{12}^3 m_{12}^3 \epsilon_f (\theta_{2}-\theta_{1})+15 \theta_{12}^2 m_{12}^2 \epsilon_f^2 \left(\theta_{1}^2-4 \theta_{1} \theta_{2}+\theta_{2}^2\right)+10 \theta_{12} \theta_{2} m_{12} \epsilon_f^3 (\theta_{1}-\theta_{2})+\theta_{2}^2 \epsilon_f^4\bigg)\\\\
  \end{dmath*}

  \begin{dmath*}
   \varepsilon^{(2)}_{2} = 
  \end{dmath*}
  
  \begin{dmath*}
   -\frac{\mu_{21}\beta^{(3)}_f \theta_{2}^2}{480}\bigg(3 \theta_{12}^4 m_{12}^4 \left(40 \theta_{1}^4+280 \theta_{1}^3 \theta_{2}+539 \theta_{1}^2 \theta_{2}^2+424 \theta_{1} \theta_{2}^3+120 \theta_{2}^4\right)+6 \theta_{12}^3 \theta_{2} m_{12}^3 \epsilon_f (12 \theta_{1}+7 \theta_{2}) \left(5 \theta_{1}^2+7 \theta_{1} \theta_{2}+4 \theta_{2}^2\right)+3 \theta_{12}^2 \theta_{2}^2 m_{12}^2 \epsilon_f^2 \left(61 \theta_{1}^2+44 \theta_{1} \theta_{2}+13 \theta_{2}^2\right)+2 \theta_{12} \theta_{2}^3 m_{12} \epsilon_f^3 (13 \theta_{1}+3 \theta_{2})+\theta_{2}^4 \epsilon_f^4\bigg)\\\\
  \end{dmath*}

  \begin{dmath*}
   \varepsilon^{(2)}_{3} = 
  \end{dmath*}
  
  \begin{dmath*}
   \frac{\beta^{(3)}_r \theta_{3}^2}{184320 \mu_{21}^3}\bigg(384 \mu_{21}^4 \left(\theta_{34}^4 m_{34}^4 \left(-15 \theta_{3}^2 \theta_{4}^2+32 \theta_{34}^2 \theta_{4} \mu_{34}^2 (15 \theta_{3}+2 \theta_{4})+32 \theta_{34} \theta_{4}^2 \mu_{34} (5 \theta_{3}+2 \theta_{4})+5760 \theta_{34}^4 \mu_{34}^4-2304 \theta_{34}^3 \theta_{4} \mu_{34}^3\right)+2 \theta_{34}^3 m_{34}^3 \epsilon_r \left(4 \theta_{34}^2 \theta_{4} \mu_{34}^2 (35 \theta_{3}-12 \theta_{4})+4 \theta_{34} \theta_{4}^2 \mu_{34} (25 \theta_{3}+4 \theta_{4})+15 \theta_{3} \theta_{4}^2 (\theta_{4}-\theta_{3})+1344 \theta_{34}^4 \mu_{34}^4-640 \theta_{34}^3 \theta_{4} \mu_{34}^3\right)+\theta_{34}^2 m_{34}^2 \epsilon_r^2 \left(15 \theta_{4}^2 \left(\theta_{3}^2-4 \theta_{3} \theta_{4}+\theta_{4}^2\right)+8 \theta_{34}^2 \theta_{4} \mu_{34}^2 (10 \theta_{3}-7 \theta_{4})+8 \theta_{34} \theta_{4}^2 \mu_{34} (10 \theta_{3}+\theta_{4})+624 \theta_{34}^4 \mu_{34}^4-352 \theta_{34}^3 \theta_{4} \mu_{34}^3\right)+2 \theta_{34} m_{34} \epsilon_r^3 (\theta_
{4}-2 \theta_{34} \mu_{34})^2 \left(5 \theta_{4} (\theta_{3}-\theta_{4})+12 \theta_{34}^2 \mu_{34}^2+4 \theta_{34} \theta_{4} \mu_{34}\right)+\epsilon_r^4 (\theta_{4}-2 \theta_{34} \mu_{34})^4\right)
  \end{dmath*}

  \begin{dmath*}
   4 \theta_{34}^5 \mu_{12}^2 m_{34}^3 e^{\frac{\epsilon_r}{2 \theta_{34} m_{34}}} \left(180 \theta_{3}^2 \theta_{34} \mu_{12}^2 m_{34} \Omega^{(7,4)}_{\epsilon_r}-60 \theta_{3}^2 \theta_{34} \mu_{12}^2 m_{34} \Omega^{(9,4)}_{\epsilon_r}-200 \theta_{3}^2 \theta_{34} \mu_{21}^2 \mu_{34}^2 m_{34} \Omega^{(9,2)}_{\epsilon_r}+10 \theta_{3}^2 \theta_{34} \mu_{21}^2 \mu_{34}^2 m_{34} \Omega^{(11,2)}_{\epsilon_r}+40 \mu_{21}^2 \Omega^{(7,2)}_{\epsilon_r} \left(3 \theta_{4} (-10 \theta_{3} \theta_{34} m_{34}+25 \theta_{34} \theta_{4} m_{34}+\theta_{3} (-\epsilon_r)+6 \theta_{4} \epsilon_r)+20 \theta_{34}^2 \mu_{34}^2 (3 \theta_{34} m_{34}+2 \epsilon_r)-28 \theta_{34} \theta_{4} \mu_{34} (5 \theta_{34} m_{34}+2 \epsilon_r)\right)+360 \theta_{3} \theta_{34} \theta_{4} \mu_{12}^2 m_{34} \Omega^{(7,4)}_{\epsilon_r}-120 \theta_{3} \theta_{34} \theta_{4} \mu_{12}^2 m_{34} \Omega^{(9,4)}_{\epsilon_r}-400 \theta_{3} \theta_{34} \theta_{4} \mu_{21}^2 \mu_{34}^2 m_{34} \Omega^{(9,2)}_{\epsilon_r}+20 \theta_{3} \theta_{34} \
theta_{4} \mu_{21}^2 \mu_{34}^2 m_{34} \Omega^{(11,2)}_{\epsilon_r}+560 \theta_{3} \theta_{34} \theta_{4} \mu_{21}^2 \mu_{34} m_{34} \Omega^{(9,2)}_{\epsilon_r}-20 \theta_{3} \theta_{34} \theta_{4} \mu_{21}^2 \mu_{34} m_{34} \Omega^{(11,2)}_{\epsilon_r}+60 \theta_{3} \theta_{34} \theta_{4} \mu_{21}^2 m_{34} \Omega^{(9,2)}_{\epsilon_r}+12 \theta_{34}^3 \mu_{12}^2 m_{34} \Omega^{(11,4)}_{\epsilon_r}+180 \theta_{34} \theta_{4}^2 \mu_{12}^2 m_{34} \Omega^{(7,4)}_{\epsilon_r}-60 \theta_{34} \theta_{4}^2 \mu_{12}^2 m_{34} \Omega^{(9,4)}_{\epsilon_r}-200 \theta_{34} \theta_{4}^2 \mu_{21}^2 \mu_{34}^2 m_{34} \Omega^{(9,2)}_{\epsilon_r}+10 \theta_{34} \theta_{4}^2 \mu_{21}^2 \mu_{34}^2 m_{34} \Omega^{(11,2)}_{\epsilon_r}+560 \theta_{34} \theta_{4}^2 \mu_{21}^2 \mu_{34} m_{34} \Omega^{(9,2)}_{\epsilon_r}-20 \theta_{34} \theta_{4}^2 \mu_{21}^2 \mu_{34} m_{34} \Omega^{(11,2)}_{\epsilon_r}-360 \theta_{34} \theta_{4}^2 \mu_{21}^2 m_{34} \Omega^{(9,2)}_{\epsilon_r}+10 \theta_{34} \theta_{4}^2 \mu_{21}^2 m_{34} \Omega^{(11,
2)}_{\epsilon_r}+120 \theta_{3}^2 \mu_{12}^2 \epsilon_r \Omega^{(7,4)}_{\epsilon_r}-6 \theta_{3}^2 \mu_{12}^2 \epsilon_r \Omega^{(9,4)}_{\epsilon_r}-20 \theta_{3}^2 \mu_{21}^2 \mu_{34}^2 \epsilon_r \Omega^{(9,2)}_{\epsilon_r}+240 \theta_{3} \theta_{4} \mu_{12}^2 \epsilon_r \Omega^{(7,4)}_{\epsilon_r}-12 \theta_{3} \theta_{4} \mu_{12}^2 \epsilon_r \Omega^{(9,4)}_{\epsilon_r}-40 \theta_{3} \theta_{4} \mu_{21}^2 \mu_{34}^2 \epsilon_r \Omega^{(9,2)}_{\epsilon_r}+40 \theta_{3} \theta_{4} \mu_{21}^2 \mu_{34} \epsilon_r \Omega^{(9,2)}_{\epsilon_r}-1440 \theta_{34}^2 \mu_{12}^2 \epsilon_r \Omega^{(5,4)}_{\epsilon_r}+120 \theta_{4}^2 \mu_{12}^2 \epsilon_r \Omega^{(7,4)}_{\epsilon_r}-6 \theta_{4}^2 \mu_{12}^2 \epsilon_r \Omega^{(9,4)}_{\epsilon_r}-20 \theta_{4}^2 \mu_{21}^2 \mu_{34}^2 \epsilon_r \Omega^{(9,2)}_{\epsilon_r}+40 \theta_{4}^2 \mu_{21}^2 \mu_{34} \epsilon_r \Omega^{(9,2)}_{\epsilon_r}-20 \theta_{4}^2 \mu_{21}^2 \epsilon_r \Omega^{(9,2)}_{\epsilon_r}\right)-9600 \theta_{34}^5 \theta_{4} \mu_{12}^2 \mu_{21}
^2 m_{34}^3 \epsilon_r (3 \theta_{3}-4 \theta_{4}) \Omega^{(3,2)}_{\epsilon_r} e^{\frac{\epsilon_r}{2 \theta_{34} m_{34}}}+1600 \theta_{34}^5 \mu_{12}^2 \mu_{21}^2 m_{34}^3 \Omega^{(5,2)}_{\epsilon_r} e^{\frac{\epsilon_r}{2 \theta_{34} m_{34}}} \left(3 \theta_{4} (3 \theta_{3} \theta_{34} m_{34}-4 \theta_{34} \theta_{4} m_{34}+2 \theta_{3} \epsilon_r-5 \theta_{4} \epsilon_r)-12 \theta_{34}^2 \mu_{34}^2 \epsilon_r+28 \theta_{34} \theta_{4} \mu_{34} \epsilon_r\right)\bigg)
  \end{dmath*}

  \begin{dmath*}
   \varepsilon^{(2)}_{4} = 
  \end{dmath*}

  \begin{dmath*}
   \frac{\beta^{(3)}_r \theta_{4}^2}{184320 \mu_{21}^3}\bigg(384 \mu_{21}^4 \left(\theta_{34}^4 m_{34}^4 \left(8 \theta_{3}^4 (\mu_{34}-1) (\mu_{34} (9 \mu_{34} (5 \mu_{34}-11)+65)-15)+8 \theta_{3}^3 \theta_{4} (\mu_{34}-1) (\mu_{34} (36 \mu_{34} (5 \mu_{34}-12)+343)-105)+\theta_{3}^2 \theta_{4}^2 (16 \mu_{34} (\mu_{34} (27 \mu_{34} (5 \mu_{34}-18)+664)-415)+1617)+24 \theta_{3} \theta_{4}^3 (\mu_{34}-1)^2 (12 \mu_{34} (5 \mu_{34}-9)+53)+360 \theta_{4}^4 (\mu_{34}-1)^4\right)+2 \theta_{34}^3 m_{34}^3 \epsilon_r \left(4 \theta_{34}^2 \mu_{34}^2 \left(252 \theta_{3}^2+803 \theta_{3} \theta_{4}+504 \theta_{4}^2\right)+3 \theta_{4} (12 \theta_{3}+7 \theta_{4}) \left(5 \theta_{3}^2+7 \theta_{3} \theta_{4}+4 \theta_{4}^2\right)-4 \theta_{34} \mu_{34} \left(40 \theta_{3}^3+312 \theta_{3}^2 \theta_{4}+419 \theta_{3} \theta_{4}^2+168 \theta_{4}^3\right)-128 \theta_{34}^3 \mu_{34}^3 (16 \theta_{3}+21 \theta_{4})+1344 \theta_{34}^4 \mu_{34}^4\right)+\theta_{34}^2 m_{34}^2 \epsilon_r^2 \left(8 \theta_{34}^2 \mu_{34}^2 \
\left(44 \theta_{3}^2+178 \theta_{3} \theta_{4}+117 \theta_{4}^2\right)-8 \theta_{34} \theta_{4} \mu_{34} \left(64 \theta_{3}^2+94 \theta_{3} \theta_{4}+39 \theta_{4}^2\right)+3 \theta_{4}^2 \left(61 \theta_{3}^2+44 \theta_{3} \theta_{4}+13 \theta_{4}^2\right)-32 \theta_{34}^3 \mu_{34}^3 (28 \theta_{3}+39 \theta_{4})+624 \theta_{34}^4 \mu_{34}^4\right)+2 \theta_{34} m_{34} \epsilon_r^3 (\theta_{4}-2 \theta_{34} \mu_{34})^2 \left(-4 \theta_{34} \mu_{34} (4 \theta_{3}+3 \theta_{4})+\theta_{4} (13 \theta_{3}+3 \theta_{4})+12 \theta_{34}^2 \mu_{34}^2\right)+\epsilon_r^4 (\theta_{4}-2 \theta_{34} \mu_{34})^4\right)
  \end{dmath*}
  
  \begin{dmath*}
   -1600 \theta_{34}^5 \mu_{12}^2 \mu_{21}^2 m_{34}^3 \Omega^{(5,2)}_{\epsilon_r} e^{\frac{\epsilon_r}{2 \theta_{34} m_{34}}} \left(3 \theta_{3} \theta_{34} m_{34} (4 \theta_{3}-3 \theta_{4})+\epsilon_r \left(\theta_{3}^2 (\mu_{34}+2) (3 \mu_{34}+2)+2 \theta_{3} \theta_{4} \left(3 \mu_{34}^2+\mu_{34}-7\right)+3 \theta_{4}^2 (\mu_{34}-1)^2\right)\right)+4 \theta_{34}^5 \mu_{12}^2 m_{34}^3 e^{\frac{\epsilon_r}{2 \theta_{34} m_{34}}} \left(40 \mu_{21}^2 \Omega^{(7,2)}_{\epsilon_r} \left(5 \theta_{34} m_{34} \left(\theta_{3}^2 (\mu_{34}+2) (3 \mu_{34}+2)+2 \theta_{3} \theta_{4} \left(3 \mu_{34}^2+\mu_{34}-7\right)+3 \theta_{4}^2 (\mu_{34}-1)^2\right)+\epsilon_r \left(8 \theta_{34} \mu_{34} (2 \theta_{3}-5 \theta_{4})+\theta_{4} (10 \theta_{4}-11 \theta_{3})+40 \theta_{34}^2 \mu_{34}^2\right)\right)+180 \theta_{3}^2 \theta_{34} \mu_{12}^2 m_{34} \Omega^{(7,4)}_{\epsilon_r}-60 \theta_{3}^2 \theta_{34} \mu_{12}^2 m_{34} \Omega^{(9,4)}_{\epsilon_r}-200 \theta_{3}^2 \theta_{34} \mu_{21}^2 \mu_{34}^2 m_{34} \Omega^{(9,
2)}_{\epsilon_r}+10 \theta_{3}^2 \theta_{34} \mu_{21}^2 \mu_{34}^2 m_{34} \Omega^{(11,2)}_{\epsilon_r}-160 \theta_{3}^2 \theta_{34} \mu_{21}^2 \mu_{34} m_{34} \Omega^{(9,2)}_{\epsilon_r}+360 \theta_{3} \theta_{34} \theta_{4} \mu_{12}^2 m_{34} \Omega^{(7,4)}_{\epsilon_r}-120 \theta_{3} \theta_{34} \theta_{4} \mu_{12}^2 m_{34} \Omega^{(9,4)}_{\epsilon_r}-400 \theta_{3} \theta_{34} \theta_{4} \mu_{21}^2 \mu_{34}^2 m_{34} \Omega^{(9,2)}_{\epsilon_r}+20 \theta_{3} \theta_{34} \theta_{4} \mu_{21}^2 \mu_{34}^2 m_{34} \Omega^{(11,2)}_{\epsilon_r}+240 \theta_{3} \theta_{34} \theta_{4} \mu_{21}^2 \mu_{34} m_{34} \Omega^{(9,2)}_{\epsilon_r}-20 \theta_{3} \theta_{34} \theta_{4} \mu_{21}^2 \mu_{34} m_{34} \Omega^{(11,2)}_{\epsilon_r}+220 \theta_{3} \theta_{34} \theta_{4} \mu_{21}^2 m_{34} \Omega^{(9,2)}_{\epsilon_r}+12 \theta_{34}^3 \mu_{12}^2 m_{34} \Omega^{(11,4)}_{\epsilon_r}+180 \theta_{34} \theta_{4}^2 \mu_{12}^2 m_{34} \Omega^{(7,4)}_{\epsilon_r}-60 \theta_{34} \theta_{4}^2 \mu_{12}^2 m_{34} \Omega^{(9,4)}_{\
epsilon_r}-200 \theta_{34} \theta_{4}^2 \mu_{21}^2 \mu_{34}^2 m_{34} \Omega^{(9,2)}_{\epsilon_r}+10 \theta_{34} \theta_{4}^2 \mu_{21}^2 \mu_{34}^2 m_{34} \Omega^{(11,2)}_{\epsilon_r}+400 \theta_{34} \theta_{4}^2 \mu_{21}^2 \mu_{34} m_{34} \Omega^{(9,2)}_{\epsilon_r}-20 \theta_{34} \theta_{4}^2 \mu_{21}^2 \mu_{34} m_{34} \Omega^{(11,2)}_{\epsilon_r}-200 \theta_{34} \theta_{4}^2 \mu_{21}^2 m_{34} \Omega^{(9,2)}_{\epsilon_r}+10 \theta_{34} \theta_{4}^2 \mu_{21}^2 m_{34} \Omega^{(11,2)}_{\epsilon_r}+120 \theta_{3}^2 \mu_{12}^2 \epsilon_r \Omega^{(7,4)}_{\epsilon_r}-6 \theta_{3}^2 \mu_{12}^2 \epsilon_r \Omega^{(9,4)}_{\epsilon_r}-20 \theta_{3}^2 \mu_{21}^2 \mu_{34}^2 \epsilon_r \Omega^{(9,2)}_{\epsilon_r}+240 \theta_{3} \theta_{4} \mu_{12}^2 \epsilon_r \Omega^{(7,4)}_{\epsilon_r}-12 \theta_{3} \theta_{4} \mu_{12}^2 \epsilon_r \Omega^{(9,4)}_{\epsilon_r}-40 \theta_{3} \theta_{4} \mu_{21}^2 \mu_{34}^2 \epsilon_r \Omega^{(9,2)}_{\epsilon_r}+40 \theta_{3} \theta_{4} \mu_{21}^2 \mu_{34} \epsilon_r \Omega^{(9,2)}_{\
epsilon_r}-1440 \theta_{34}^2 \mu_{12}^2 \epsilon_r \Omega^{(5,4)}_{\epsilon_r}+120 \theta_{4}^2 \mu_{12}^2 \epsilon_r \Omega^{(7,4)}_{\epsilon_r}-6 \theta_{4}^2 \mu_{12}^2 \epsilon_r \Omega^{(9,4)}_{\epsilon_r}-20 \theta_{4}^2 \mu_{21}^2 \mu_{34}^2 \epsilon_r \Omega^{(9,2)}_{\epsilon_r}+40 \theta_{4}^2 \mu_{21}^2 \mu_{34} \epsilon_r \Omega^{(9,2)}_{\epsilon_r}-20 \theta_{4}^2 \mu_{21}^2 \epsilon_r \Omega^{(9,2)}_{\epsilon_r}\right)++9600 \theta_{3} \theta_{34}^5 \mu_{12}^2 \mu_{21}^2 m_{34}^3 \epsilon_r (4 \theta_{3}-3 \theta_{4}) \Omega^{(3,2)}_{\epsilon_r} e^{\frac{\epsilon_r}{2 \theta_{34} m_{34}}}\bigg)\\\\\
  \end{dmath*}

\end{dgroup*}

\subsubsection{Component-3}
\begin{dgroup*}
 \begin{dmath*}
  \chi^{(3)} = 
 \end{dmath*}

 \begin{dmath*}
  -\beta_r \theta_{3}^2 \mu_{34} \left(\theta_{34}^2 m_{34}^2 \left(24 \theta_{3}^2+40 \theta_{3} \theta_{4}+15 \theta_{4}^2\right)+2 \theta_{3} \theta_{34} m_{34} \epsilon_r (4 \theta_{3}+5 \theta_{4})+\theta_{3}^2 \epsilon_r^2\right)
 \end{dmath*}

 \begin{dmath*}
  +\frac{\beta_f \mu_{34}}{96}\bigg(4 \theta_{12}^3 m_{12} e^{\frac{\epsilon_f}{2 \theta_{12} m_{12}}} \left(10 \theta_{12} m_{12} \Omega^{(7,2)}_{\epsilon_f} (\theta_{1} (\mu_{21}-1)+\theta_{2} \mu_{21})^2-20 \Omega^{(5,2)}_{\epsilon_f} \left(\epsilon_f (\theta_{1}-2 \theta_{12} \mu_{21})^2-3 \theta_{1} \theta_{12} \theta_{2} m_{12}\right)+12 \theta_{12}^3 m_{12} \Omega^{(7,4)}_{\epsilon_f}-120 \theta_{1} \theta_{2} \epsilon_f \Omega^{(3,2)}_{\epsilon_f}-24 \theta_{12}^2 \epsilon_f \Omega^{(5,4)}_{\epsilon_f}\right)+96 \left(\theta_{12}^2 m_{12}^2 \left(-32 \theta_{1}^2 \theta_{12} \mu_{21} (6 \theta_{1}+5 \theta_{2})+\theta_{1}^2 \left(24 \theta_{1}^2+40 \theta_{1} \theta_{2}+15 \theta_{2}^2\right)-768 \theta_{1} \theta_{12}^3 \mu_{21}^3+32 \theta_{1} \theta_{12}^2 \mu_{21}^2 (18 \theta_{1}+5 \theta_{2})+384 \theta_{12}^4 \mu_{21}^4\right)+2 \theta_{12} m_{12} \epsilon_f (\theta_{1} (\mu_{21}-1)+\theta_{2} \mu_{21})^2 \left(-16 \theta_{1} \theta_{12} \mu_{21}+\theta_{1} (4 \theta_{1}+5 \theta_{2})+16 \
theta_{12}^2 \mu_{21}^2\right)+\epsilon_f^2 (\theta_{1} (\mu_{21}-1)+\theta_{2} \mu_{21})^4\right)\bigg)\\\\
 \end{dmath*}

  \begin{dmath*}
   \varepsilon^{(3)}_{1} = 
  \end{dmath*}
  
  \begin{dmath*}
   \frac{\beta^{(3)}_f \theta_{1}^2 \mu_{34}}{184320}\bigg(384 \left(\theta_{12}^4 m_{12}^4 \left(360 \theta_{1}^4 (\mu_{21}-1)^4+24 \theta_{1}^3 \theta_{2} (\mu_{21}-1)^2 (12 \mu_{21} (5 \mu_{21}-9)+53)+\theta_{1}^2 \theta_{2}^2 (16 \mu_{21} (\mu_{21} (27 \mu_{21} (5 \mu_{21}-18)+664)-415)+1617)+8 \theta_{1} \theta_{2}^3 (\mu_{21}-1) (\mu_{21} (36 \mu_{21} (5 \mu_{21}-12)+343)-105)+8 \theta_{2}^4 (\mu_{21}-1) (\mu_{21} (9 \mu_{21} (5 \mu_{21}-11)+65)-15)\right)+2 \theta_{12}^3 m_{12}^3 \epsilon_f \left(4 \theta_{12}^2 \mu_{21}^2 \left(504 \theta_{1}^2+803 \theta_{1} \theta_{2}+252 \theta_{2}^2\right)+3 \theta_{1} (7 \theta_{1}+12 \theta_{2}) \left(4 \theta_{1}^2+7 \theta_{1} \theta_{2}+5 \theta_{2}^2\right)-4 \theta_{12} \mu_{21} \left(168 \theta_{1}^3+419 \theta_{1}^2 \theta_{2}+312 \theta_{1} \theta_{2}^2+40 \theta_{2}^3\right)-128 \theta_{12}^3 \mu_{21}^3 (21 \theta_{1}+16 \theta_{2})+1344 \theta_{12}^4 \mu_{21}^4\right)+\theta_{12}^2 m_{12}^2 \epsilon_f^2 \left(8 \theta_{12}^2 \mu_{21}^2 \left(117 \
theta_{1}^2+178 \theta_{1} \theta_{2}+44 \theta_{2}^2\right)-8 \theta_{1} \theta_{12} \mu_{21} \left(39 \theta_{1}^2+94 \theta_{1} \theta_{2}+64 \theta_{2}^2\right)+3 \theta_{1}^2 \left(13 \theta_{1}^2+44 \theta_{1} \theta_{2}+61 \theta_{2}^2\right)-32 \theta_{12}^3 \mu_{21}^3 (39 \theta_{1}+28 \theta_{2})+624 \theta_{12}^4 \mu_{21}^4\right)+2 \theta_{12} m_{12} \epsilon_f^3 (\theta_{1} (\mu_{21}-1)+\theta_{2} \mu_{21})^2 \left(-4 \theta_{12} \mu_{21} (3 \theta_{1}+4 \theta_{2})+\theta_{1} (3 \theta_{1}+13 \theta_{2})+12 \theta_{12}^2 \mu_{21}^2\right)+(\theta_{1} \epsilon_f-2 \theta_{12} \mu_{21} \epsilon_f)^4\right)
  \end{dmath*}
  
  \begin{dmath*}
   +4 \theta_{12}^5 m_{12}^3 e^{\frac{\epsilon_f}{2 \theta_{12} m_{12}}} \left(-400 \Omega^{(5,2)}_{\epsilon_f} \left(3 \theta_{12} \theta_{2} m_{12} (4 \theta_{2}-3 \theta_{1})+\epsilon_f \left(3 \theta_{1}^2 (\mu_{21}-1)^2+2 \theta_{1} \theta_{2} \left(3 \mu_{21}^2+\mu_{21}-7\right)+\theta_{2}^2 (\mu_{21}+2) (3 \mu_{21}+2)\right)\right)+600 \theta_{1}^2 \theta_{12} \mu_{21}^2 m_{12} \Omega^{(7,2)}_{\epsilon_f}-200 \theta_{1}^2 \theta_{12} \mu_{21}^2 m_{12} \Omega^{(9,2)}_{\epsilon_f}+10 \theta_{1}^2 \theta_{12} \mu_{21}^2 m_{12} \Omega^{(11,2)}_{\epsilon_f}-1200 \theta_{1}^2 \theta_{12} \mu_{21} m_{12} \Omega^{(7,2)}_{\epsilon_f}+400 \theta_{1}^2 \theta_{12} \mu_{21} m_{12} \Omega^{(9,2)}_{\epsilon_f}-20 \theta_{1}^2 \theta_{12} \mu_{21} m_{12} \Omega^{(11,2)}_{\epsilon_f}+600 \theta_{1}^2 \theta_{12} m_{12} \Omega^{(7,2)}_{\epsilon_f}+180 \theta_{1}^2 \theta_{12} m_{12} \Omega^{(7,4)}_{\epsilon_f}-200 \theta_{1}^2 \theta_{12} m_{12} \Omega^{(9,2)}_{\epsilon_f}-60 \theta_{1}^2 \theta_{12} m_{12} \Omega^{(9,
4)}_{\epsilon_f}+10 \theta_{1}^2 \theta_{12} m_{12} \Omega^{(11,2)}_{\epsilon_f}+1200 \theta_{1} \theta_{12} \theta_{2} \mu_{21}^2 m_{12} \Omega^{(7,2)}_{\epsilon_f}-400 \theta_{1} \theta_{12} \theta_{2} \mu_{21}^2 m_{12} \Omega^{(9,2)}_{\epsilon_f}+20 \theta_{1} \theta_{12} \theta_{2} \mu_{21}^2 m_{12} \Omega^{(11,2)}_{\epsilon_f}+400 \theta_{1} \theta_{12} \theta_{2} \mu_{21} m_{12} \Omega^{(7,2)}_{\epsilon_f}+240 \theta_{1} \theta_{12} \theta_{2} \mu_{21} m_{12} \Omega^{(9,2)}_{\epsilon_f}-20 \theta_{1} \theta_{12} \theta_{2} \mu_{21} m_{12} \Omega^{(11,2)}_{\epsilon_f}-2800 \theta_{1} \theta_{12} \theta_{2} m_{12} \Omega^{(7,2)}_{\epsilon_f}+360 \theta_{1} \theta_{12} \theta_{2} m_{12} \Omega^{(7,4)}_{\epsilon_f}+220 \theta_{1} \theta_{12} \theta_{2} m_{12} \Omega^{(9,2)}_{\epsilon_f}-120 \theta_{1} \theta_{12} \theta_{2} m_{12} \Omega^{(9,4)}_{\epsilon_f}+12 \theta_{12}^3 m_{12} \Omega^{(11,4)}_{\epsilon_f}+600 \theta_{12} \theta_{2}^2 \mu_{21}^2 m_{12} \Omega^{(7,2)}_{\epsilon_f}-200 \theta_{12} \theta_
{2}^2 \mu_{21}^2 m_{12} \Omega^{(9,2)}_{\epsilon_f}+10 \theta_{12} \theta_{2}^2 \mu_{21}^2 m_{12} \Omega^{(11,2)}_{\epsilon_f}+1600 \theta_{12} \theta_{2}^2 \mu_{21} m_{12} \Omega^{(7,2)}_{\epsilon_f}-160 \theta_{12} \theta_{2}^2 \mu_{21} m_{12} \Omega^{(9,2)}_{\epsilon_f}+800 \theta_{12} \theta_{2}^2 m_{12} \Omega^{(7,2)}_{\epsilon_f}+180 \theta_{12} \theta_{2}^2 m_{12} \Omega^{(7,4)}_{\epsilon_f}-60 \theta_{12} \theta_{2}^2 m_{12} \Omega^{(9,4)}_{\epsilon_f}+400 \theta_{1}^2 \mu_{21}^2 \epsilon_f \Omega^{(7,2)}_{\epsilon_f}-20 \theta_{1}^2 \mu_{21}^2 \epsilon_f \Omega^{(9,2)}_{\epsilon_f}-800 \theta_{1}^2 \mu_{21} \epsilon_f \Omega^{(7,2)}_{\epsilon_f}+40 \theta_{1}^2 \mu_{21} \epsilon_f \Omega^{(9,2)}_{\epsilon_f}-360 \theta_{1}^2 \epsilon_f \Omega^{(5,4)}_{\epsilon_f}+400 \theta_{1}^2 \epsilon_f \Omega^{(7,2)}_{\epsilon_f}+120 \theta_{1}^2 \epsilon_f \Omega^{(7,4)}_{\epsilon_f}-20 \theta_{1}^2 \epsilon_f \Omega^{(9,2)}_{\epsilon_f}-6 \theta_{1}^2 \epsilon_f \Omega^{(9,4)}_{\epsilon_f}+800 \theta_{1} \
theta_{2} \mu_{21}^2 \epsilon_f \Omega^{(7,2)}_{\epsilon_f}-40 \theta_{1} \theta_{2} \mu_{21}^2 \epsilon_f \Omega^{(9,2)}_{\epsilon_f}-480 \theta_{1} \theta_{2} \mu_{21} \epsilon_f \Omega^{(7,2)}_{\epsilon_f}+40 \theta_{1} \theta_{2} \mu_{21} \epsilon_f \Omega^{(9,2)}_{\epsilon_f}+2400 \theta_{2} \epsilon_f (4 \theta_{2}-3 \theta_{1}) \Omega^{(3,2)}_{\epsilon_f}-720 \theta_{1} \theta_{2} \epsilon_f \Omega^{(5,4)}_{\epsilon_f}-440 \theta_{1} \theta_{2} \epsilon_f \Omega^{(7,2)}_{\epsilon_f}+240 \theta_{1} \theta_{2} \epsilon_f \Omega^{(7,4)}_{\epsilon_f}-12 \theta_{1} \theta_{2} \epsilon_f \Omega^{(9,4)}_{\epsilon_f}+400 \theta_{2}^2 \mu_{21}^2 \epsilon_f \Omega^{(7,2)}_{\epsilon_f}-20 \theta_{2}^2 \mu_{21}^2 \epsilon_f \Omega^{(9,2)}_{\epsilon_f}+320 \theta_{2}^2 \mu_{21} \epsilon_f \Omega^{(7,2)}_{\epsilon_f}-360 \theta_{2}^2 \epsilon_f \Omega^{(5,4)}_{\epsilon_f}+120 \theta_{2}^2 \epsilon_f \Omega^{(7,4)}_{\epsilon_f}-6 \theta_{2}^2 \epsilon_f \Omega^{(9,4)}_{\epsilon_f}\right)\bigg)\\\\
  \end{dmath*}

  \begin{dmath*}
   \varepsilon^{(3)}_{2} = 
  \end{dmath*}
  
  \begin{dmath*}
   \frac{\beta^{(3)}_f \theta_{2}^2 \mu_{34}}{184320}\bigg(384 \left(\theta_{12}^4 m_{12}^4 \left(32 \theta_{1}^2 \theta_{12} \mu_{21} (2 \theta_{1}+5 \theta_{2})-15 \theta_{1}^2 \theta_{2}^2-2304 \theta_{1} \theta_{12}^3 \mu_{21}^3+32 \theta_{1} \theta_{12}^2 \mu_{21}^2 (2 \theta_{1}+15 \theta_{2})+5760 \theta_{12}^4 \mu_{21}^4\right)+2 \theta_{12}^3 m_{12}^3 \epsilon_f \left(4 \theta_{1}^2 \theta_{12} \mu_{21} (4 \theta_{1}+25 \theta_{2})+15 \theta_{1}^2 \theta_{2} (\theta_{1}-\theta_{2})-640 \theta_{1} \theta_{12}^3 \mu_{21}^3-4 \theta_{1} \theta_{12}^2 \mu_{21}^2 (12 \theta_{1}-35 \theta_{2})+1344 \theta_{12}^4 \mu_{21}^4\right)+\theta_{12}^2 m_{12}^2 \epsilon_f^2 \left(8 \theta_{1}^2 \theta_{12} \mu_{21} (\theta_{1}+10 \theta_{2})+15 \theta_{1}^2 \left(\theta_{1}^2-4 \theta_{1} \theta_{2}+\theta_{2}^2\right)-352 \theta_{1} \theta_{12}^3 \mu_{21}^3-8 \theta_{1} \theta_{12}^2 \mu_{21}^2 (7 \theta_{1}-10 \theta_{2})+624 \theta_{12}^4 \mu_{21}^4\right)+2 \theta_{12} m_{12} \epsilon_f^3 (\theta_{1} (\mu_{21}-
1)+\theta_{2} \mu_{21})^2 \left(4 \theta_{1} \theta_{12} \mu_{21}+5 \theta_{1} (\theta_{2}-\theta_{1})+12 \theta_{12}^2 \mu_{21}^2\right)+(\theta_{1} \epsilon_f-2 \theta_{12} \mu_{21} \epsilon_f)^4\right)
  \end{dmath*}

  \begin{dmath*}
   +4 \theta_{12}^5 m_{12}^3 e^{\frac{\epsilon_f}{2 \theta_{12} m_{12}}} \left(600 \theta_{1}^2 \theta_{12} \mu_{21}^2 m_{12} \Omega^{(7,2)}_{\epsilon_f}-200 \theta_{1}^2 \theta_{12} \mu_{21}^2 m_{12} \Omega^{(9,2)}_{\epsilon_f}+10 \theta_{1}^2 \theta_{12} \mu_{21}^2 m_{12} \Omega^{(11,2)}_{\epsilon_f}-2800 \theta_{1}^2 \theta_{12} \mu_{21} m_{12} \Omega^{(7,2)}_{\epsilon_f}+560 \theta_{1}^2 \theta_{12} \mu_{21} m_{12} \Omega^{(9,2)}_{\epsilon_f}-20 \theta_{1}^2 \theta_{12} \mu_{21} m_{12} \Omega^{(11,2)}_{\epsilon_f}+3000 \theta_{1}^2 \theta_{12} m_{12} \Omega^{(7,2)}_{\epsilon_f}+180 \theta_{1}^2 \theta_{12} m_{12} \Omega^{(7,4)}_{\epsilon_f}-360 \theta_{1}^2 \theta_{12} m_{12} \Omega^{(9,2)}_{\epsilon_f}-60 \theta_{1}^2 \theta_{12} m_{12} \Omega^{(9,4)}_{\epsilon_f}+10 \theta_{1}^2 \theta_{12} m_{12} \Omega^{(11,2)}_{\epsilon_f}-400 \Omega^{(5,2)}_{\epsilon_f} \left(3 \theta_{1} (4 \theta_{1} \theta_{12} m_{12}-3 \theta_{12} \theta_{2} m_{12}+5 \theta_{1} \epsilon_f-2 \theta_{2} \epsilon_f)-28 \theta_{1} \
theta_{12} \mu_{21} \epsilon_f+12 \theta_{12}^2 \mu_{21}^2 \epsilon_f\right)+1200 \theta_{1} \theta_{12} \theta_{2} \mu_{21}^2 m_{12} \Omega^{(7,2)}_{\epsilon_f}-400 \theta_{1} \theta_{12} \theta_{2} \mu_{21}^2 m_{12} \Omega^{(9,2)}_{\epsilon_f}+20 \theta_{1} \theta_{12} \theta_{2} \mu_{21}^2 m_{12} \Omega^{(11,2)}_{\epsilon_f}-2800 \theta_{1} \theta_{12} \theta_{2} \mu_{21} m_{12} \Omega^{(7,2)}_{\epsilon_f}+560 \theta_{1} \theta_{12} \theta_{2} \mu_{21} m_{12} \Omega^{(9,2)}_{\epsilon_f}-20 \theta_{1} \theta_{12} \theta_{2} \mu_{21} m_{12} \Omega^{(11,2)}_{\epsilon_f}-1200 \theta_{1} \theta_{12} \theta_{2} m_{12} \Omega^{(7,2)}_{\epsilon_f}+360 \theta_{1} \theta_{12} \theta_{2} m_{12} \Omega^{(7,4)}_{\epsilon_f}+60 \theta_{1} \theta_{12} \theta_{2} m_{12} \Omega^{(9,2)}_{\epsilon_f}-120 \theta_{1} \theta_{12} \theta_{2} m_{12} \Omega^{(9,4)}_{\epsilon_f}+12 \theta_{12}^3 m_{12} \Omega^{(11,4)}_{\epsilon_f}+600 \theta_{12} \theta_{2}^2 \mu_{21}^2 m_{12} \Omega^{(7,2)}_{\epsilon_f}-200 \theta_{12} \theta_{2}
^2 \mu_{21}^2 m_{12} \Omega^{(9,2)}_{\epsilon_f}+10 \theta_{12} \theta_{2}^2 \mu_{21}^2 m_{12} \Omega^{(11,2)}_{\epsilon_f}+180 \theta_{12} \theta_{2}^2 m_{12} \Omega^{(7,4)}_{\epsilon_f}-60 \theta_{12} \theta_{2}^2 m_{12} \Omega^{(9,4)}_{\epsilon_f}+400 \theta_{1}^2 \mu_{21}^2 \epsilon_f \Omega^{(7,2)}_{\epsilon_f}-20 \theta_{1}^2 \mu_{21}^2 \epsilon_f \Omega^{(9,2)}_{\epsilon_f}-1120 \theta_{1}^2 \mu_{21} \epsilon_f \Omega^{(7,2)}_{\epsilon_f}+40 \theta_{1}^2 \mu_{21} \epsilon_f \Omega^{(9,2)}_{\epsilon_f}-360 \theta_{1}^2 \epsilon_f \Omega^{(5,4)}_{\epsilon_f}+720 \theta_{1}^2 \epsilon_f \Omega^{(7,2)}_{\epsilon_f}+120 \theta_{1}^2 \epsilon_f \Omega^{(7,4)}_{\epsilon_f}-20 \theta_{1}^2 \epsilon_f \Omega^{(9,2)}_{\epsilon_f}-6 \theta_{1}^2 \epsilon_f \Omega^{(9,4)}_{\epsilon_f}+800 \theta_{1} \theta_{2} \mu_{21}^2 \epsilon_f \Omega^{(7,2)}_{\epsilon_f}-40 \theta_{1} \theta_{2} \mu_{21}^2 \epsilon_f \Omega^{(9,2)}_{\epsilon_f}-1120 \theta_{1} \theta_{2} \mu_{21} \epsilon_f \Omega^{(7,2)}_{\epsilon_f}+40 \
theta_{1} \theta_{2} \mu_{21} \epsilon_f \Omega^{(9,2)}_{\epsilon_f}+2400 \theta_{1} \epsilon_f (4 \theta_{1}-3 \theta_{2}) \Omega^{(3,2)}_{\epsilon_f}-720 \theta_{1} \theta_{2} \epsilon_f \Omega^{(5,4)}_{\epsilon_f}-120 \theta_{1} \theta_{2} \epsilon_f \Omega^{(7,2)}_{\epsilon_f}+240 \theta_{1} \theta_{2} \epsilon_f \Omega^{(7,4)}_{\epsilon_f}-12 \theta_{1} \theta_{2} \epsilon_f \Omega^{(9,4)}_{\epsilon_f}+400 \theta_{2}^2 \mu_{21}^2 \epsilon_f \Omega^{(7,2)}_{\epsilon_f}-20 \theta_{2}^2 \mu_{21}^2 \epsilon_f \Omega^{(9,2)}_{\epsilon_f}-360 \theta_{2}^2 \epsilon_f \Omega^{(5,4)}_{\epsilon_f}+120 \theta_{2}^2 \epsilon_f \Omega^{(7,4)}_{\epsilon_f}-6 \theta_{2}^2 \epsilon_f \Omega^{(9,4)}_{\epsilon_f}\right)\bigg)\\\\
  \end{dmath*}

  \begin{dmath*}
   \varepsilon^{(3)}_{3} = 
  \end{dmath*}
  
  \begin{dmath*}
   -\frac{\mu_{34}\beta^{(3)}_r \theta_{3}^2}{480}\bigg(3 \theta_{34}^4 m_{34}^4 \left(120 \theta_{3}^4+424 \theta_{3}^3 \theta_{4}+539 \theta_{3}^2 \theta_{4}^2+280 \theta_{3} \theta_{4}^3+40 \theta_{4}^4\right)+6 \theta_{3} \theta_{34}^3 m_{34}^3 \epsilon_r (7 \theta_{3}+12 \theta_{4}) \left(4 \theta_{3}^2+7 \theta_{3} \theta_{4}+5 \theta_{4}^2\right)+3 \theta_{3}^2 \theta_{34}^2 m_{34}^2 \epsilon_r^2 \left(13 \theta_{3}^2+44 \theta_{3} \theta_{4}+61 \theta_{4}^2\right)+2 \theta_{3}^3 \theta_{34} m_{34} \epsilon_r^3 (3 \theta_{3}+13 \theta_{4})+\theta_{3}^4 \epsilon_r^4\bigg)\\\\
  \end{dmath*}

  \begin{dmath*}
   \varepsilon^{(3)}_{4} = 
  \end{dmath*}
  
  \begin{dmath*}
   -\frac{\mu_{34}}{480} \beta^{(3)}_r \theta_{3}^2 \theta_{4}^2\bigg(-15 \theta_{34}^4 \theta_{4}^2 m_{34}^4+30 \theta_{34}^3 \theta_{4} m_{34}^3 \epsilon_r (\theta_{3}-\theta_{4})+15 \theta_{34}^2 m_{34}^2 \epsilon_r^2 \left(\theta_{3}^2-4 \theta_{3} \theta_{4}+\theta_{4}^2\right)+10 \theta_{3} \theta_{34} m_{34} \epsilon_r^3 (\theta_{4}-\theta_{3})+\theta_{3}^2 \epsilon_r^4\bigg)\\\\
  \end{dmath*}

\end{dgroup*}

\subsubsection{Component-4}
\begin{dgroup*}
 \begin{dmath*}
  \chi^{(4)} = 
 \end{dmath*}

  \begin{dmath*}
   \frac{\beta_f}{96 \mu_{43}^3}\bigg(4 \theta_{12}^3 \mu_{34}^2 m_{12} e^{\frac{\epsilon_f}{2 \theta_{12} m_{12}}} \left(20 \mu_{43}^2 \Omega^{(5,2)}_{\epsilon_f} \left(3 \theta_{1} \theta_{12} \theta_{2} m_{12}-\epsilon_f (\theta_{2}-2 \theta_{12} \mu_{12})^2\right)+12 \theta_{12}^3 \mu_{34}^2 m_{12} \Omega^{(7,4)}_{\epsilon_f}+10 \theta_{12} \mu_{43}^2 m_{12} \Omega^{(7,2)}_{\epsilon_f} (\theta_{2}-2 \theta_{12} \mu_{12})^2-120 \theta_{1} \theta_{2} \mu_{43}^2 \epsilon_f \Omega^{(3,2)}_{\epsilon_f}-24 \theta_{12}^2 \mu_{34}^2 \epsilon_f \Omega^{(5,4)}_{\epsilon_f}\right)+96 \mu_{43}^4 \left(\theta_{12}^2 m_{12}^2 \left(\theta_{2}^2 \left(15 \theta_{1}^2+40 \theta_{1} \theta_{2}+24 \theta_{2}^2\right)+32 \theta_{12}^2 \theta_{2} \mu_{12}^2 (5 \theta_{1}+18 \theta_{2})-32 \theta_{12} \theta_{2}^2 \mu_{12} (5 \theta_{1}+6 \theta_{2})+384 \theta_{12}^4 \mu_{12}^4-768 \theta_{12}^3 \theta_{2} \mu_{12}^3\right)+2 \theta_{12} m_{12} \epsilon_f (\theta_{1} \mu_{12}+\theta_{2} (\mu_{12}-1))^2 \left(\theta_{2} (5 \
theta_{1}+4 \theta_{2})+16 \theta_{12}^2 \mu_{12}^2-16 \theta_{12} \theta_{2} \mu_{12}\right)+\epsilon_f^2 (\theta_{1} \mu_{12}+\theta_{2} (\mu_{12}-1))^4\right)\bigg)
  \end{dmath*}

   \begin{dmath*}
    -\beta_r \theta_{4}^2 \mu_{43} \left(\theta_{34}^2 m_{34}^2 \left(15 \theta_{3}^2+40 \theta_{3} \theta_{4}+24 \theta_{4}^2\right)+2 \theta_{34} \theta_{4} m_{34} \epsilon_r (5 \theta_{3}+4 \theta_{4})+\theta_{4}^2 \epsilon_r^2\right)\\\\
   \end{dmath*}

  \begin{dmath*}
   \varepsilon^{(4)}_{1} = 
  \end{dmath*}
  
  \begin{dmath*}
   \frac{\beta^{(3)}_f \theta_{1}^2}{184320 \mu_{43}^3} \bigg(384 \mu_{43}^4 \left(\theta_{12}^4 m_{12}^4 \left(-15 \theta_{1}^2 \theta_{2}^2+32 \theta_{12}^2 \theta_{2} \mu_{12}^2 (15 \theta_{1}+2 \theta_{2})+32 \theta_{12} \theta_{2}^2 \mu_{12} (5 \theta_{1}+2 \theta_{2})+5760 \theta_{12}^4 \mu_{12}^4-2304 \theta_{12}^3 \theta_{2} \mu_{12}^3\right)+2 \theta_{12}^3 m_{12}^3 \epsilon_f \left(4 \theta_{12}^2 \theta_{2} \mu_{12}^2 (35 \theta_{1}-12 \theta_{2})+4 \theta_{12} \theta_{2}^2 \mu_{12} (25 \theta_{1}+4 \theta_{2})+15 \theta_{1} \theta_{2}^2 (\theta_{2}-\theta_{1})+1344 \theta_{12}^4 \mu_{12}^4-640 \theta_{12}^3 \theta_{2} \mu_{12}^3\right)+\theta_{12}^2 m_{12}^2 \epsilon_f^2 \left(15 \theta_{2}^2 \left(\theta_{1}^2-4 \theta_{1} \theta_{2}+\theta_{2}^2\right)+8 \theta_{12}^2 \theta_{2} \mu_{12}^2 (10 \theta_{1}-7 \theta_{2})+8 \theta_{12} \theta_{2}^2 \mu_{12} (10 \theta_{1}+\theta_{2})+624 \theta_{12}^4 \mu_{12}^4-352 \theta_{12}^3 \theta_{2} \mu_{12}^3\right)+2 \theta_{12} m_{12} \epsilon_f^3 (\
theta_{2}-2 \theta_{12} \mu_{12})^2 \left(5 \theta_{2} (\theta_{1}-\theta_{2})+12 \theta_{12}^2 \mu_{12}^2+4 \theta_{12} \theta_{2} \mu_{12}\right)+\epsilon_f^4 (\theta_{2}-2 \theta_{12} \mu_{12})^4\right)
  \end{dmath*}

  \begin{dmath*}
   +4 \theta_{12}^5 \mu_{34}^2 m_{12}^3 e^{\frac{\epsilon_f}{2 \theta_{12} m_{12}}} \left(-200 \theta_{1}^2 \theta_{12} \mu_{12}^2 \mu_{43}^2 m_{12} \Omega^{(9,2)}_{\epsilon_f}+10 \theta_{1}^2 \theta_{12} \mu_{12}^2 \mu_{43}^2 m_{12} \Omega^{(11,2)}_{\epsilon_f}+180 \theta_{1}^2 \theta_{12} \mu_{34}^2 m_{12} \Omega^{(7,4)}_{\epsilon_f}-60 \theta_{1}^2 \theta_{12} \mu_{34}^2 m_{12} \Omega^{(9,4)}_{\epsilon_f}+40 \mu_{43}^2 \Omega^{(7,2)}_{\epsilon_f} \left(3 \theta_{2} (-10 \theta_{1} \theta_{12} m_{12}+25 \theta_{12} \theta_{2} m_{12}+\theta_{1} (-\epsilon_f)+6 \theta_{2} \epsilon_f)+20 \theta_{12}^2 \mu_{12}^2 (3 \theta_{12} m_{12}+2 \epsilon_f)-28 \theta_{12} \theta_{2} \mu_{12} (5 \theta_{12} m_{12}+2 \epsilon_f)\right)-400 \theta_{1} \theta_{12} \theta_{2} \mu_{12}^2 \mu_{43}^2 m_{12} \Omega^{(9,2)}_{\epsilon_f}+20 \theta_{1} \theta_{12} \theta_{2} \mu_{12}^2 \mu_{43}^2 m_{12} \Omega^{(11,2)}_{\epsilon_f}+560 \theta_{1} \theta_{12} \theta_{2} \mu_{12} \mu_{43}^2 m_{12} \Omega^{(9,2)}_{\epsilon_f}-20 \
theta_{1} \theta_{12} \theta_{2} \mu_{12} \mu_{43}^2 m_{12} \Omega^{(11,2)}_{\epsilon_f}+360 \theta_{1} \theta_{12} \theta_{2} \mu_{34}^2 m_{12} \Omega^{(7,4)}_{\epsilon_f}-120 \theta_{1} \theta_{12} \theta_{2} \mu_{34}^2 m_{12} \Omega^{(9,4)}_{\epsilon_f}+60 \theta_{1} \theta_{12} \theta_{2} \mu_{43}^2 m_{12} \Omega^{(9,2)}_{\epsilon_f}+12 \theta_{12}^3 \mu_{34}^2 m_{12} \Omega^{(11,4)}_{\epsilon_f}-200 \theta_{12} \theta_{2}^2 \mu_{12}^2 \mu_{43}^2 m_{12} \Omega^{(9,2)}_{\epsilon_f}+10 \theta_{12} \theta_{2}^2 \mu_{12}^2 \mu_{43}^2 m_{12} \Omega^{(11,2)}_{\epsilon_f}+560 \theta_{12} \theta_{2}^2 \mu_{12} \mu_{43}^2 m_{12} \Omega^{(9,2)}_{\epsilon_f}-20 \theta_{12} \theta_{2}^2 \mu_{12} \mu_{43}^2 m_{12} \Omega^{(11,2)}_{\epsilon_f}+180 \theta_{12} \theta_{2}^2 \mu_{34}^2 m_{12} \Omega^{(7,4)}_{\epsilon_f}-60 \theta_{12} \theta_{2}^2 \mu_{34}^2 m_{12} \Omega^{(9,4)}_{\epsilon_f}-360 \theta_{12} \theta_{2}^2 \mu_{43}^2 m_{12} \Omega^{(9,2)}_{\epsilon_f}+10 \theta_{12} \theta_{2}^2 \mu_{43}^2 m_{12} \Omega^{(
11,2)}_{\epsilon_f}-20 \theta_{1}^2 \mu_{12}^2 \mu_{43}^2 \epsilon_f \Omega^{(9,2)}_{\epsilon_f}+120 \theta_{1}^2 \mu_{34}^2 \epsilon_f \Omega^{(7,4)}_{\epsilon_f}-6 \theta_{1}^2 \mu_{34}^2 \epsilon_f \Omega^{(9,4)}_{\epsilon_f}-40 \theta_{1} \theta_{2} \mu_{12}^2 \mu_{43}^2 \epsilon_f \Omega^{(9,2)}_{\epsilon_f}+40 \theta_{1} \theta_{2} \mu_{12} \mu_{43}^2 \epsilon_f \Omega^{(9,2)}_{\epsilon_f}+240 \theta_{1} \theta_{2} \mu_{34}^2 \epsilon_f \Omega^{(7,4)}_{\epsilon_f}-12 \theta_{1} \theta_{2} \mu_{34}^2 \epsilon_f \Omega^{(9,4)}_{\epsilon_f}-1440 \theta_{12}^2 \mu_{34}^2 \epsilon_f \Omega^{(5,4)}_{\epsilon_f}-20 \theta_{2}^2 \mu_{12}^2 \mu_{43}^2 \epsilon_f \Omega^{(9,2)}_{\epsilon_f}+40 \theta_{2}^2 \mu_{12} \mu_{43}^2 \epsilon_f \Omega^{(9,2)}_{\epsilon_f}+120 \theta_{2}^2 \mu_{34}^2 \epsilon_f \Omega^{(7,4)}_{\epsilon_f}-6 \theta_{2}^2 \mu_{34}^2 \epsilon_f \Omega^{(9,4)}_{\epsilon_f}-20 \theta_{2}^2 \mu_{43}^2 \epsilon_f \Omega^{(9,2)}_{\epsilon_f}\right)-9600 \theta_{12}^5 \theta_{2} \mu_{34}^2 \mu_{
43}^2 m_{12}^3 \epsilon_f (3 \theta_{1}-4 \theta_{2}) \Omega^{(3,2)}_{\epsilon_f} e^{\frac{\epsilon_f}{2 \theta_{12} m_{12}}}+1600 \theta_{12}^5 \mu_{34}^2 \mu_{43}^2 m_{12}^3 \Omega^{(5,2)}_{\epsilon_f} e^{\frac{\epsilon_f}{2 \theta_{12} m_{12}}} \left(3 \theta_{2} (3 \theta_{1} \theta_{12} m_{12}-4 \theta_{12} \theta_{2} m_{12}+2 \theta_{1} \epsilon_f-5 \theta_{2} \epsilon_f)-12 \theta_{12}^2 \mu_{12}^2 \epsilon_f+28 \theta_{12} \theta_{2} \mu_{12} \epsilon_f\right)\bigg)\\\\
  \end{dmath*}

  \begin{dmath*}
   \varepsilon^{(4)}_{2} = 
  \end{dmath*}
  
  \begin{dmath*}
   \frac{\beta^{(3)}_f \theta_{2}^2}{184320 \mu_{43}^3}\bigg(384 \mu_{43}^4 \left(\theta_{12}^4 m_{12}^4 \left(8 \theta_{1}^4 (\mu_{12}-1) (\mu_{12} (9 \mu_{12} (5 \mu_{12}-11)+65)-15)+8 \theta_{1}^3 \theta_{2} (\mu_{12}-1) (\mu_{12} (36 \mu_{12} (5 \mu_{12}-12)+343)-105)+\theta_{1}^2 \theta_{2}^2 (16 \mu_{12} (\mu_{12} (27 \mu_{12} (5 \mu_{12}-18)+664)-415)+1617)+24 \theta_{1} \theta_{2}^3 (\mu_{12}-1)^2 (12 \mu_{12} (5 \mu_{12}-9)+53)+360 \theta_{2}^4 (\mu_{12}-1)^4\right)+2 \theta_{12}^3 m_{12}^3 \epsilon_f \left(4 \theta_{12}^2 \mu_{12}^2 \left(252 \theta_{1}^2+803 \theta_{1} \theta_{2}+504 \theta_{2}^2\right)+3 \theta_{2} (12 \theta_{1}+7 \theta_{2}) \left(5 \theta_{1}^2+7 \theta_{1} \theta_{2}+4 \theta_{2}^2\right)-4 \theta_{12} \mu_{12} \left(40 \theta_{1}^3+312 \theta_{1}^2 \theta_{2}+419 \theta_{1} \theta_{2}^2+168 \theta_{2}^3\right)-128 \theta_{12}^3 \mu_{12}^3 (16 \theta_{1}+21 \theta_{2})+1344 \theta_{12}^4 \mu_{12}^4\right)+\theta_{12}^2 m_{12}^2 \epsilon_f^2 \left(8 \theta_{12}^2 \mu_{12}^2 \
\left(44 \theta_{1}^2+178 \theta_{1} \theta_{2}+117 \theta_{2}^2\right)-8 \theta_{12} \theta_{2} \mu_{12} \left(64 \theta_{1}^2+94 \theta_{1} \theta_{2}+39 \theta_{2}^2\right)+3 \theta_{2}^2 \left(61 \theta_{1}^2+44 \theta_{1} \theta_{2}+13 \theta_{2}^2\right)-32 \theta_{12}^3 \mu_{12}^3 (28 \theta_{1}+39 \theta_{2})+624 \theta_{12}^4 \mu_{12}^4\right)+2 \theta_{12} m_{12} \epsilon_f^3 (\theta_{2}-2 \theta_{12} \mu_{12})^2 \left(-4 \theta_{12} \mu_{12} (4 \theta_{1}+3 \theta_{2})+\theta_{2} (13 \theta_{1}+3 \theta_{2})+12 \theta_{12}^2 \mu_{12}^2\right)+\epsilon_f^4 (\theta_{2}-2 \theta_{12} \mu_{12})^4\right)
  \end{dmath*}

  \begin{dmath*}
   -1600 \theta_{12}^5 \mu_{34}^2 \mu_{43}^2 m_{12}^3 \Omega^{(5,2)}_{\epsilon_f} e^{\frac{\epsilon_f}{2 \theta_{12} m_{12}}} \left(3 \theta_{1} \theta_{12} m_{12} (4 \theta_{1}-3 \theta_{2})+\epsilon_f \left(\theta_{1}^2 (\mu_{12}+2) (3 \mu_{12}+2)+2 \theta_{1} \theta_{2} \left(3 \mu_{12}^2+\mu_{12}-7\right)+3 \theta_{2}^2 (\mu_{12}-1)^2\right)\right)+4 \theta_{12}^5 \mu_{34}^2 m_{12}^3 e^{\frac{\epsilon_f}{2 \theta_{12} m_{12}}} \left(40 \mu_{43}^2 \Omega^{(7,2)}_{\epsilon_f} \left(5 \theta_{12} m_{12} \left(\theta_{1}^2 (\mu_{12}+2) (3 \mu_{12}+2)+2 \theta_{1} \theta_{2} \left(3 \mu_{12}^2+\mu_{12}-7\right)+3 \theta_{2}^2 (\mu_{12}-1)^2\right)+\epsilon_f \left(8 \theta_{12} \mu_{12} (2 \theta_{1}-5 \theta_{2})+\theta_{2} (10 \theta_{2}-11 \theta_{1})+40 \theta_{12}^2 \mu_{12}^2\right)\right)-200 \theta_{1}^2 \theta_{12} \mu_{12}^2 \mu_{43}^2 m_{12} \Omega^{(9,2)}_{\epsilon_f}+10 \theta_{1}^2 \theta_{12} \mu_{12}^2 \mu_{43}^2 m_{12} \Omega^{(11,2)}_{\epsilon_f}-160 \theta_{1}^2 \theta_{12} \mu_{12} \mu_{43}
^2 m_{12} \Omega^{(9,2)}_{\epsilon_f}+180 \theta_{1}^2 \theta_{12} \mu_{34}^2 m_{12} \Omega^{(7,4)}_{\epsilon_f}-60 \theta_{1}^2 \theta_{12} \mu_{34}^2 m_{12} \Omega^{(9,4)}_{\epsilon_f}-400 \theta_{1} \theta_{12} \theta_{2} \mu_{12}^2 \mu_{43}^2 m_{12} \Omega^{(9,2)}_{\epsilon_f}+20 \theta_{1} \theta_{12} \theta_{2} \mu_{12}^2 \mu_{43}^2 m_{12} \Omega^{(11,2)}_{\epsilon_f}+240 \theta_{1} \theta_{12} \theta_{2} \mu_{12} \mu_{43}^2 m_{12} \Omega^{(9,2)}_{\epsilon_f}-20 \theta_{1} \theta_{12} \theta_{2} \mu_{12} \mu_{43}^2 m_{12} \Omega^{(11,2)}_{\epsilon_f}+360 \theta_{1} \theta_{12} \theta_{2} \mu_{34}^2 m_{12} \Omega^{(7,4)}_{\epsilon_f}-120 \theta_{1} \theta_{12} \theta_{2} \mu_{34}^2 m_{12} \Omega^{(9,4)}_{\epsilon_f}+220 \theta_{1} \theta_{12} \theta_{2} \mu_{43}^2 m_{12} \Omega^{(9,2)}_{\epsilon_f}+12 \theta_{12}^3 \mu_{34}^2 m_{12} \Omega^{(11,4)}_{\epsilon_f}-200 \theta_{12} \theta_{2}^2 \mu_{12}^2 \mu_{43}^2 m_{12} \Omega^{(9,2)}_{\epsilon_f}+10 \theta_{12} \theta_{2}^2 \mu_{12}^2 \mu_{43}^2 m_{12} \
Omega^{(11,2)}_{\epsilon_f}+400 \theta_{12} \theta_{2}^2 \mu_{12} \mu_{43}^2 m_{12} \Omega^{(9,2)}_{\epsilon_f}-20 \theta_{12} \theta_{2}^2 \mu_{12} \mu_{43}^2 m_{12} \Omega^{(11,2)}_{\epsilon_f}+180 \theta_{12} \theta_{2}^2 \mu_{34}^2 m_{12} \Omega^{(7,4)}_{\epsilon_f}-60 \theta_{12} \theta_{2}^2 \mu_{34}^2 m_{12} \Omega^{(9,4)}_{\epsilon_f}-200 \theta_{12} \theta_{2}^2 \mu_{43}^2 m_{12} \Omega^{(9,2)}_{\epsilon_f}+10 \theta_{12} \theta_{2}^2 \mu_{43}^2 m_{12} \Omega^{(11,2)}_{\epsilon_f}-20 \theta_{1}^2 \mu_{12}^2 \mu_{43}^2 \epsilon_f \Omega^{(9,2)}_{\epsilon_f}+120 \theta_{1}^2 \mu_{34}^2 \epsilon_f \Omega^{(7,4)}_{\epsilon_f}-6 \theta_{1}^2 \mu_{34}^2 \epsilon_f \Omega^{(9,4)}_{\epsilon_f}-40 \theta_{1} \theta_{2} \mu_{12}^2 \mu_{43}^2 \epsilon_f \Omega^{(9,2)}_{\epsilon_f}+40 \theta_{1} \theta_{2} \mu_{12} \mu_{43}^2 \epsilon_f \Omega^{(9,2)}_{\epsilon_f}+240 \theta_{1} \theta_{2} \mu_{34}^2 \epsilon_f \Omega^{(7,4)}_{\epsilon_f}-12 \theta_{1} \theta_{2} \mu_{34}^2 \epsilon_f \Omega^{(9,4)}_{\epsilon_f}
-1440 \theta_{12}^2 \mu_{34}^2 \epsilon_f \Omega^{(5,4)}_{\epsilon_f}-20 \theta_{2}^2 \mu_{12}^2 \mu_{43}^2 \epsilon_f \Omega^{(9,2)}_{\epsilon_f}+40 \theta_{2}^2 \mu_{12} \mu_{43}^2 \epsilon_f \Omega^{(9,2)}_{\epsilon_f}+120 \theta_{2}^2 \mu_{34}^2 \epsilon_f \Omega^{(7,4)}_{\epsilon_f}-6 \theta_{2}^2 \mu_{34}^2 \epsilon_f \Omega^{(9,4)}_{\epsilon_f}-20 \theta_{2}^2 \mu_{43}^2 \epsilon_f \Omega^{(9,2)}_{\epsilon_f}\right)++9600 \theta_{1} \theta_{12}^5 \mu_{34}^2 \mu_{43}^2 m_{12}^3 \epsilon_f (4 \theta_{1}-3 \theta_{2}) \Omega^{(3,2)}_{\epsilon_f} e^{\frac{\epsilon_f}{2 \theta_{12} m_{12}}}\bigg)\\\\
  \end{dmath*}

  \begin{dmath*}
   \varepsilon^{(4)}_{3} = 
  \end{dmath*}
  
  \begin{dmath*}
   -\frac{\mu_{43}}{480} \beta^{(3)}_r \theta_{3}^2 \theta_{4}^2\big(-15 \theta_{3}^2 \theta_{34}^4 m_{34}^4+30 \theta_{3} \theta_{34}^3 m_{34}^3 \epsilon_r (\theta_{4}-\theta_{3})+15 \theta_{34}^2 m_{34}^2 \epsilon_r^2 \left(\theta_{3}^2-4 \theta_{3} \theta_{4}+\theta_{4}^2\right)+10 \theta_{34} \theta_{4} m_{34} \epsilon_r^3 (\theta_{3}-\theta_{4})+\theta_{4}^2 \epsilon_r^4\bigg)\\\\
  \end{dmath*}

  \begin{dmath*}
   \varepsilon^{(4)}_{4} = 
  \end{dmath*}

  \begin{dmath*}
   -\frac{\mu_{43}\beta^{(3)}_r \theta_{4}^2}{480}\bigg(3 \theta_{34}^4 m_{34}^4 \left(40 \theta_{3}^4+280 \theta_{3}^3 \theta_{4}+539 \theta_{3}^2 \theta_{4}^2+424 \theta_{3} \theta_{4}^3+120 \theta_{4}^4\right)+6 \theta_{34}^3 \theta_{4} m_{34}^3 \epsilon_r (12 \theta_{3}+7 \theta_{4}) \left(5 \theta_{3}^2+7 \theta_{3} \theta_{4}+4 \theta_{4}^2\right)+3 \theta_{34}^2 \theta_{4}^2 m_{34}^2 \epsilon_r^2 \left(61 \theta_{3}^2+44 \theta_{3} \theta_{4}+13 \theta_{4}^2\right)+2 \theta_{34} \theta_{4}^3 m_{34} \epsilon_r^3 (13 \theta_{3}+3 \theta_{4})+\theta_{4}^4 \epsilon_r^4\bigg)\\\\
  \end{dmath*}

\end{dgroup*}
The coefficients $\beta^{(3)}_f$ and $\beta^{(3)}_r$ are given as:
\begin{gather*}
  \beta^{(3)}_f = \left(\frac{\sqrt{\pi } d_{12}^2 M n_1 n_2 \text{sf}^2 }{\theta_{12}^{15/2} m_{12}^4}\right)\exp\left[{-\frac{\epsilon_f}{2 \theta_{12} m_{12}}}\right], \quad  \beta^{(3)}_r = \left(\frac{\sqrt{\pi } d_{34}^2 M n_3 n_4 \text{sr}^2 }{\theta_{34}^{15/2} m_{34}^4}\right) \exp\left[{-\frac{\epsilon_r}{2 \theta_{34} m_{34}}}\right] 
\end{gather*}

\end{document}